\documentclass[adp2,fleqn]{w-art}
\usepackage{times}
\usepackage{w-thm}
\usepackage[]{graphicx}
\usepackage{amssymb}
\chardef\bslash=`\\ % p. 424, TeXbook

\begin{document}
%%
%%  Most of the following commands will be completed by the publisher.
%%
%%  The copyrightyear is defined in the .clo file as the first argument
%%  of the copyrightinfo command. If the copyrightyear differs from that
%%  value it might be adjusted by the following definition:
%%
%% \renewcommand{\copyrightyear}{2002}% uncomment to change the copyrightyear.
%%
\DOIsuffix{theDOIsuffix}
%%
%% issueinfo for header and copyright line
%%
% \Volume{12}
% \Issue{1}
% \Copyrightissue{01}
% \Month{01}
% \Year{2003}
%%
%%  First and last pagenumber of the article. If the option
%%  'autolastpage' is set (default) the second argument may be left empty.
% \pagespan{1}{}
%%
%%    Dates will be filled in by the publisher. The 'reviseddate' and
%%    'dateposted' (Published online) entry may be left empty.
% \Receiveddate{15 November 1900}
%%
%% \Reviseddate{30 November 1900}
%%
% \Accepteddate{2 December 1900}
%% \Dateposted{3 December 1900}
%%
%%
 % up to three, separated by commas

%% \pretitle{Editor's Choice}

%% We have a short and a long form for the title. The short form
%% (optional argument) goes into the running head.

\title{Interacting bosons in an optical lattice}

%% Please do not enter footnotes or \inst{}-notes into the optional
%% argument of the author command. The optional argument will go into
%% the header. If there is only one address the marker \inst{x} may be
%% omitted.

%%   Information for the first author.
\author{Ch. Moseley\footnote{Corresponding author:
    e-mail: {\sf christopher.moseley@physik.uni-augsburg.de}, Phone: +40\,821\,598\,3221, Fax:
    +40\,821\,598\,3262}\inst{1}} 
    \address[\inst{1}]{Institut f\"ur Physik, Universit\"at Augsburg, Germany}

%%    Information for second author
\author{O. Fialko\inst{1}}

%%    Information for third author
\author{K. Ziegler\inst{1}}

\begin{abstract}
Several models of a strongly interacting Bose gas in an optical lattice are studied within the 
functional-integral approach. The one-dimensional Bose gas is briefly discussed. 
Then the Bose-Einstein condensate and the Mott insulator of a three-dimensional Bose gas
are described in mean-field approximation, and the corresponding phase
diagrams are evaluated. Other characteristic quantities, like the spectrum of quasiparticle
excitations and the static structure factor, are obtained from Gaussian fluctuations around
the mean-field solutions. We discuss the role of quantum and thermal fluctuations, and
determine the behavior of physical quantities in terms of density and temperature of
the Bose gas. In particular, we study the dilute limit, where the mean-field
equation becomes the Gross-Pitaevskii equation. This allows us to extend the 
Gross-Pitaevskii equation to the dense regime by introducing renormalized parameters
in the latter.
\end{abstract}
\maketitle                   

%% If there is not enough space inside the running head
%% for all authors including the title you may provide
%% the leftmark in one of the following three forms:

%% \renewcommand{\leftmark}
%% {F.\ Author: Short Title}

%% \renewcommand{\leftmark}
%% {F.\ Author and S.\ Author: Short Title}

%% \renewcommand{\leftmark}
%% {F.\ Author et al.: Short Title}

\tableofcontents  % Produces the table of contents.

\section{Introduction}

% \subsection{Definition of the condensate density}

The quantum statistics of non-interacting particles was established by
S. N. Bose in 1924 \cite{bose1}. Bose was able to deduce Planck's
radiation law on the assumption that each quantum state can be occupied by an
arbitrary number of indistinguishable photons. By applying this idea to the
quantum statistics of an ideal gas of $N_{\rm tot}$ atoms enclosed in a volume $V$, 
A. Einstein predicted the occurrence of a phase transition \cite{einstein1}: Below a 
critical temperature $T_c$, a certain fraction of atoms would ``condense'' in the
ground state of the system. This phenomenon is called Bose-Einstein condensation (BEC).

In a homogeneous ideal Bose-gas (i.e., in the absence of an external potential), the
critical temperature of the ideal Bose gas is given as 
\cite{Bhuang,Bpitaevskii,leggett1,stringari3,yukalov1}
\begin{equation}
 k_{\rm B} T_c = 
 \frac{2\pi\hbar^2}m \left(\frac{n_{\rm tot}}{\zeta\left(\frac 32\right)}\right)^{\frac 23} \; ,
 \label{Int1_Tc}
\end{equation}
where $k_{\rm B}$ is Boltzmann's constant, $\hbar$ is the reduced Planck's constant, $n_{\rm tot}=N/V$ is the particle density, 
$m$ is the mass of the particles, and 
$\zeta(x)$ is Riemann's Zeta-Function. The condensate fraction is given as
\begin{equation}
 \frac {n_0}{n_{\rm tot}} = \left\{ \begin{array}{l@{\quad}l}
 0 & \mbox{if } T>T_c \\
 1 - \left(\frac{T}{T_c}\right)^{\frac 32} & \mbox{if } T<T_c \end{array} \right. \; ,
 \label{Int1_cond-frac}
\end{equation}
where $n_0$ is the condensate density.

Historically,
the first candidate for a possible realization of Bose-Einstein condensation was
superfluid $^4$He, discovered by P. L. Kapitza in 1934 below $T_c=2.2\rm K$.
Although superfluid Helium is far away from the ideal Bose gas considered by Einstein
because of strong interactions between the Helium atoms, the phenomena of superfluidity
and BEC are related. Superfluidity
was first explained by L. D. Landau in 1941 by an argument which is based on the idea that 
the viscosity of a fluid depends on the existence of quasiparticle excitations. Those excitations
are created by friction between the fluid and a wall of the container. 
When the fluid has a velocity
$\bf v$ relative to the wall, these excitations are relevant only if their creation 
at momentum $\bf k$ is
energetically profitable, i. e. if the excitation energy is negative \cite{Bpitaevskii}:
\begin{displaymath}
 E_{\bf k} + \hbar{\bf k\cdot v} < 0 \; .
\end{displaymath}
Here $E_{\bf k}$ is the quasiparticle spectrum.
In other words, the superfluid is destroyed by excitations if the
velocity $|{\bf v}|$ exceeds a critical value $v_c$ with
\begin{displaymath}
 v_c = {\rm min}_{\bf k} \frac{E_{\bf k}}{\hbar k} \; ,
\end{displaymath}
where the minimum is calculated over all the values of $\bf k$. If the spectrum is 
linear for small momenta, a non-zero value of $v_c$ is found. It is important to
notice that superfluidity and BEC are not identical. For instance, an ideal Bose gas
can condense, but it is not superfluid due to Landau's principle, because the excitation spectrum
is quadratic in $k$ and therefore $v_c$ is zero.
On the other hand, a weakly-interacting two-dimensional Bose gas satisfies Landau's criterion
for superfluidity, but long-range order cannot appear due to the Mermin-Wagner theorem
\cite{merminwagner,hohenberg,posazhennikova}, therefore there is no BEC.

In an interacting Bose gas of uncharged atoms, the main contribution to the
interparticle interaction comes from $s$-wave
scattering between two particles. The characteristic length scale here is the 
scattering length $a_s$. We assume $a_s$ to be positive, although 
it can also be negative in trapped Bose gases (without trapping
potential a Bose gas with negative $a_s$ is instable \cite{Bpitaevskii}).
For theoretical description, usually two-body interaction is assumed.
Approximately, the two-body interaction potential can be written in the form
of a $\delta$-potential:
\begin{equation}
 V_{\rm int}({\bf r}-{\bf r'}) \approx g \, \delta({\bf r}-{\bf r'}) \; .
 \label{Int1_interaction}
\end{equation}
Here, $g$ is the strength of the repulsive interaction 
between two bosons. It is connected to the $s$-wave scattering length by the relation
\cite{Bpitaevskii}
\begin{equation}
 g = \frac{4\pi a_s \hbar^2}m \; .
 \label{Int1_g}
\end{equation}
This approximation is justified if the $a_s$ is small compared to the thermal de Broglie
wavelength, the interparticle spacing, and the characteristic length scale of the
trapping potential \cite{leggett1}. It is possible to
tune the scattering length over a large range of values (positive as well as negative) to
reach the strongly interacting regime, where Bogoliubov theory is not applicable anymore 
\cite{Bpitaevskii,wieman2,wieman3}. These  magnetic Feshbach resonances became possible after 
the development of optical trapping as an alternative to magnetic trapping.

After the introduction of an external potential $V_{\rm ext}$, the full Hamiltonian of
the Bose system in terms of bosonic field operators is
\begin{equation}
 \hat{H} = \int {\rm d}^3r \left[ \hat\psi^+({\bf r})\left(-\frac{\hbar^2}{2m} \nabla^2 + 
 V_{\rm ext}({\bf r}) \right)\hat\psi({\bf r}) + 
 \frac g2 \, \hat\psi^+({\bf r})\hat\psi^+({\bf r})\hat\psi({\bf r})\hat\psi({\bf r}) \right]\; .
 \label{Int1_H}
\end{equation}
The ground state of this interacting many-body system is not known, therefore the condensate
density cannot be defined by the population density of the ground state like in
the ideal Bose gas. An appropriate definition for a homogeneous system is the concept of 
``off-diagonal long-range order'' which was
developed in the 1950's \cite{Bpitaevskii,leggett1,penrose1}. The condensate density is given
by the long-range behavior of the one-particle correlation function
\begin{equation}
 n_0 := \lim_{{\bf r-r'}\rightarrow\infty} \langle 
 \hat\psi^+({\bf r}) \hat\psi({\bf r'}) \rangle \; .
 \label{Int1_odlro}
\end{equation}
If the one-particle correlation function decays exponentially or algebraically, the condensate 
density is zero. An algebraic decay is found in a two-dimensional Bose gas at low temperature
and in a one-dimensional Bose gas at zero temperature \cite{Bpopov}.

\subsection{Dilute Bose gas}

When the mean distance between atoms is large compared to their scattering length, which is the
case when $n_{\rm tot}a_s^3\ll 1$, the system is said to be in the dilute regime. In this case,
the effect of interaction is small.
A consistent mean-field theory of a dilute Bose gas which is valid for low temperatures 
$T\ll T_c$ was given by N. N. Bogoliubov
in 1947 \cite{Bhuang,Bpitaevskii}. The condensed phase is described by replacing 
the bosonic field-operators by the sum of a complex condensate
order parameter $\Phi_0$ and fluctuations out of the condensate as
\begin{equation}
 \hat\psi({\bf r},t) = \Phi_0({\bf r},t) + \tilde\psi({\bf r},t) \; ,
 \label{Int1_field-operators}
\end{equation}
where the field operators $\tilde\psi$ of the fluctuations fulfill bosonic 
commutation relations.
This theory gives elementary excitations out of the condensate which have the energy spectrum
\begin{equation}
 E_{\bf k} = \sqrt{\frac{\hbar^2 k^2}{2m} \left(
 2gn_0 + \frac{\hbar^2 k^2}{2m} \right) }
  \label{Int1_bogoliubov-spectrum}
\end{equation}
where $\bf k$ is the wave vector.
It is linear for small momenta (``phonon spectrum'') and therefore satisfies 
Landau's criterion for superfluidity, 
in contrast to Einstein's non-interacting Bose gas with a quadratic energy spectrum.
An important feature of an interacting Bose gases is the ground state depletion, which
means that even at $T=0$ the condensate fraction is smaller than $1$. This is also found
in Bogoliubov theory. In a dilute Bose gas, the condensate depletion is small.

The condensate order parameter $\Phi_0$ is connected to the breaking of the global
$U(1)$ symmetry, which reflects the fact that the replacement
\begin{equation}
 \Phi_0({\bf r},t)\rightarrow e^{i\alpha}\Phi_0({\bf r},t) \; ,
 \label{Int1_global-U1}
\end{equation}
where $\alpha$ is a global phase, does not change the physics of the system.
The phase $\alpha$ can be chosen arbitrarily, but once it has been chosen, the symmetry is 
broken. This is the case in the BEC phase.
This phase $\alpha$ is responsible for the fact that the quasiparticle spectrum
in Eq. (\ref{Int1_bogoliubov-spectrum}) vanishes for ${\bf k}=0$: The Goldstone-theorem
states that the existence of a broken $U(1)$ phase symmetry leads to a gapless excitation 
spectrum \cite{goldstone}.

The order parameter is interpreted as a macroscopic wave function and can be split into
its modulus and phase:
\begin{equation}
 \Phi_0({\bf r},t) = |\Phi_0({\bf r},t)| \, e^{{\rm i}\theta({\bf r},t)} \; .
\end{equation}
The local condensate density is related to the modulus squared of the order parameter
\begin{equation}
 n_0({\bf r},t) = |\Phi_0({\bf r},t)|^2 \; ,
 \label{Int1_n0}
\end{equation}
and the gradient of its phase, $\nabla\theta({\bf r},t)$, is associated with the velocity
field of the condensed atoms.
Gross and Pitaevskii have independently derived an equation to describe the dynamics of
the order parameter, which is known as the Gross-Pitaevskii (GP)
equation
\cite{Bpitaevskii,stringari3,leggett1}:
\begin{equation}
 \left(- \frac{\hbar^2}{2m} \nabla^2 + V_{\rm ext}({\bf r}) +
 g |\Phi_0({\bf r},t)|^2 \right) \Phi_0({\bf r},t) =
 {\rm i}\hbar\frac\partial{\partial t}\Phi_0({\bf r},t) \; .
 \label{Int1_GPE}
\end{equation}
The third order term in $\Phi_0$, which is proportional to the interaction constant $g$, can be
interpreted as the coupling of the order parameter to the local particle density as given
in Eq. (\ref{Int1_n0}).
For stationary solutions of the GP equation we use the ansatz
$\Phi_0({\bf r},t)=\Phi_0({\bf r})\exp(-i\mu t/\hbar)$, where $\mu$ is the chemical potential.
The GP equation then reduces to the stationary form
\begin{equation}
 \left( - \frac{\hbar^2}{2m} \nabla^2 + V_{\rm ext}({\bf r}) - \mu +
 g |\Phi_0({\bf r})|^2 \right) \Phi_0({\bf r}) = 0 \; .
 \label{Int1_GP-stationary}
\end{equation}

\subsection{Trapped Bose gas}

The experimental realisation of a weakly interacting BEC in a magnetic trap achieved
in 1995 by E. Cornell and C. Wiemann at Boulder and W. Ketterle at MIT in vapors of
$^{87}$Rb ($a_s=5.77$nm) and $^{23}$Na ($a_s=2.75$nm). 
This became possible by a combination of evaporative cooling and
laser cooling. These systems are well described by Bogoliubov theory and the GP equation. 

For models of the trapped condensates as those realized in experiments, usually a harmonic
trap potential of the general form
\begin{equation}
 V_{\rm ext}({\bf r}) = V_{\rm tr}({\bf r}) = 
 \frac m2 (\omega_x^2 x^2 + \omega_y^2 y^2 + \omega_z^2 z^2)
 \label{Int1_harmonic-trap}
\end{equation}
is assumed. For an ideal Bose gas, the critical temperature is given as \cite{Bpitaevskii}
\begin{equation}
 k_{\rm B}T = \hbar\omega_{\rm ho} \left(\frac{N_{\rm tot}}{\zeta(3)}\right)^{\frac 13} \; ,
 \omega_{\rm ho} = \left(\omega_x\omega_y\omega_z\right)^{\frac 13} \; ,
 \label{Int1_Tc-trap}
\end{equation}
in contrast to the critical temperature of a homogeneous BEC in Eq. (\ref{Int1_Tc}).
Instead of Eq. (\ref{Int1_cond-frac}), the condensate fraction in a trapped condensate is
\begin{equation}
 \frac {n_0}{n_{\rm tot}} = \left\{ \begin{array}{l@{\quad}l}
 0 & \mbox{if } T>T_c \\
 1 - \left(\frac{T}{T_c}\right)^3 & \mbox{if } T<T_c \end{array} \right. \; .
 \label{Int1_cond-frac-trap}
\end{equation}

In rotating BECs, quantized vortices and vortex lattices have been observed, a
phenomenon which is also known in type-II superconductors and superfluid $^4$He
\cite{fetter2,tempere1}. 
Vortices are observed by absorption imaging \cite{dalibard3}. 

If the condensate is in rotational equilibrium at angular velocity $\Omega$ around the
$z$-axis, the critical angular velocity $\Omega_c$, at which the creation of a vortex occurs, 
as well as the stability and dynamics of vortex cores and vortex lattices
have, can be calculated by minimizing the free energy within the GP approach
\cite{stringari2,feder1,fetter1,dalibard1,fetter3}.

\subsection{Light scattering and structure factor}

Light scattering experiments on BECs allow the study of density fluctuations.
In so-called Bragg scattering experiments, light scattering is studied as a stimulated 
process, induced by two laser beams which illuminate the atomic sample \cite{ketterle1}.
In scattering events elementary excitations are created, and the momentum and energy transfer
is pre-determined by the angle and frequency difference between the incident beams.

The most important quantity here is the dynamic structure factor
$S({\bf q},\omega)$, which is proportional to the excitation rate 
per particle.  Here, ${\bf q}={\bf q}_{\rm f}-{\bf q}_{\rm i}$, and ${\bf q}_{\rm i}$ is the
wave vector of the incoming, ${\bf q}_{\rm f}$ is the wave vector of the reflected light beam,
and $\omega$ is the frequency difference between the two laser beams.

The dynamic structure factor describes a correlation between a density
fluctuation at time $t_0=0$ and at time $t_1=t$ and is defined as the expectation 
value \cite{Bpines}
\begin{equation}
 S({\bf q},\omega) = \frac 1{N_{\rm tot}} \int
 \left\langle \hat\rho_{\bf q}(t) \hat\rho^+_{\bf q}(0) \right\rangle \, e^{{\rm i}\omega t}
 {\rm d}t \; ,
 \label{FI2_S1}
\end{equation}
with the density operator in momentum space, which is given as
\begin{equation}
 \hat\rho^+_{\bf q} = \int {\hat n}_{\bf r} \, e^{\rm i\bf q\cdot r} \, {\rm d}^dr =
 \sum_{\bf k} \hat{a}_{\bf k+q}^+ \hat{a}_{\bf k} \; ,
 \label{FI2_density-operator}
\end{equation}
in Schr\"odinger representation and
\begin{equation}
 \hat\rho_{\bf q}(t) = e^{-{\rm i}(\hat{H}-\mu\hat{N})t/\hbar} \hat\rho_{\bf q} \,
 e^{{\rm i}(\hat{H}-\mu\hat{N})t/\hbar} \; .
 \label{FI2_density-heisenberg}
\end{equation}
in Heisenberg representation, where
and $\hat{a}_{\bf k}$,  $\hat{a}^+_{\bf k}$ fulfil bosonic commutation relations. 
Integrating over all frequencies $\omega$ one obtains the static structure factor
\begin{equation}
 S({\bf q}) = \int S({\bf q},\omega) \, {\rm d}\omega \; ,
 \label{Int1_S-static}
\end{equation}
which is equivalent to the line strength of the Bragg resonance.
The static structure factor is then given by Eq. (\ref{Int1_S-static}) as
\begin{equation}
 S({\bf q}) = \frac 1{N_{\rm tot}} \,
 \left\langle \hat\rho_{\bf q}(0) \hat\rho^+_{\bf q}(0) \right\rangle \; .
 \label{FI2_Sstatic1}
\end{equation}
In the ground state of a non-interacting condensate, the static structure factor is unity,
and in the Bogoliubov ground state, it is given as
\begin{equation}
 S({\bf q}) = \frac{\hbar^2{\bf q}^2}{2m\,E_{\bf q}} \; ,
 \label{Int1_S-bogoliubov}
\end{equation}
where $E_{\bf q}$ is the quasiparticle spectrum given in 
(\ref{Int1_bogoliubov-spectrum}). This result has been originally derived by R. Feynman 
for the static structure factor of superfluid $^4$He \cite{feynman1}, and will be
reproduced in chapter \ref{Chap_weak}.
In the regime of long wave lengths this becomes
\begin{equation}
 S({\bf q}) = \frac{\hbar|{\bf q}|}{2m c} + {\cal O}(q^2) \; ,
 \label{Int1_Feynman}
\end{equation}
where $c$ is the sound velocity.

\subsection{Optical lattices\label{Sec_optical-lattices}}

Recently, ultracold gases were superimposed by optical lattices, which are created by
standing waves of laser fields \cite{bloch2}.
There are one-, two- and three-dimensional optical lattices. The lattice potential 
of a three-dimensional cubic optical lattice created of three perpendicular laser
beams parallel to the coordinate axes, is of the general form
\begin{equation}
 V_{\rm latt}({\bf r}) = V_x \sin^2(q_0 x) + V_y \sin^2(q_0 y) + V_z \sin^2 (q_0 z) \; ,
 \label{Int1_V-latt}
\end{equation}
where the amplitudes $V_x$, $V_y$, $V_z$ are proportional to the intensity of the laser field.
Together with the harmonic trap potential given in Eq. (\ref{Int1_harmonic-trap}) the
external potential of the atoms is 
$V_{\rm ext}({\bf r})=V_{\rm tr}({\bf r})+V_{\rm latt}({\bf r})$. 

A one-dimensional Bose gas, where the movement of atoms is only possible in one
direction (e.g. the $z$-direction), can be created by tightly confining the
particle motion in two directions (the $x$- and $y$-direction) to zero point 
oscillations. This can be done by increasing the amplitude $V_x$
and $V_y$ until tunneling of atoms through the lattice wells is prohibited.
If $V_z=0$, the Bose gas is trapped in one-dimensional tubes, and if $V_z\neq 0$
but small compared to $V_x$ and $V_y$, a one-dimensional lattice is created where
atoms can only tunnel between neighboring lattice-sites in the $z$-direction
\cite{paredes2}.

The conventional model for a single-component system of bosons in an optical lattice
is the Bose-Hubbard model. Assuming a $d$-dimensional simple-cubic lattice potential with 
$q_x=q_y=q_z\equiv q$ and $V_x=V_y=V_z\equiv V_0/3$, it has the form \cite{fisher1,kampf1,jaksch1}
\begin{equation}
 \hat{H}_{\rm BH} = -\frac J{2d} \sum_{\langle {\bf r,r'}\rangle} \hat{a}^+_{\bf r} \hat{a}_{\bf r'} +
 \sum_{\bf r} V_{\bf r} \, \hat{a}^+_{\bf r} \hat{a}_{\bf r} + 
 \frac U2 \sum_{\bf r} \hat{a}^+_{\bf r} \hat{a}^+_{\bf r} \hat{a}_{\bf r} \hat{a}_{\bf r} \; ,
 \label{Int1_bose-hubbard}
\end{equation}
where $\bf r,r'$ denote the discrete positions of the lattice sites, $\hat{a}$ and $\hat{a}^+$ 
are bosonic annihilation
and creation operators and the sum of the kinetic term runs over nearest
neighbor sites only. The position ${\bf r}_i$ of site $i$ is at a minimum
of the lattice potential, i.e. $V_{\rm latt}({\bf r}_i)=0$.

%---------------------------------------------------------------------------
\begin{figure}
\includegraphics{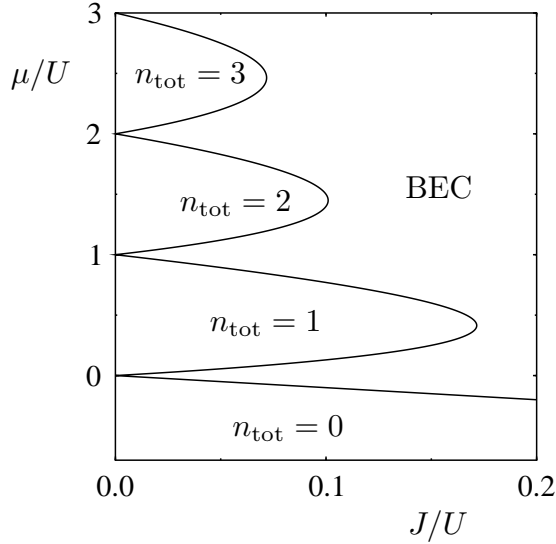}
\caption{Zero temperature phase diagram of the Bose-Hubbard model 
calculated in mean-field theory.}
\label{Fig_Int1_lobes}
\end{figure}
%---------------------------------------------------------------------------
The Bose-Hubbard model can describe a new phase, the Mott-insulator (MI). 
It is characterized by a complete loss
of phase coherence between different lattice sites and an integer number of bosons at each
lattice site (``lobes'' in the phase diagram in Fig. \ref{Fig_Int1_lobes}). 
The loss of phase coherence has been shown in experiments \cite{bloch2}.
The MI is favored if the on-site interaction $U$ dominates the hopping $J$.

In the hard-core boson model, which will be discussed in the following sections, each 
lattice site cannot be occupied by more than one boson. Contrary, the Bose-Hubbard model which
allows multiple occupation to the price of the interaction energy $U$. The existence
of BEC phase in the three-dimensional hard-core boson model has been proven rigorously \cite{lieb1}.

The Hamiltonian of the hard-core boson model can be written in terms of creation and
annihilation operators $\hat{a}^+_{\bf r}$ and $\hat{a}_{\bf r}$ with the usual bosonic commutation
relations $[\hat{a}_{\bf r},\hat{a}^+_{\bf r'}]=0$ for different sites $\bf r\neq r'$. 
They have the additional hard-core property
\begin{equation}
 \hat{a}_{\bf r}^2 = (\hat{a}^+_{\bf r})^2 = 0 \; ,
 \label{Int1_hard-core-condition}
\end{equation}
which limits the occupation number at lattice site $\bf r$ to $0$ and $1$.
With those operators, the Hamiltonian is \cite{Blieb,ziegler4}
\begin{equation}
 \hat{H}_{\rm hc} = -\frac J{2d} \sum_{\langle {\bf r,r'}\rangle} \hat{a}^+_{\bf r} \hat{a}_{\bf r'} +
 \sum_{\bf r} V_{\bf r} \, \hat{a}^+_{\bf r} \hat{a}_{\bf r} \; .
 \label{Int1_H-hardcore}
\end{equation}

\begin{figure}
\centering
\scalebox{0.8}{\includegraphics{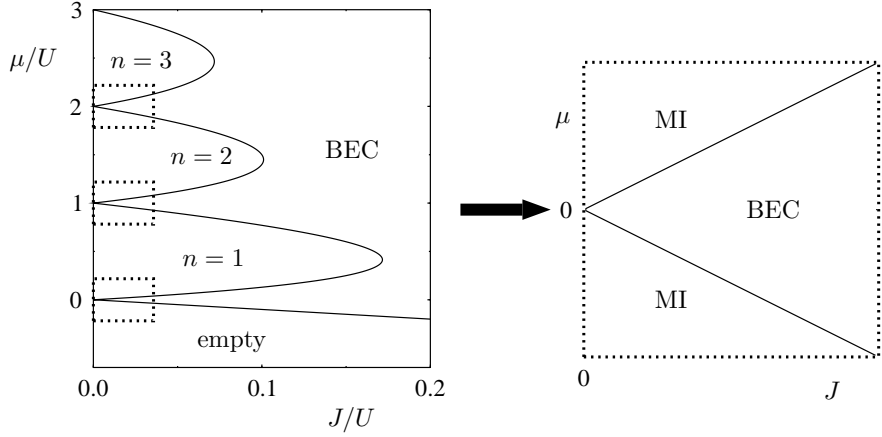}}
\caption{A projection of the phase diagram of the Bose-Hubbard model in the vicinity of the point, 
where the two Mott lobes meet. $\mu$ and $J$ are in arbitrary energy units after the projection.  }
\label{projection}
\end{figure}

The hard-core boson model can be understood
as a projection of the more general Bose-Hubbard model in the vicinity
of those points of the phase diagram, where two adjacent Mott lobes meet
(Fig. \ref{projection}).
This is similar to the picture which was applied to the tips of the Mott lobes in a recent paper
by Huber et al. \cite{huber1}. It is based on the following idea.
The number of bosons per site is fixed in the Mott state.
For adjacent Mott lobes this means that the corresponding Mott states differ exactly
by one boson per site. Now we consider two adjacent lobes with $n$ and $n+1$ ($n\ge0$ bosons per
site), respectively and assume that the chemical potential is fixed such that the ground
state is the Mott state with $n$ particles per site.
Low-energy excitations of this state for a grand-canonical system are
states, where one or a few sites (e.g. $k\ge1$ sites) have $n+1$ bosons,
all other sites have $n$ bosons. The $k$ excessive bosons are relatively free to move from site
to site on top of the $n$ Mott state.
Therefore, the physics of these excitations can be described approximately by the tunneling
of the $k$ excessive bosons alone.
Due to the repulsion of order $U$, assumed to be not too small, it is unlikely that a
site with $n+2$ bosons is created. Consequently, these excessive bosons form a hard-core
Bose gas. 
%------------------------------ FIGURE -------------------------------------
%---------------------------------------------------------------------------

\subsection{Outline of the following sections}

In section \ref{Chap_FI} the functional integral representation is introduced in the form
as it is applied to the models which are reviewed. It is shown that all physical quantities 
can be derived from of the functional integral representation of the grand canonical 
partition function.

In section \ref{Chap_non}, exactly solvable models are presented, namely the ideal Bose gas
and a one dimensional hard-core Bose gas an optical lattice.
Section \ref{Chap_weak} presents a summary of the results of the weakly interacting Bose gas
on the level of Gaussian fluctuations around the mean-field solutions. It leads to the well-known
results of Bogoliubov theory.
Two approaches to the dense regime of strongly interacting bosons are provided in section
\ref{Chap_strong}.The first one will be called the paired-fermion model, and the second is based
on the slave-boson approach.

%++++++++++++++++++++++++++++++++++++++++++++++++++++++++++++++++++++++++++++++++++++++++
%++++++++++++++++++++++++++++++++++++++++++++++++++++++++++++++++++++++++++++++++++++++++
%++++++++++++++++++++++++++++++++++++++++++++++++++++++++++++++++++++++++++++++++++++++++

\section{Functional integral method\label{Chap_FI}}

\subsection{Grand canonical partition function as functional integral}

The grand canonical partition function $Z$ of a many-body system contains all information
about the thermodynamic equilibrium properties of that system \cite{Bhuang}. 
For given Hamiltonian $\hat H$ it is given as the trace of the density operator $\rho$:
\begin{equation}
 \hat\rho = e^{-\beta({\hat H}-\mu{\hat N}_{\rm tot})} \; , \quad Z = {\rm Tr} \, (\hat\rho)
 \label{FI1_density-matrix}
\end{equation}
Here, $\beta=1/(k_{\rm B}T)$ is the inverse temperature, $\mu$ is the chemical potential
and the particle number operator is 
${\hat N}_{\rm tot}=\sum_\alpha\hat{a}^+_\alpha\hat{a}_\alpha$.
It is possible to write a grand canonical partition function in terms of a functional
integral \cite{Bnegele,Bpopov}.

\subsubsection{Bosonic functional integral}

Consider a bosonic many-body system given by the Hamiltonian 
$\hat{H}(\hat{a}^+_\alpha,\hat{a}_\alpha)$, where the creation and annihilation
operators $\hat{a}^+_\alpha$ and $\hat{a}_\alpha$ fulfil bosonic commutation
relations:
\begin{equation}
 [\hat{a}_\alpha,\hat{a}^+_\beta]_- = \delta_{\alpha\beta} \; ; \quad
 [\hat{a}_\alpha,\hat{a}_\beta]_- = [\hat{a}^+_\alpha,\hat{a}^+_\beta]_- = 0 \; .
 \label{FI1_boson-commutator}
\end{equation}
The index $\alpha$ denotes the states $|\alpha\rangle$ of an arbitrary single-particle basis,
e.g. $\alpha$ can denote a lattice site or a wave vector.
The grand canonical partition function is given as a functional integral over
the complex field $\phi$:
\begin{equation}
 Z = \lim_{M\rightarrow\infty} \int e^{-A(\phi^\ast,\phi)}
 \prod_{n=1}^M \prod_\alpha \frac{{\rm d}\phi^\ast_{\alpha,n}{\rm d}\phi_{\alpha,n}}{
 2\pi\rm i}
 \label{FI1_Z-boson-discrete}
\end{equation}
with the action
\begin{equation}
 A(\phi^\ast,\phi) = \frac{\beta}M \sum_{n=1}^{M} \left\{ \sum_\alpha
 \phi^\ast_{\alpha,n+1} \left[ \frac M{\beta} \left(\phi_{\alpha,n+1}-\phi_{\alpha,n}\right) -
 \mu \phi_{\alpha,n} \right] + H(\phi^\ast_{\alpha,n+1},\phi_{\alpha,n}) \right\} \; .
 \label{FI1_S-boson-discrete}
\end{equation}
We require for bosons the periodic boundary conditions 
$\phi_{\alpha,1}=\phi_{\alpha,M+1}$ and $\phi^\ast_{\alpha,1}=\phi^\ast_{\alpha,M+1}$.
The function $H(\phi^\ast_{\alpha,n+1},\phi_{\alpha,n})$ is obtained from the Hamiltonian
$\hat{H}(\hat{a}^+_\alpha,\hat{a}_\alpha)$ by making the replacements
$\hat{a}^+_\alpha\rightarrow\phi^\ast_{\alpha,n+1}$ and 
$\hat{a}_\alpha\rightarrow\phi_{\alpha,n}$. After performing the limit $M\rightarrow\infty$,
$n$ plays the role of a continuous imaginary time variable. Using $\tau:=n\hbar\beta/M$ we
can write
\begin{equation}
 Z = \int e^{-A(\phi^\ast,\phi)} {\cal D}(\phi^\ast(\tau)\phi(\tau)) \; ,
 \quad {\cal D}(\phi^\ast(\tau)\phi(\tau)) :=
 \lim_{M\rightarrow\infty} 
 \prod_{n=1}^M \prod_\alpha \frac{{\rm d}\phi^\ast_{\alpha,n}{\rm d}\phi_{\alpha,n}}{2\pi\rm i}
 \label{FI1_Z-boson}
\end{equation}
and
\begin{equation}
 A(\phi^\ast,\phi) = \frac 1\hbar \int_0^{\hbar\beta} {\rm d}\tau 
 \left\{ \sum_\alpha \phi^\ast_\alpha(\tau) 
 \left( \hbar\frac\partial{\partial\tau}-\mu\right) \phi_\alpha(\tau) +
 H(\phi^\ast_\alpha(\tau),\phi^\ast_\alpha(\tau)) \right\} \; .
 \label{FI1_S-boson}
\end{equation}
In the following we keep $M$ finite during the calculations and the limit $M \rightarrow\infty $ is performed in the end.

\subsubsection{Fermionic functional integral}

In the case of a fermionic many-body Hamiltonian
$\hat{H}(\hat{c}^+_\alpha,\hat{c}_\alpha)$, the creation and annihilation operators
fulfil the anti-commutation relations
\begin{equation}
 [\hat{c}_\alpha,\hat{c}^+_\beta]_+ = \delta_{\alpha\beta} \; ; \quad
 [\hat{c}_\alpha,\hat{c}_\beta]_+ = [\hat{c}^+_\alpha,\hat{c}^+_\beta]_+ = 0 \; .
 \label{FI1_fermion-commutator}
\end{equation}
A functional integral of a fermionic system is given as an integral of conjugate
Grassmann variables.
The definition of a Grassmann algebra can be found in
refs. \cite{Bnegele,Bpopov,Bitzykson}. Here it shall only be mentioned that the variables
of conjugate Grassmann fields $\bar\psi,\psi$ are anti-commuting, i. e.
$$
 \psi_{\alpha,n}\psi_{\beta,m} = -\psi_{\beta,m}\psi_{\alpha,n} \; , \quad
 \bar\psi_{\alpha,n}\bar\psi_{\beta,m} = -\bar\psi_{\beta,m}\bar\psi_{\alpha,n} \; , \quad
 \bar\psi_{\alpha,n}\psi_{\beta,m} = -\psi_{\beta,m}\bar\psi_{\alpha,n} \; ,
$$
and a Grassmann integral gives unity only if it is performed over a full product of all
variables, and zero otherwise:
\begin{equation}
 \int \bar\psi_{\alpha,n} \psi_{\alpha,n} \, 
 {\rm d}\psi_{\alpha,n} {\rm d}\bar\psi_{\alpha,n} = 1 \; ,
 \label{FI1_Grassmann-integral1}
\end{equation}
\begin{equation}
 \int {\rm d}\psi_{\alpha,n} {\rm d}\bar\psi_{\alpha,n} =
 \int \bar\psi_{\alpha,n} \, {\rm d}\psi_{\alpha,n} {\rm d}\bar\psi_{\alpha,n} =
 \int \psi_{\alpha,n} \, {\rm d}\psi_{\alpha,n} {\rm d}\bar\psi_{\alpha,n} = 0 \; .
 \label{FI1_Grassmann-integral2}
\end{equation}

Using these rules, the functional integral of the fermionic grand canonical partition function
can be constructed by analogy with Eq. (\ref{FI1_Z-boson-discrete}) as
\begin{equation}
 Z = \lim_{M\rightarrow\infty} \int e^{-A(\bar\psi,\psi)}
 \prod_{n=1}^M \prod_\alpha \, {\rm d}\bar\psi_{\alpha,n}{\rm d}\psi_{\alpha,n} \; .
 \label{FI1_Z-fermion-discrete}
\end{equation}
In the action (\ref{FI1_S-boson-discrete}), the complex variables
$\phi^\ast_{\alpha,n},\phi_{\alpha,n}$ have to be replaced by the Grassmann variables
$\bar\psi_{\alpha,n},\psi_{\alpha,n}$, and the periodic boundary conditions have to be replaced by anti-periodic boundary conditions
$\psi_{\alpha,1}=-\psi_{\alpha,M+1}$ and $\bar\psi_{\alpha,1}=-\bar\psi_{\alpha,M+1}$.
The same replacements can be done in the imaginary time functional integral defined 
by Eqs. (\ref{FI1_Z-boson}) and (\ref{FI1_S-boson}), then the integration measure
in (\ref{FI1_Z-boson}) is replaced by 
\begin{equation}
 {\cal D}(\bar\psi(\tau)\psi(\tau)) :=
 \lim_{M\rightarrow\infty} \prod_{n=1}^M \prod_\alpha 
 {\rm d}\bar\psi_{\alpha,n}{\rm d}\psi_{\alpha,n}
\end{equation}
for the Grassmann fields. (For the construction of the functional integral for bosons and
fermions with coherent states see Appendix \ref{App_coherentstates})

\subsection{Correlation functions\label{Chap_FI1_CF}}

Physical quantities can be written in terms of expectation values. The expectation value
of an arbitrary operator $\hat{X}$ is given by the relation
\begin{equation}
 \langle \hat{X} \rangle = \frac 1Z \, {\rm Tr} \, \left( \hat{X} \, \hat\rho \right)
 \label{FI1_expectation-operator}
\end{equation}
with the density operator (\ref{FI1_density-matrix}).
The general static $n$-particle correlation function (CF) is defined as a product of $n$ 
creation and $n$ annihilation operators:
\begin{equation}
 C_n(\alpha_1,\ldots,\alpha_n;\beta_n,\ldots,\beta_1) :=
 \langle \hat{a}^+_{\alpha_1} \cdots \hat{a}^+_{\alpha_n}
 \hat{a}_{\beta_n} \cdots \hat{a}_{\beta_1} \rangle \; .
 \label{FI1_CF}
\end{equation}
In the functional integral representation of a bosonic system, an expectation value of some
function $f(\phi^\ast,\phi)$, which depends on the complex field variables, is defined as
\begin{equation}
 \langle f(\phi^\ast,\phi) \rangle = \frac 1Z \,
 \int f(\phi^\ast,\phi) \, e^{-A(\phi^\ast,\phi)} {\cal D}(\phi^\ast(\tau)\phi(\tau)) \; .
 \label{FI1_expectation-FI}
\end{equation}
Note that in a fermionic system, the complex fields have to be replaced by Grassmann fields,
otherwise there is no difference in the formalism.
To translate the static CF (\ref{FI1_CF}) to an expectation value in terms of
a functional integral, it is necessary to introduce a dynamic $n$-particle CF, 
which depends on the imaginary time variable $\tau$. Therefore we introduce the imaginary 
time Heisenberg representation of the bosonic creation and annihilation operators 
$\hat{a}^+_\alpha$ and $\hat{a}_\alpha$:
\begin{eqnarray}
 \hat{a}_\alpha^+(\tau) &=& e^{\tau(\hat{H}-\mu\hat{N}_{\rm tot})/\hbar} \hat{a}^+_\alpha
  {\rm e}^{-\tau(\hat{H}-\mu\hat{N}_{\rm tot})/\hbar} \\
 \hat{a}_\alpha(\tau) &=& e^{\tau(\hat{H}-\mu\hat{N}_{\rm tot})/\hbar} \hat{a}_\alpha
  {\rm e}^{-\tau(\hat{H}-\mu\hat{N}_{\rm tot})/\hbar} \; .
\end{eqnarray}
The dynamic $n$-particle CF can now be defined as
\begin{equation}
 C_n(\alpha_1\tau_1,\ldots,\alpha_n\tau_n;\beta_n\tau_{n+1},\ldots,\beta_1\tau_{2n}) :=
 \langle \hat{a}^+_{\alpha_1}(\tau_1) \cdots \hat{a}^+_{\alpha_n}(\tau_n)
 \hat{a}_{\beta_n}(\tau_{n+1}) \cdots \hat{a}_{\beta_1}(\tau_{2n}) \rangle \; .
 \label{FI1_CF-dyn}
\end{equation}
An expectation value of the complex field variables is given as an expectation value of
a time ordered product of the creation and annihilation operators in the Heisenberg
representation \cite{Bnegele}. The time ordering in the imaginary time variable is 
indicated by the time
ordering operator $\hat T$. The ordering begins with the largest imaginary time and ends 
with the smallest. The rule for a translation of an expectation value of a time ordered
product of operators into an expectation value of a product of complex field variables is simply
$$
 \langle \phi^\ast_{\alpha_1}(\tau_1) \cdots \phi^\ast_{\alpha_n}(\tau_n)
 \phi_{\alpha_{n+1}}(\tau_{n+1}) \cdots \phi_{\alpha_{2n}}(\tau_{2n}) \rangle =
$$
\begin{equation}
 \langle \hat{T} \hat{a}^+_{\alpha_1}(\tau_1) \cdots \hat{a}^+_{\alpha_n}(\tau_n)
 \hat{a}_{\alpha_{n+1}}(\tau_{n+1}) \cdots \hat{a}_{\alpha_{2n}}(\tau_{2n}) \rangle \; .
 \label{FI1_time-ordering}
\end{equation}
Introducing a time-slice $\varepsilon>0$, the static $n$-particle CF (\ref{FI1_CF}) can thus 
be constructed by
\begin{samepage}
$$
 C_n(\alpha_1,\ldots,\alpha_n;\beta_n,\ldots,\beta_1) = 
$$
$$
 \lim_{\varepsilon\rightarrow 0} \; 
 \langle \hat{a}^+_{\alpha_1}(\tau+(2n-1)\varepsilon) \cdots 
 \hat{a}^+_{\alpha_n}(\tau+n\varepsilon) \hat{a}_{\beta_n}(\tau+(n-1)\varepsilon) \cdots 
 \hat{a}_{\beta_1}(\tau) \rangle =
$$
\begin{equation}
 \lim_{\varepsilon\rightarrow 0} \; 
 \langle \phi^\ast_{\alpha_1}(\tau+(2n-1)\varepsilon) \cdots 
 \phi^\ast_{\alpha_n}(\tau+n\varepsilon) \phi_{\beta_n}(\tau+(n-1)\varepsilon) \cdots 
 \phi_{\beta_1}(\tau) \rangle
 \label{FI1_CF-FI}
\end{equation}
\end{samepage}
Note that this expression is independent of $\tau$. Because the imaginary time is periodic
with periodicity $\hbar\beta$, it does not matter which point $\tau$ is regarded
as the beginning of a period, thus in particular we can assume $\tau=0$. In general, it is not
possible to replace the limit $\varepsilon\rightarrow 0$ simply by putting $\varepsilon=0$,
because the limits for $\varepsilon>0$ and $\varepsilon<0$ are not necessarily the same.
This feature reflects the fact that the creation and annihilation operators do not commute
in the operator formalism.

Some relevant physical quantities which can be calculated from correlation functions
shall be mentioned here:

\subsubsection{Total particle number}

The total particle number is derived from the grand canonical partition
function by \cite{Bhuang}
\begin{equation}
 N_{\rm tot} = \frac 1{\beta} \, \frac\partial{\partial\mu} \log Z \; .
 \label{FI1_N-total}
\end{equation}
Applying Eq. (\ref{FI1_N-total}) to $Z$ as it is given in Eqs. (\ref{FI1_Z-boson})
and (\ref{FI1_S-boson}), we get
$$
 N_{\rm tot} = \lim_{\varepsilon\rightarrow 0} 
 \frac 1{\beta} \, \frac 1Z \, \int \left[\sum_\alpha \, \int_0^{\hbar\beta}
 \phi^\ast_\alpha(\tau+\varepsilon) \phi_\alpha(\tau) {\rm d}\tau \right]
 e^{-A(\phi^\ast,\phi)} {\cal D}(\phi^\ast(\tau)\phi(\tau)) \; .
$$
Because of the independence of the CFs of $\tau$, we have
\begin{equation}
 N_{\rm tot} = \lim_{\varepsilon\rightarrow 0} \,
 \sum_\alpha \langle \phi^\ast_\alpha(\varepsilon) \phi_\alpha(0) \rangle \; .
 \label{FI1_N-CF}
\end{equation}
The particle occupation number in state $\alpha$ is
\begin{equation}
 n_\alpha = \lim_{\varepsilon\rightarrow 0} \,
 \langle \phi^\ast_\alpha(\varepsilon) \phi_\alpha(0) \rangle \; .
 \label{FI1_n-alpha-CF}
\end{equation}
If $\alpha$ denotes a position in space or a lattice site, $n_\alpha$
is a local particle density, if $\alpha$ is a momentum index,  $n_\alpha$ is the
momentum distribution of particles.

As has been mentioned before, it is not allowed to put the time-slice $\varepsilon=0$ in general,
because in the discrete-time definition of the
action (\ref{FI1_S-boson-discrete}), the $\mu$-dependent term is given by
\begin{equation}
 -\frac{\beta} M \sum_{n=0}^{M-1} \sum_\alpha \mu \phi^\ast_{\alpha,n+1} \phi_{\alpha,n}
 \label{FI1_mu-off-diagonal}
\end{equation}
and therefore occupies the off-diagonal matrix elements in the imaginary time index. 
It should be noted here, that it is also possible to construct the 
functional integral with
the $\mu$-dependent term being on the diagonal matrix elements, i. e.
\begin{equation}
 -\frac{\beta}M \sum_{n=0}^{M-1} \sum_\alpha \mu \phi^\ast_{\alpha,n} \phi_{\alpha,n} \; .
 \label{FI1_mu-diagonal}
\end{equation}
In this case the occupation number would be 
$n_\alpha =\langle \phi^\ast_\alpha(0) \phi_\alpha(0) \rangle$, which means that the
expressions for the physical quantities significantly depend on the definition of
the functional integral, which in some cases might be more convenient. 
However, in this chapter we will keep the off-diagonal representation
given in (\ref{FI1_mu-off-diagonal}).

\subsubsection{Condensate density}

The condensate density of a BEC is a measure for the off-diagonal long range
order of the one-particle CF. It has to do with
the spacial range of the one-particle CF and thus $\alpha$ should denote a position
vector (in a continuous system) or a lattice site (in an optical lattice).
In terms of complex variables, the definition (\ref{Int1_odlro}) of the condensate density 
in a system without confining potential is
\begin{equation}
 n_0 := \lim_{{\bf r-r'}\rightarrow\infty}   \lim_{\varepsilon\rightarrow 0} 
 \langle \phi^\ast_{\bf r}(\varepsilon)
 \phi_{\bf r'}(0) \rangle \; .
 \label{FI1_n0}
\end{equation}

\subsubsection{Density-density correlation function}

The density-density CF is a two-particle CF. It describes the spacial
behaviour of density correlations, which means that here $\alpha$ denotes a position
index as well. In terms of field operators it is defined as
\begin{equation}
 D({\bf r-r'}) = \langle \hat{n}_{\bf r} \hat{n}_{\bf r'} \rangle =
 \langle \hat\psi^+_{\bf r} \hat\psi_{\bf r}
 \hat\psi^+_{\bf r'} \hat\psi_{\bf r'} \rangle \, ,
 \label{FI1_dd-operator}
\end{equation}
and in terms of complex field variables it is given as
\begin{equation}
 D({\bf r-r'}) = \lim_{\varepsilon\rightarrow 0} 
 \left\langle \phi^\ast_{\bf r}(\varepsilon) \phi_{\bf r}(0)
 \phi^\ast_{\bf r'}(\varepsilon) \phi_{\bf r'}(0) \right\rangle
 \label{FI1_dd-FI}
\end{equation}
A good physical quantity, which describes correlations of density fluctuations
is the truncated density-density CF
\begin{equation}
 D_{\rm trunc}({\bf r-r'}) = \langle \hat{n}_{\bf r} \hat{n}_{\bf r'} \rangle -
 \langle \hat{n}_{\bf r} \rangle \langle \hat{n}_{\bf r'} \rangle \; .
\end{equation}
The Fourier transform of the density-density CF is called the static structure factor
\begin{equation}
 S({\bf q}) = \frac 1{N_{\rm tot}} \sum_{\bf r,r'} D({\bf r-r'}) e^{{\rm i}\bf k\cdot(r-r')} \; .
 \label{FI1_sf}
\end{equation}

%++++++++++++++++++++++++++++++++++++++++++++++++++++++++++++++++++++++++++++++++++++++++
%++++++++++++++++++++++++++++++++++++++++++++++++++++++++++++++++++++++++++++++++++++++++
%++++++++++++++++++++++++++++++++++++++++++++++++++++++++++++++++++++++++++++++++++++++++

\section{Exactly solvable models\label{Chap_non}}

\subsection{Ideal Bose gas\label{Chap_ideal}}

\subsubsection{Hamiltonian and partition function\label{Sec_ideal-gas-cf}}

In this chapter we will survey the basic results of the previously mentioned quantities
for an ideal Bose gas. This seems to be reasonable, because it allows us to introduce the
methods we will apply for an interacting hard-core Bose gas as well. Contrary to the
interacting system, exact analytic results can be found for the non-interacting case
of the ideal Bose gas.

A non-interacting Bose gas in a $d$-dimensional cubic lattice with nearest-neighbour hopping $J$ and lattice
constant $a$ is given by the Hamiltonian
\begin{equation}
 \hat{H} = J -\frac J{2d} \sum_{\langle {\bf r},{\bf r'}\rangle} 
 \hat{a}^+_{\bf r} \hat{a}_{\bf r'} + J \sum_{\bf r} \hat{a}^+_{\bf r} \hat{a}_{\bf r} \; \,
\end{equation}
with the dispersion relation 
\begin{equation}
 \epsilon_{\bf k} = J -\frac J{d} \sum_{\nu=1}^d \cos (ak_\nu) \; ,
 \label{FI2_dispersion-latt}
\end{equation}
where $k_\nu$ is the $\nu$-th component of the $d$-dimensional wave vector $\bf k$.
Note that the sum over nearest neighbors $\langle{\bf r}_i,{\bf r}_j\rangle$ means, 
that the index $i$ runs
over the entire lattice and the index $j$ runs over all sites, which are nearest neighbours
of $j$. This means, that each bond appears twice in the sum, once with a hopping process from
site $i$ to site $j$ and vice versa. For small wave vectors $\bf k$, the lattice dispersion can
be approximated by the translation invariant counterpart
\begin{equation}
 \epsilon_{\bf k} = \frac{\hbar^2 {\bf k}^2}{2m^\ast} + {\cal O}({\bf k}^4) \; , \quad
 m^\ast := \frac{d\hbar^2}{Ja^2} \; ,
 \label{FI2_band-mass}
\end{equation}
where $m^\ast$ is the band mass.

We apply the discrete time action given in
Eq. (\ref{FI1_S-boson-discrete}) and perform the limit $M\rightarrow\infty$ at the very end.
It is possible to write the functional integral (\ref{FI1_S-boson-discrete})
in the form
\begin{equation}
 Z = \lim_{M\rightarrow\infty} \int
 \exp \left[ -\sum_{\bf k} \sum_{n,m=1}^M \phi^\ast_{{\bf k},n} \hat{A}_{nm}^{({\bf k})} 
 \phi_{{\bf k},m} \right] \, \prod_{\bf k} \prod_{n=1}^M 
 d\phi^\ast_{{\bf k},n} d\phi_{{\bf k},n} \; ,
 \label{FI2_Z-matrix}
\end{equation}
where the relation between the complex fields in real space and in momentum space is
\begin{equation}
 \phi_{{\bf r},n} = \frac 1{\sqrt N} \sum_{\bf k} e^{{\rm i}\bf k\cdot r} \phi_{{\bf k},n} \; ,
\end{equation}
and the matrix elements of $\hat{A}^{({\bf k})}$ represent the structure of the discrete
imaginary time variable:
\begin{equation}
 \hat{A}^{({\bf k})} = \left[
 \begin{array}{cccccc}
  1 & 0 & \quad & \cdots & 0 & -b_{\bf k} \\
  -b_{\bf k} & 1 & 0 & \quad & \quad & \quad \\
  0 & -b_{\bf k} & 1 & \ddots & & \vdots \\
  \quad & 0 & -b_{\bf k} & \ddots & 0 & \quad \\
  \vdots & \quad & 0 & \ddots & 1 & 0 \\
  0 & \quad & \quad & \cdots & -b_{\bf k} & 1
 \end{array} \right] \; , \quad b_{\bf k}=1-\frac\beta M (\epsilon_{\bf k}-\mu) \; .
 \label{FI2_time-matrix}
\end{equation}
The entry in the upper right corner is necessary to realize the periodic boundary
conditions. The Gaussian integral can be integrated out and we get
\begin{equation*}
 Z = \lim_{M\rightarrow\infty} \prod_{\bf k} \det \hat{A}^{({\bf k})} =
 \lim_{M\rightarrow\infty} \prod_{\bf k} \left[ 1 - \left(
 1 - \frac{\beta(\epsilon_{\bf k}-\mu)}M \right)^M \right]^{-1}
\end{equation*}
If we now, as a final step, perform the limit $M\rightarrow\infty$, we get the correct
form of the grand canonical partition function of an ideal Bose gas \cite{Bnegele}:
\begin{equation}
 Z = \prod_{\bf k} \left[ 1 - e^{-\beta(\epsilon_{\bf k}-\mu)} \right]^{-1} \; .
\end{equation}

\subsubsection{One-particle correlation function\label{Sec_ideal-matrix}}

As already discussed in section \ref{Chap_FI1_CF}, the momentum distribution
and the condensate density in a Bose gas can both be described by the one-particle
correlation function, cf. Eqs. (\ref{FI1_N-CF}) and (\ref{FI1_n0}).
Thus we should at first calculate the one-particle CF for an ideal Bose gas in
general to determine those quantities. To achieve this we again start with the 
discrete time functional integral and take the limit $M\rightarrow\infty$ at the end of the calculations.
In this sense, we define the imaginary time dependent one-particle CF in momentum space as
$$
 C({\bf k}_1,\tau_1;{\bf k}_2,\tau_2) =
 \langle \phi^\ast_{{\bf k}_1,n_1} \phi_{{\bf k}_2,n_2} \rangle = 
$$
\begin{equation}
 \lim_{M\rightarrow\infty} \frac 1Z \int \phi^\ast_{{\bf k}_1,n_1} \phi_{{\bf k}_2,n_2}
 \exp \left[ -\sum_{\bf k} \sum_{n,m=1}^M \phi^\ast_{{\bf k},n} \hat{A}_{nm}^{({\bf k})} 
 \phi_{{\bf k},m} \right] \, \prod_{\bf k} \prod_{n=1}^M 
 d\phi^\ast_{{\bf k},n} d\phi_{{\bf k},n} \; ,
 \label{FI2_1pCF-ideal-BG}
\end{equation}
where the indices $n_1,n_2$ are defined such that
\begin{equation}
 \frac\beta M (n_{1,2}-1)<\tau_{1,2}<\frac\beta M n_{1,2} \; .
 \label{FI2_discrete-time}
\end{equation}
The Gaussian integral (\ref{FI2_1pCF-ideal-BG}) picks out a matrix element of the 
inverse matrix $\hat{A}^{-1}$:
\begin{equation}
 C({\bf k}_1,\tau_1;{\bf k}_2,\tau_2) = \lim_{M\rightarrow\infty}
 (\hat{A}^{({\bf k}_1)})^{-1}_{n_2,n_1} \, \delta_{{\bf k}_1,{\bf k}_2} \; .
 \label{FI2_C-discrete1}
\end{equation}
Therefore it is necessary to determine the matrix elements of $\hat{A}^{-1}$.
By means of the unitary transformation matrices
\begin{equation}
 U_{nm} = \frac 1{\sqrt M} \, e^{\frac{2\pi\rm i}M\, nm} \; , \quad
 U^+_{nm} = \frac 1{\sqrt M} \, e^{-\frac{2\pi\rm i}M\, nm} \; ,
 \label{FI2_unitary-trafo}
\end{equation}
we can diagonalize the matrix to get
$$
 (U(\hat{A}^{({\bf k})})^{-1}U^+)_{jn} = \frac{\delta_{jn}}{1-b_{\bf k} e^{\frac{2\pi\rm i}M n}} \; ,
$$
$$
 (\hat{A}^{({\bf k})})^{-1}_{jn} = [U^+(U(\hat{A}^{({\bf k})})^{-1}U^+)U]_{jn} =
 \sum_{l=1}^M \frac 1M \frac{e^{-\frac{2\pi\rm i}M l(j-n)}}{1-b_{\bf k}e^{\frac{2\pi\rm i}M l}} \; .
$$
This sum is given in the Appendix. The result is
\begin{equation}
 (\hat{A}^{({\bf k})})^{-1}_{jn} = \frac 1{1-b_{\bf k}^M} \times \left\{
 \begin{array}{l@{\quad \mbox{if} \quad}l}
 b_{\bf k}^{j-n} & j\ge n \\ b_{\bf k}^{M+n-j} & j<n
 \end{array} \right. \; .
\end{equation}
Performing the limit $M\rightarrow\infty$ in (\ref{FI2_discrete-time}) 
we get
\begin{equation}
 C({\bf k}_1,\tau_1;{\bf k}_2,\tau_2) = \frac{\delta_{{\bf k}_1,{\bf k}_2} }{
 1-e^{-\beta \left(\epsilon_{\bf k}-\mu\right)} } \times \left\{
 \begin{array}{l@{\quad \mbox{if} \quad}l}
 e^{\left(\tau_2-\tau_1\right)\left(\epsilon_{\bf k}-\mu\right)/\hbar} & \tau_1\ge\tau_2 \\
 e^{\left(\tau_1-\tau_2-\hbar\beta\right)\left(\epsilon_{\bf k}-\mu\right)/\hbar} & \tau_1<\tau_2
 \end{array} \right. \; .
\end{equation}
Using this result and the definition (\ref{FI1_CF-FI}), the one-particle CF in momentum
space for an ideal Bose gas is
\begin{equation}
 C_1({\bf k};{\bf k'}) = \lim_{\epsilon\rightarrow 0}
 \left\langle \phi^\ast_{\bf k}(\epsilon) \phi_{\bf k'}(0) \right\rangle =
 \frac {\delta_{{\bf k},{\bf k'}} }{e^{\beta \left(\epsilon_{\bf k}-\mu\right)}-1} = 
 \delta_{{\bf k},{\bf k'}} N_{{\bf k}}\; ,
 \label{FI2_CF-ideal-BG}
\end{equation}
where $n_{{\bf k}}$ is the usual momentum distribution of an ideal Bose gas.

In the condensed phase, where the chemical potential takes the value $\mu=0$, the
momentum distribution function diverges at ${\bf k}=0$. In this case, the lowest
momentum state ${\bf k}=0$ is macroscopically occupied and builds the condensate.
The condensate density in this case is given by
\begin{equation}
 n_0 = \frac{N_0}{\cal N} \; .
 \label{FI2_ground-state}
\end{equation}
The normaliation with the number of lattice sites $\cal N$ is necessary, because in the BEC
phase the ground state is the only macroscopically occupied state, whereas all other
occupation numbers are of the order of unity.
The total particle density in the condensed phase is the sum of the condensate density
and the particle density of all excited states. In the thermodynamic limit, the sum  becomes
an integral:
\begin{equation}
 n_{\rm tot} = n_0 + \int N_{\bf k} \, \frac{{\rm d}^3k}{(2\pi)^3} \; .
 \label{FI2_ntot-ideal-BG}
\end{equation}
It should be noted here, that in one and two dimensions a condensate cannot exist.
The reason is, that the integral (\ref{FI2_ntot-ideal-BG}) is divergent in 
these cases if $\mu=0$, because $n_{\bf k}$ behaves like $k^{-2}$ for small momenta.

This definition of the condensate density in an ideal Bose gas is also compatible
with the more general definition via off-diagonal long range order given in Eq.
(\ref{FI1_n0}):
\begin{equation*}
 \lim_{{\bf r-r'}\rightarrow\infty}  \lim_{\epsilon\rightarrow 0}  
 \langle \phi^\ast_{\bf r}(\epsilon)
 \phi_{\bf r'}(0) \rangle = \lim_{{\bf r-r'}\rightarrow\infty} C({\bf r};{\bf r'}) =
 \lim_{{\bf r-r'}\rightarrow\infty}
 \int\frac{{\rm d}^3k}{(2\pi)^3} N_{{\bf k}} e^{\rm i\bf k\dot(r-r')}
\end{equation*}

\subsubsection{Structure factor}

\begin{figure}
\centering
\includegraphics{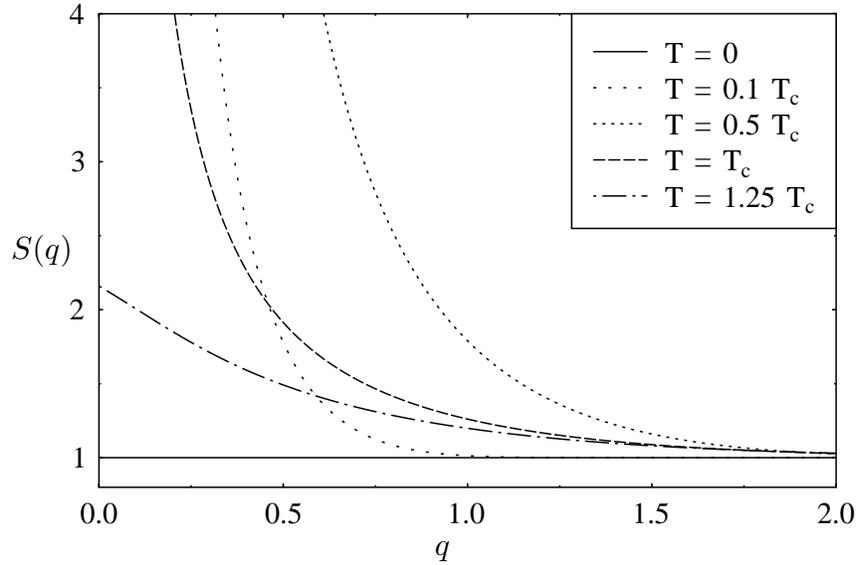}
\caption{Static structure factor of an ideal Bose gas of free particles. At $T=0$, $S$ is
constantly unity and has a $\delta$-peak at $q=0$. At $0<T<T_c$ it diverges, and at $T>T_c$
it reaches a constant near $q=0$. All cases are characterised by the relation
$\lim_{q\rightarrow\infty}S(q)=1$.}
\label{Fig_FI3_S-ideal}
\end{figure}

From Eqs. (\ref{FI1_dd-FI}) and (\ref{FI1_sf}) the static structure factor can be obtained.
The fourth-order correlation function can be calculated using Wick's theorem 
(Appendix \ref{App_Wick}):
$$
 \lim_{\varepsilon\rightarrow 0} \left\langle \phi^\ast_{\bf k}(\varepsilon) 
 \phi^\ast_{\bf k'+q}(0) \phi_{\bf k+q}(\varepsilon) \phi_{\bf k'}(0) \right\rangle =
$$
$$
 \lim_{\varepsilon\rightarrow 0} [
 \langle \phi^\ast_{\bf k}(\varepsilon) \phi_{\bf k+q}(\varepsilon) \rangle
 \langle \phi^\ast_{\bf k'+q}(0) \phi_{\bf k'}(0) \rangle +
 \langle \phi^\ast_{\bf k}(\varepsilon) \phi_{\bf k'}(0) \rangle
 \langle \phi^\ast_{\bf k'+q}(0) \phi_{\bf k+q}(\varepsilon) \rangle ] =
$$
$$
 N_{\bf k} \delta_{\bf k,k+q} N_{\bf k'} \delta_{\bf k'+q,k'} +
 N_{\bf k} \delta_{\bf k,k'} \left( N_{\bf k+q}+1 \right) \delta_{\bf k'+q,k+q} \, .
$$
For ${\bf q}\neq 0$, the first term vanishes. Thus we find the result
\begin{equation}
 S({\bf q}) = \frac 1{N_{\rm tot}} \sum_{\bf k} N_{\bf k} (N_{\bf k+q}+1) \; .
 \label{FI3_S-ideal-general}
\end{equation}
In the BEC by separating the ground state and excited states, we get
\begin{equation}
 S({\bf q}) = 1 + 2n_0 N_{\bf q} +
 \frac 1{N_{\rm tot}} \, \sum_{{\bf k}\neq\{0,{-\bf q}\}} N_{\bf k} N_{\bf k+q} \; .
 \label{FI3_S-ideal-final}
\end{equation}
Instead of Eq. (\ref{FI1_dd-FI}) one can use the
more convenient definition in terms of expectation values without time slices
\begin{equation}
 S({\bf q}) = 1 + \frac 1{N_{\rm tot}} \sum_{\bf k,k'} \, \left\langle 
 \phi^\ast_{\bf k}(0) \phi^\ast_{\bf k'+q}(0) 
 \phi_{\bf k+q}(0) \phi_{\bf k'}(0) \right\rangle \; ,
 \label{FI3_S-without-slice}
\end{equation}
which leads to Eq. (\ref{FI3_S-ideal-final}) as well.
Graphs for different temperature regimes are shown in Fig. \ref{Fig_FI3_S-ideal}.

%------------------------------ FIGURE -------------------------------------
%---------------------------------------------------------------------------

\subsubsection{Random walk expansion and world-lines\label{Sec_worldlines-ideal-bosons}}

In this section a very intuitive method of diagrammatically visualizing a grand canonical 
partition function shall be introduced for an ideal Bose gas in an optical lattice, namely the
random walk expansion \cite{Bglimm,ziegler7}. We will perform the same expansion in the following
chapters for a system of hard-core bosons, in order to demonstrate the effect of the
hard-core condition. 

The grand canonical partition function of an ideal Bose gas in a $d$-dimensional cubic
lattice is given by the functional integral Eq. (\ref{FI2_Z-matrix}). But here we use
the real-space representation. The time structure of the
matrix $\hat A$ is the same as in Eq. (\ref{FI2_time-matrix}), but instead of the dispersion
relation $\epsilon_{\bf k}$ we use the hopping matrix
\begin{equation}
 \hat{J}_{\bf rr'} := \left\{ \begin{array}{ll}
 -J/2d & \mbox{ if ${\bf r,r'}$ nearest neighbours} \\
 0   & \mbox{ otherwise} \end{array} \right. \; ,
 \label{FI3_hoppingJ}
\end{equation}
which establishes the spacial structure of $\hat A$, and make use of
\begin{equation}
 \hat\epsilon_{\bf rr'}:= \hat{J}_{\bf rr'} + J \, \delta_{\bf rr'} \; .
 \label{FI3_hopping}
\end{equation}
Thus we can write
\begin{equation}
 \hat{A}_{{\bf rr'};nm} :=  \delta_{nm} \delta_{\bf rr'} - 
 (\delta_{n,m+1} + \delta_{n1}\delta_{mM})
 \left[\delta_{\bf rr'}-\frac\beta M(\hat\epsilon_{\bf rr'}-\mu\, \delta_{\bf rr'})\right] \; ,
 \label{FI3_S-lattice}
\end{equation}
where the term $\delta_{n1}\delta_{mM}$ accounts for the upper right matrix element in
(\ref{FI2_time-matrix}) which arises from the periodicity in imaginary time.

The idea of the random walk expansion is to expand the off-diagonal part of the 
exponential in the functional integral expression in terms of the field variables:
\begin{equation*}
 \exp \left[ -\sum_{\bf r,r'} \sum_{n,m=1}^M \phi^\ast_{{\bf r},n} \hat{A}_{{\bf rr'};nm}
 \phi_{{\bf r'},m} \right] =
\end{equation*}
\begin{equation*}
 \exp \left[ -\sum_{\bf r} \sum_{n=1}^M \phi^\ast_{{\bf r},n} \phi_{{\bf r},n} \right]
 \sum_{\{l_{{\bf rr'},n}\ge 0\}} \, \frac 1{l_{{\bf rr'},n}!} \left[
 \prod_{{\bf r,r'},n} \phi^\ast_{{\bf r},n} \underbrace{\left(
 \delta_{\bf rr'} - \frac\beta M\left(\hat\epsilon_{\bf rr'}-\mu\delta_{\bf rr'}\right)
 \right)}_{\displaystyle =:\hat{u}_{\bf rr'}} \phi_{{\bf r'},n-1} \right]^{l_{{\bf rr'},n}}
\end{equation*}
The abbreviation $\hat{u}_{\bf rr'}$ has been introduced for convenience.
The functional integral can be solved by using the identities
\begin{equation}
 \prod_{{\bf r,r'},n} 
 \left(\phi^\ast_{{\bf r},n}\phi'_{{\bf r},n-1}\right)^{l_{{\bf rr'},n}} = 
 \prod_{{\bf r},n} \left[(\phi^\ast_{{\bf r},n})^{m_{{\bf r},n}}
 (\phi_{{\bf r},n})^{m'_{{\bf r},n}} \right] \; ,
 \label{FI3_index}
\end{equation}
$$
 \mbox{where } m_{{\bf r},n} := \sum_{\bf r'} l_{{\bf rr'},n} \mbox{ and }
 m'_{{\bf r},n} := \sum_{\bf r'} l_{{\bf r'r},n+1}
$$
and
\begin{equation}
 \int (\phi^\ast)^m \phi^{m'} \, e^{-\phi^\ast\phi} \, 
 \frac{{\rm d}\phi^\ast{\rm d}\phi}{2\pi\rm i} = m! \, \delta_{mm'} \; .
 \label{FI3_gauss}
\end{equation}
This results in the following form of the grand canonical partition function as a 
sum over all indices $l_{{\bf rr'},n}$:
\begin{equation}
 Z = \sum_{\{l_{{\bf rr'},n}\ge 0\}} \prod_{{\bf r},n} \left(
 m_{{\bf r},n}! \, \delta_{m_{{\rm r},n},m'_{{\rm r},n}} \right) \prod_{{\bf r,r'},n}
 \left[\frac{(\hat{u}_{\bf rr'})^{l_{{\bf rr'},n}}}{l_{{\bf rr'},n}!}\right] \; .
 \label{FI3_Z-random-walk}
\end{equation}
Note that it is necessary to define $(\hat{u}_{\bf rr'})^0\equiv 1$ here, even for
the vanishing matrix elements of $\hat u$.

%------------------------------ FIGURE -------------------------------------
\begin{figure}
\centering
\includegraphics{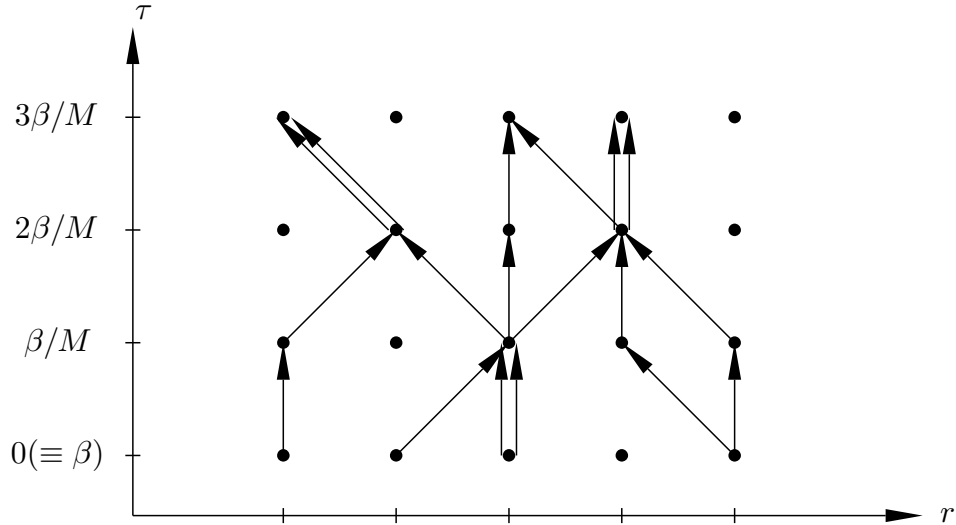}
\caption{Random walk expansion of an ideal Bose gas; world-line diagram.}
\label{Fig_FI3_RandomWalk}
\end{figure}
%---------------------------------------------------------------------------
One possible interpretation of this expression is as follows: Each term of the sum can be
represented by a diagram, where a particle propagation from site $\bf r$ at imaginary time
$\tau$ to site $\bf r'$ at time $\tau+\hbar\beta/M$ is indicated by an arrow. So each particle
is characterised by a ``world-line'' showing its movement through the lattice in imaginary time.
The contribution of a certain diagram is defined by the following properties:
\begin{itemize}
\item The number of particles (arrows) propagated from site $\bf r'$ at time 
$(n-1)\hbar\beta/M$ to site $\bf r$ at time $n\hbar\beta/M$ is 
given by $l_{{\bf rr'},n}$. In the case of nearest neighbour hopping, particle propagation
in one time step $\hbar\beta/M$ is only possible between neighbouring sites, or the particle stays
at the same site.
\item The number of particles (arrows) which are propagated to site $\bf r$ at time $n\hbar\beta/M$ 
from the previous time step is $m_{{\bf r},n}$.
\item The number of particles (arrows) propagating from site $r$ at time $n\hbar\beta/M$ to 
the next time step is $m'_{{\bf r},n}$.
\item Particle conservation is assured by the $\delta$-function in Eq. (\ref{FI3_Z-random-walk}),
such that $m_{{\bf r},n}=m'_{{\bf r},n}$ is equal to the number of particles at site $\bf r$
and time $n\hbar\beta/M$. 
\item There is a periodicity in imaginary time: Time $\tau=\hbar\beta$ is equivalent to time 
$\tau=0$, so the diagrams have to be periodic in time.
\end{itemize}
Note that in the ideal Bose gas $m_{{\rm r},n}>1$ is possible, i.e. more than one particle
can occupy the same lattice site at the same time.
This will be excluded to establish the hard-core interaction in a Bose gas.

%++++++++++++++++++++++++++++++++++++++++++++++++++++++++++++++++++++++++++++++++++++++++
%++++++++++++++++++++++++++++++++++++++++++++++++++++++++++++++++++++++++++++++++++++++++
%++++++++++++++++++++++++++++++++++++++++++++++++++++++++++++++++++++++++++++++++++++++++

\subsection{Hard-core bosons in 1$D$\label{Chap_d1}}

\subsubsection{General remarks}

The main feature of the one-dimensional hard-core
Bose gas is, that the particles cannot interchange their position. An interesting 
consequence of this property is the equivalence 
to an ideal non-interacting one-dimensional Fermi gas. However, it is important to 
mention, that this equivalence does not hold for all physical
quantities in momentum space, namely for those which are given by one-particle 
correlation functions like the momentum distribution
\cite{olshanii1,shlyapnikov1,papen1,girardeau2,gangardt1}.
It is possible to calculate the momentum distribution by means of a Jordan-Wigner
transformation (see e. g. refs. \cite{Bnegele,Bfradkin}).
This approach has been used in a couple of works \cite{paredes2,muramatsu1,paredes1}.
However, this problem will not be addressed here.
On the other hand, quantities given by
two-particle correlation functions like the density-density correlation
function and the dynamic structure factor are the same for hard-core bosons and for
ideal fermions.

The zero temperature phase diagram of a hard-core Bose gas in a one-dimensional optical 
lattice shows three phases \cite{ates1}: An empty phase (EP), an incommensurate phase (ICP) 
with a particle number per lattice site of $0<n_{\rm tot}<1$, and a Mott insulator (MI) with 
$n_{\rm tot}=1$. 
Here we will especially be interested in the phase transition between the ICP and the MI phase
for zero and non-zero temperatures. Again, the quantity we chose for investigating 
this transition is the static structure factor.
It has also been considered in other works about one-dimensional Bose gases, in the weakly
interacting regime as well as in the strongly interacting regime 
\cite{stringari4,astra1,ana1,ana2}.

As has been demonstrated
for the ideal Bose gas, a random walk expansion leads to a world-line picture. To make 
the mapping to a system of ideal fermions possible, it has to be assured that world lines cannot
intersect each other. So instead of constructing the functional integral by starting from the
Hamiltonian, we choose a different way and construct it by starting out from the random-walk
picture directly. 

When the random walk expansion for a system of ideal spinless fermions is performed, one obtains a
sum which is analogous to the sum in Eq. (\ref{FI3_Z-random-walk}) with two important differences:
Because of the nilpotent property of the Grassmann variables, the fermionic analog to Eq.
(\ref{FI3_gauss}) reads
\begin{equation}
 \int \bar\psi^m \psi^{m'} \, e^{-\bar\psi\psi} \, 
 {\rm d}\psi{\rm d}\bar\psi = \delta_{mm'} (\delta_{m,0}+\delta_{m,1})\; .
\end{equation}
This means that all terms, where the particle number
$m_{{\rm r},n}$ or $m'_{{\rm r},n}$ is larger than $1$ at lattice site $r$, do not contribute.
This reflects the Pauli principle or in the case of hard-core bosons, the hard-core property.
The second is that the Grassmann variable analog
to Eq. (\ref{FI3_index}) gets an additional sign because of the anti-commutation property.
To avoid this problem it is possible to construct a world-line model where world-lines do
not intersect. For this purpose we adopt an approach to the statistics of directed
polymers in two dimensions \cite{forgacs1}.

\subsubsection{Particle density and phase diagram}

It has been shown that the grand canonical partition function is given by the functional 
integral \cite{ates1}
\begin{equation}
 Z = \lim_{M\rightarrow\infty} \int \exp\left[ 
 -\sum_{k} \, \sum_{n=1}^M \, \sum_{j,j'=1}^2 \bar\psi_{k,n,j}
 \frac{[\hat{G}_n^{-1}(k)]_{jj'}}{1-\frac\beta{M} \mu} 
 \psi_{k,n,j'} \right] \prod_{k,n,j} {\rm d}\psi_{k,n,j} \, {\rm d}\bar\psi_{k,n,j}
 \label{d11_Z-hcb1}
\end{equation}
with the $2\times 2$ matrix
\begin{equation}
 \hat{G}_n^{-1}(k) = \left( \begin{array}{cc}
 -e^{\frac{2\pi\rm i}M\left(n-\frac 12\right)} +
 1-\frac\beta{M} \mu & -\frac\beta M \frac J2 
 e^{\frac{2\pi\rm i}M\left(n-\frac 12\right)} (1+e^{ik}) \\
 -\frac\beta M \frac J2 (1+e^{-ik}) &
 -e^{\frac{2\pi\rm i}M\left(n-\frac 12\right)} + 1-\frac\beta{M} \mu
 \end{array} \right) \; .
\end{equation}
This integral can be performed and it yields
\begin{equation}
 Z = \lim_{M\rightarrow\infty} \left(1-\frac\beta{M} \mu\right)^{-2M\cal N} \det \hat{G}^{-1} \; ,
 \label{d11_Z-hcb2}
\end{equation}
where $\cal N$ is the number of lattice sites.
The one-particle correlation function of the fermions at equal times can be calculated as
\begin{equation}
 C(k) = \lim_{M\rightarrow\infty}
 \frac 1M \sum_{n,m=1}^M \hat{G}_{11}(k)_{nm}
 \label{d11_Ck}
\end{equation}
This sum is performed in Appendix \ref{App_sumCk}. After performing the limit
$M\rightarrow\infty$, the result is
\begin{equation}
 C(k) = \frac 12 \left( \frac 1{1+e^{-\beta\left(J\cos\frac k2-\mu\right)}} +
 \frac 1{1+e^{-\beta\left(-J\cos\frac k2-\mu\right)}}\right) \; .
 \label{d11_sum-result}
\end{equation}
%------------------------------ FIGURE -------------------------------------
\begin{figure}
\centering
\includegraphics{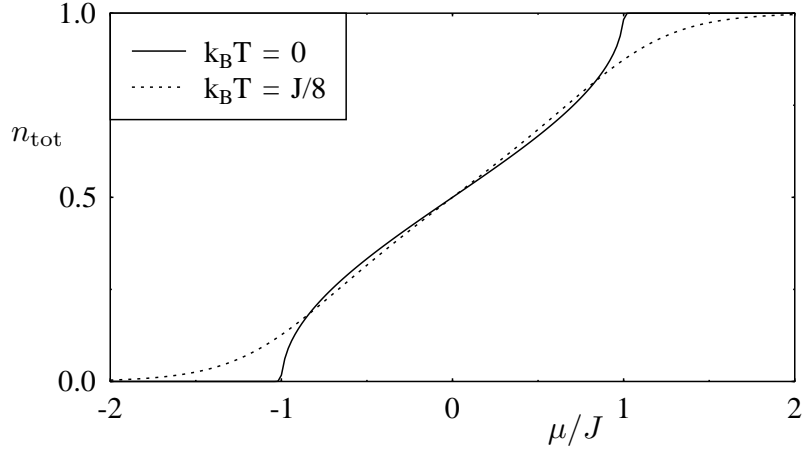}
\caption{Total particle density of a hard-core Bose gas in a one-dimensional optical lattice
calculated from Eq. (\ref{d11_ntot-finite-T}), for both zero temperature (solid line) and finite
temperature (dashed line).}
\label{Fig_d11_ntot1}
\end{figure}
%---------------------------------------------------------------------------
\begin{figure}
\includegraphics{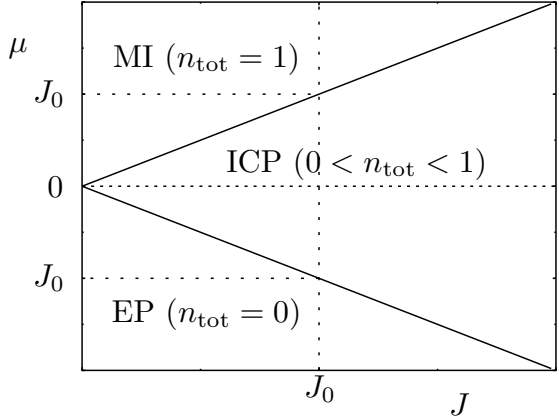}
\caption{Phase diagram of the one-dimensional hard-core Bose gas at 
zero temperature with an empty phase (EP), an incommensurate phase (ICP), and a 
Mott insulator (MI).}
\label{Fig_d11_phase-diagram1}
\end{figure}
%---------------------------------------------------------------------------
As was mentioned before, the one-particle correlation function does not lead to the momentum
distribution. However, the total particle density of the bosons is given 
by taking the fermionic one-particle correlation function in real space
\begin{equation}
 C(r,r') = \int_0^{2\pi} C(k) \, e^{{\rm i}k(r-r')} \, \frac{{\rm d}k}{2\pi} 
 \label{d11_Fourier}
\end{equation} 
at $r=r'$. This can
be shown by applying the expression (\ref{FI1_N-total}) of the total particle number to the
partition function (\ref{d11_Z-hcb1}). We need an additional factor of $1/2$ because of
our special construction:
$$
 N_{\rm tot} = \frac 1{2\beta} \, \frac\partial{\partial\mu} \log Z = \lim_{M\rightarrow\infty}
 \frac 1{2\beta}\, \frac\partial{\partial\mu} \left[-2M{\cal N}\log\left(1-\frac\beta M \mu\right) -
 \log\, \det \hat{G}\right]
$$
$$
 = {\cal N} - \frac 1{2\beta Z} \lim_{M\rightarrow\infty} \frac \beta M \sum_{r,n,j}
 \langle \bar\psi_{r,t,j} \psi_{r,t,j} \rangle
$$
So, because of $\langle\bar\psi_{r,n,1}\psi_{r,n,1}\rangle=
\langle\bar\psi_{r,n,2}\psi_{r,n,2}\rangle=C(r,r)$, we find the result
\begin{equation}
 n_{\rm tot} = \frac{N_{\rm tot}}{\cal N} = 1 - C(r,r)
 \label{d11_ntot-finite-T}
\end{equation}
for the total particle density. Note that the time slice $\varepsilon$, which was necessary for
the definition of the total particle density for weakly interacting bosons
(see Eq. (\ref{FI1_N-CF})), is absent here, because of the construction of the Green's matrix.
The zero temperature result is
\begin{equation}
 \lim_{\beta\rightarrow\infty} n_{\rm tot} =\left\{ \begin{array}{l@{\mbox{ if }}l}
 0 & \mu<-J \\ 1-\frac 1\pi \arccos\left(\frac\mu{J}\right) & -J<\mu<J \\
 1 & \mu<J \end{array} \right. \; .
\end{equation}
Graphs for zero temperature and finite temperature are 
plotted in Fig. \ref{Fig_d11_ntot1}. Both graphs are symmetric to the point $\mu/J=0$,
$n_{\rm tot}=1/2$. This reflects the particle hole symmetry of the system: Because of 
the Pauli principle a given configuration of the system is symmetric to the configuration, 
in which each occupied site is empty and vice versa. 
Further one can see that the system is empty 
($n_{\rm tot}=0$) if $\mu/J<-1$, and it is a Mott-insulator ($n_{\rm tot}=1$) if $\mu/J>1$.
The phase transitions between the EP and the incommensurate phase with $0<n_{\rm tot}<1$,
and between the ICP and the MI, are characterised by a diverging slope of the curve
at the transition points.
At non-zero temperatures the sharp phase transition is smeared out.
The zero temperature phase diagram is depicted schematically in
Fig. \ref{Fig_d11_phase-diagram1}.

\subsubsection{Density correlations and static structure factor}

We define the truncated density-density CF of the hard-core Bose gas as
\begin{equation}
 D(r-r') = \left\langle \bar\psi_{r,n,1}\psi_{r,n,1}
 \bar\psi_{r',n,1}\psi_{r',n,1} \right\rangle - \underbrace{
 \left\langle \bar\psi_{r,n,1}\psi_{r,n,1}\right\rangle
 \left\langle \bar\psi_{r',n,1}\psi_{r',n,1}\right\rangle}_{\displaystyle = n_{\rm tot}^2} \; .
\end{equation}
Using Wick's theorem for Grassmann variables as given in Appendix \ref{App_Wick}, we find
$$
 \left\langle \bar\psi_{r,n,1}\psi_{r,n,1}
 \bar\psi_{r',n,1}\psi_{r',n,1} \right\rangle =
 n_{\rm tot}^2 - C(r,r') C(r',r) \; ,
$$
leading to the result
\begin{equation}
 D(r-r') = - C(r,r') C(r',r) \; .
 \label{d12_D-general}
\end{equation}
The static structure factor is related to the density-density CF by means of a Fourier 
transformation which is shifted by unity, and a normalisation. We use the definition
\cite{ates1,astra1}
\begin{equation}
 S(q) = 1 + \frac{\sum_{r,r'} D(r-r') e^{{\rm i}q(r-r')}}{
 \sum_{r,r'} D(r-r')} \; .
\end{equation}
It is the analog to
the definition of the static structure factor of an ideal Bose gas
(\ref{FI3_S-without-slice}), where the term $1$ appears when the time slice is canceled in
the expectation value of the complex fields. Expressed in terms of the one-particle
CF in momentum space $C(k)$ by applying the Fourier transformation in Eq.
(\ref{d11_Fourier}), the above expression reads
\begin{equation}
 S(q) = 1 - \frac{\int_0^{2\pi} C(k) C(k+q)\,{\rm d}k}{\int_0^{2\pi} C(k)^2\,{\rm d}k} \; .
 \label{d12_S-general}
\end{equation}

We want to investigate the static structure factor at zero temperature in the ICP near
the phase transitions to the EP and the MI. Because of the particle-hole symmetry discussed
in the previous section, both transitions should be symmetric with respect to the physics of 
light scattering. Let us first discuss the region $\mu>0$.
Defining the characteristic wave vector $k^\star$ we find the result
\begin{equation}
 S(q) = \left\{ \begin{array}{l@{\quad}l}
 \frac q{2k^\star} &  \mbox{if } q<2k^\star \\
 1 & \mbox{if } 2k^\star<q<2\pi-2k^\star \\
 \frac{2\pi-q}{2k^\star} & \mbox{if } q>2\pi-2k^\star
 \end{array} \right. \; .
 \label{d12_S-result}
\end{equation}
In order to keep the particle hole symmetry for the static structure factor, in the region 
$\mu<0$ we make the substitution $C(k)\rightarrow 1-C(k)$ in the expression (\ref{d12_S-general}), 
and find the same result as in Eq. (\ref{d12_S-result}). The expression for 
the density-density CF $D(r-r')$ near both phase transitions we get from the Eqs. 
(\ref{d11_Fourier}) and (\ref{d12_D-general}). At zero temperature it is
\begin{equation}
 D(r-r') = \left( \frac{\sin(k^\star(r-r'))}{2\pi(r-r')} \right)^2 \; .
 \label{d12_D-result}
\end{equation}
The characteristic wave vector can be written in terms 
of the total particle density (\ref{d11_ntot-finite-T}):
\begin{equation}
 k^\star =  \left\{ \begin{array}{l@{\quad}l}
 2\pi n_{\rm tot} & \mbox{if } n_{\rm tot}<1/2 \\
 2\pi(1-n_{\rm tot}) & \mbox{if } n_{\rm tot}>1/2 \end{array} \right. \; .
 \label{d12_k-star2}
\end{equation}
Near the phase transitions where $\delta:=|\mu-\mu_c|/J\ll 1$, we have $\mu=(1-\delta)J$
at the ICP-MI phase transition, and $\mu=-(1-\delta)J$ at the ICP-EP transition.
Here, we can approximate
\begin{equation}
 k^\star \approx \sqrt{8\delta} \; .
\end{equation}
%------------------------------ FIGURE -------------------------------------
\begin{figure}
\centering
\includegraphics{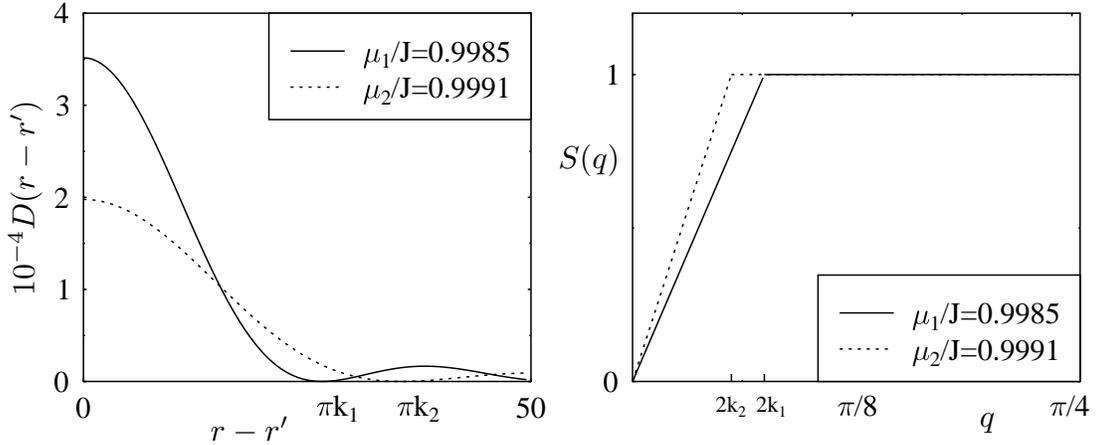}
\caption{Truncated density-density correlation function $D(r-r')$ and static structure factor
$S(q)$ in the vicinity of the ICP-MI phase transition. The transition point is at
$\mu_c=J$. For the ICP-EP phase transition, the situation is symmetrical.}
\label{Fig_d12_DundS}
\end{figure}
%---------------------------------------------------------------------------
For a homogeneous impenetrable Bose gas the role of $k^\star$ is played by the Fermi
wave vector $k_{\rm F}=\pi n_{\rm tot}$ \cite{astra1}.
In our result (\ref{d12_k-star2}), $k^\star$ depends linearly on the density as well as
in the region $n_{\rm tot}<1/2$, but the discontinuous slope of the function 
$k^\star(n_{\rm tot})$ at the point $n_{\rm tot}=1/2$ is a consequence of the optical lattice
potential. The relation (\ref{Int1_Feynman}) allows us to identify
the excitation spectrum
\begin{equation}
 \epsilon(q)=\hbar cq+{\cal O}(q^2) \; , \quad
 c=\frac{\hbar k^\star}m \; . \label{d12_Ek}
\end{equation}
which is linear for small values of $q$, where $c$ is the sound velocity.
The density-density CF and the static structure factor near the ICP-MI phase transition
are plotted in Fig. \ref{Fig_d12_DundS}.

The density-density CF shows characteristic oscillations with length $\lambda=\pi/k^\star$.
This length scale diverges at the ICP-EP and ICP-MI phase transition with $1/n_{\rm tot}$ and
$1/(1-n_{\rm  tot})$, respectively. Thus it can be used as a measure for the distance
of the system to one of the two phase transitions. In the EP and the MI phase, the
density-density CF vanishes because of the absence of particle number fluctuations,
and the static structure factor saturates to $S(q)\equiv 1$.

\subsubsection{External trap potential\label{Sec_d12-trap}}

In the previous sections a system in a translational invariant lattice was considered.
Calculations have also been made for a one-dimensional Bose gas in a harmonic trap potential
\cite{ates1}
\begin{equation}
 V(r) = \frac m2 \omega_{\rm ho}^2 (ar)^2 \; ,
\end{equation}
where again $a$ is the lattice constant,
and $\omega_{\rm ho}$ is the harmonic oscillator frequency of the trap.
%------------------------------ FIGURE -------------------------------------
\begin{figure}
\includegraphics{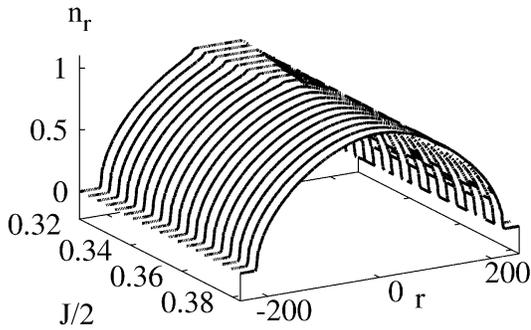}
\caption{Local particle density for system in harmonic trap potential
($\mu=0.7$, $ma^2\omega_{\rm ho}^2/2=3\times 10^{-5}$) with varying tunneling 
rate $J$. A Mott plateau appears in the center of the
trap ($r$=0) as $J$ is decreased below a critical value $J_{\rm P}\approx 0.70$.
(Fig. taken from ref. \cite{ates1}.)}
\label{Fig_d12_plateau}
\end{figure}
%---------------------------------------------------------------------------
The numerical result for the local particle density at zero temperature
is plotted in Fig. \ref{Fig_d12_plateau}, where the formation
of a Mott plateau can be seen below a critical value $J_{\rm P}$. A similar behavior was found for the one-dimensional Bose-Hubbard model with a harmonic 
trapping potential \cite{bergkvist1}.

The properties of the density-density CF and the static 
structure factor are qualitatively the same as in
the translational invariant case. $D(r)$ vanishes when $J_{\rm P}$ is reached, owing to the fact
that there are no density fluctuations within the plateau. The characteristic length scales become
larger as the Mott plateau is reached. 

%++++++++++++++++++++++++++++++++++++++++++++++++++++++++++++++++++++++++++++++++++++++++
%++++++++++++++++++++++++++++++++++++++++++++++++++++++++++++++++++++++++++++++++++++++++
%++++++++++++++++++++++++++++++++++++++++++++++++++++++++++++++++++++++++++++++++++++++++

\section{Weakly interacting bosons: Bogoliubov theory\label{Chap_weak}}

Before discussing an interacting Bose gas in an optical lattice, we begin with
the derivation of the Bogoliubov approximation for a dilute homogeneous
Bose gas. Although the Bogoliubov theory can also be applied for bosons in a lattice
potential, a Mott-insulating phase is not found within this approximation \cite{stoof1}.
Many aspects of the physics discussed in this chapter show
up in the hard-core Bose gases in optical lattices as well.

\subsection{Derivation from saddle point approximation\label{Sec_Bog-saddle-point}}

It might be interesting to derive the results of
Bogoliubov theory from
the functional integral point of view. The method which will be used here and in the
following chapters is the saddle point approximation (or: stationary phase approximation,
Gaussian approximation) \cite{Bnegele,Bma,Bnagaosa}. It allows to find a mean-field solution
plus fluctuations around the mean-field result. The mean-field solution is connected
to the condensate order parameter, while the fluctuations contain the information
about the quasiparticles and their spectrum.
The saddle-point approximation is good as long as these fluctuations are small.

The main idea of a saddle point approximation is to expand the action of the system around
its minimum up to second order in the field variables. This leads to a Gaussian integral
which can be performed. The action of a bosonic system is given in Eq. (\ref{FI1_S-boson}),
where in this case the index $\alpha$ shall denote the position vector $\bf r$.
Together with the Hamiltonian (\ref{Int1_H}) of the interacting Bose gas we have
$$
 A(\phi^\ast,\phi) = \frac 1\hbar \int_0^{\hbar\beta} {\rm d}\tau \int {\rm d}^3 r
 \Bigg\{ \phi^\ast({\bf r},\tau) \left[\left(
 \hbar\frac\partial{\partial\tau}-\mu\right) - \frac{\hbar^2}{2m}\nabla^2 + 
 V_{\rm ext}({\bf r})\right] \phi({\bf r},\tau)
$$
\begin{equation} 
 + \frac g2 \left|\phi({\bf r},\tau)\right|^4 \Bigg\} \; .
 \label{Bog2_action}
\end{equation}
By minimising $A$ with respect to the complex fields we get a mean-field equation for the 
condensate order parameter $\Phi_0({\bf r},\tau)$:
\begin{equation}
 \left( -\frac{\hbar^2}{2m}\nabla^2 + V_{\rm ext}({\bf r}) +
 g \left|\Phi_0({\bf r},\tau)\right|^2 \right) \Phi_0({\bf r},\tau) =
 -\left(\frac\partial{\partial\tau}-\mu\right) \Phi_0({\bf r},\tau) \; .
 \label{Bog2_GPE}
\end{equation}
After performing the analytic continuation $\frac\partial{\partial\tau} \rightarrow
-i\hbar \frac\partial{\partial t}$
and omitting the chemical potential term, this is identical to the time-dependent 
Gross-Pitaevskii equation (\ref{Int1_GPE}).
We recall that the invariance of the mean-field solution under a
gauge transformation (\ref{Int1_global-U1}) with global phase $\alpha$
reflects the broken global $U(1)$ symmetry of the BEC phase.

To find the results from the previous sections in this chapter, we assume a homogeneous
system, i.e. $V_{\rm ext}({\bf r})\equiv 0$ in the action (\ref{Bog2_action}).
Further we assume that the mean-field solution is constant in space and
imaginary time: $\Phi_0({\bf r},\tau)\equiv \Phi_0$. In this case, the solution of
Eq. (\ref{Bog2_GPE}) is
\begin{equation}
 |\Phi_0|^2 = n_0 = \frac\mu g \; .
 \label{Bog2_mu}
\end{equation}
We now write the complex field as the sum of the mean-field solution plus fluctuations
\begin{equation}
 \phi({\bf r},\tau) = \Phi_0 + \delta\phi({\bf r},\tau) \; , \quad
 \phi^\ast({\bf r},\tau) = \Phi^\ast_0 + \delta\phi^\ast({\bf r},\tau) \; ,
\end{equation}
where the complex field of fluctuations $\delta\phi$ is considered to be small, such that
those terms in the action which are of higher than second order in the fluctuations,
can be neglected. We split the quasiparticle field into its real and imaginary part
and write $\delta\phi({\bf r},\tau)=\delta\phi'+{\rm i}\delta\phi''$, $\delta\phi^\ast({\bf r},\tau)=\delta\phi'-{\rm i}\delta\phi''$. The expansion yields
\begin{equation}
 A \approx A_0 + \frac 1\hbar \int_0^{\hbar\beta} {\rm d}\tau \int {\rm d}^3 r \left(
 \begin{array}{l} \delta\phi' \\ \delta\phi'' \end{array} \right)
 \cdot \left( \begin{array}{cc}
 -\frac{\hbar^2}{2m}\nabla^2 & -{\rm i} \, \hbar\frac\partial{\partial\tau} \\
 {\rm i}  \hbar\frac\partial{\partial\tau} & -\frac{\hbar^2}{2m}\nabla^2 + 2\mu
 \end{array} \right)\left(
 \begin{array}{l} \delta\phi' \\ \delta\phi'' \end{array} \right) \; ,
\end{equation}
where we have already eliminated the condensate order parameter by Eq. (\ref{Bog2_mu}),
and the zeroth-order part of the action is
\begin{equation}
 A_0 = \beta V \left( -\mu|\Phi_0|^2 + \frac g2 |\Phi_0|^4 \right) 
 = -\frac{\beta V \mu^2}{2g} \; .
\end{equation}
Because $A_0$ does not depend on the field fluctuations, and the second term is of second
order in $\delta\phi$ and $\delta\phi^\ast$, the functional integral for the grand canonical
partition function
\begin{equation}
 Z = \int e^{-A(\delta\phi',\delta\phi'')} 
 {\cal D}(\delta\phi'({\bf r},\tau)\delta\phi''({\bf r},\tau))
\end{equation}
can be performed because it is Gaussian.
We Fourier transform the field of fluctuations with respect to the spacial coordinate like
\begin{eqnarray}
 \delta\phi'({\bf r},\tau) &=& \frac 1{\sqrt{2\pi V}} \sum_{\bf k}
 \delta\phi'_{\bf k}(\tau) \cos({\bf k\cdot r}) \\
 \delta\phi''({\bf r},\tau) &=& \frac 1{\sqrt{2\pi V}} \sum_{\bf k}
 \delta\phi''_{\bf k}(\tau) \cos({\bf k\cdot r}) \; ,
\end{eqnarray}
with the constraints $\delta\phi'_{\bf k}=\delta\phi'_{-\bf k}$ and 
$\delta\phi''_{\bf k}=\delta\phi''_{-\bf k}$ and thus get
\begin{equation}
 A = A_0 + \frac 1\hbar \int_0^{\hbar\beta} {\rm d}\tau \, \sum_{\bf k}\left(
 \begin{array}{l} \delta\phi'_{\bf k}(\tau) \\ \delta\phi''_{\bf k}(\tau)
 \end{array} \right)
 \cdot \left( \begin{array}{cc}
 \epsilon_{\bf k} & -{\rm i} \hbar \frac\partial{\partial\tau} \\
 {\rm i} \hbar \frac\partial{\partial\tau} & \epsilon_{\bf k} + 2\mu
 \end{array} \right)\left(
 \begin{array}{l} \delta\phi'_{\bf k}(\tau) \\ \delta\phi''_{\bf k}(\tau)
 \end{array} \right)
 \label{Bog2_continuous-time}
\end{equation}
with the free-particle dispersion relation $\epsilon_{\bf k}=\hbar^2{\bf k}^2/2m$. 
It is further possible to perform a Fourier transformation with respect to
the imaginary time coordinate as well, namely
\begin{eqnarray}
 \delta\phi'_{\bf k}(\tau) &=& \frac 1{\sqrt\beta} \sum_n
 \delta\phi'_{{\bf k},\omega_n} \cos(\omega_n \tau) \\
 \delta\phi''_{\bf k}(\tau) &=& \frac 1{\sqrt\beta} \sum_n 
 \delta\phi''_{{\bf k},\omega_n}  \cos(\omega_n \tau) \; ,
\end{eqnarray}
with the Matsubara frequencies for bosons $\omega_n=2\pi n/\hbar\beta$ and the constraints
$\delta\phi'_{{\bf k},\omega_n}=\delta\phi'_{{\bf k},-\omega_n}$ and
$\delta\phi''_{{\bf k},\omega_n}=\delta\phi''_{{\bf k},-\omega_n}$.
This leads to the form
\begin{equation}
 A = A_0 + \sum_{{\bf k},n} \left(
 \begin{array}{l} \delta\phi'_{{\bf k},\omega_n} \\ \delta\phi''_{{\bf k},\omega_n}
 \end{array} \right)
 \cdot {\cal G}^{-1}({\bf k},\omega_\nu) \, \left(
 \begin{array}{l} \delta\phi'_{{\bf k},\omega_n} \\ \delta\phi''_{{\bf k},\omega_n}
 \end{array} \right) \; ,
 \label{Bog2_S-matsubara}
\end{equation}
and allows to identify the quasiparticle Green's function 
(a $2\times 2$ matrix in this case)
\begin{equation}
 {\cal G}^{-1}({\bf k},{\rm i}\hbar\omega_n) = \left( \begin{array}{cc}
 \epsilon_{\bf k} & {\rm i} \hbar\omega_n \\
 {\rm i} \hbar\omega_n & \epsilon_{\bf k} + 2\mu \end{array} \right) \; .
 \label{Bog2_qp-Greens}
\end{equation}
The excitation energies of the quasiparticles are given by the poles of the quasiparticle
Green's function \cite{Bnegele}, which are found by solving the equation
\begin{equation}
 \det \, {\cal G}^{-1}({\bf k},{\rm i}\hbar\omega_n) = 0 \; .
 \label{Bog2_qp-pole}
\end{equation}
After performing the analytic continuation ${\rm i}\hbar\omega_n \longrightarrow E_{\bf k}$
we get
\begin{equation}
 E_{\bf k} = \sqrt{\epsilon_{\bf k} \left( 2\mu+\epsilon_{\bf k} \right)} \; ,
 \label{Bog1_bogol-spectrum}
\end{equation}
which is identical to the Bogoliubov spectrum, if the relation $n_0=\mu/g$ is inserted.

\subsection{Partition function and condensate depletion}

To find the correct expression for the grand canonical partition function as well as 
for the correlation functions, we have to perform the same steps as in section
\ref{Sec_ideal-gas-cf}, namely to start with the discrete-time functional integral
and sending the number of time steps $M$ to infinity at the end.
By analogy with Eq. (\ref{FI2_Z-matrix}), the discrete-time version of Eq.
(\ref{Bog2_continuous-time}) is
\begin{equation}
 A_{\rm discrete} = A_0 + \sum_{\bf k} \sum_{n,m=1}^M \left(
 \begin{array}{c} \delta\phi'_{{\bf k},n} \\ \delta\phi''_{{\bf k},n} \end{array} \right)
 \cdot \hat{A}_{nm}^{({\bf k})} \left(
 \begin{array}{c} \delta\phi''_{{\bf k},m} \\ \delta\phi'_{{\bf k},m} \end{array} \right) \; ,
\end{equation}
where $\hat{A}_{nm}^{({\bf k})}$ has the $M\times M$ structure
\begin{equation}
 \hat{A}^{({\bf k})} = \left[
 \begin{array}{cccccc}
  \hat{B} & -\hat{b}^\ast_{\bf k} & 0 & \cdots & 0 & -\hat{b}_{\bf k} \\
  -\hat{b}_{\bf k} & \hat{B} & -\hat{b}^\ast_{\bf k} & 0 & \quad & \quad \\
  0 & -\hat{b}_{\bf k} & \hat{B} & \ddots & & \vdots \\
  \quad & 0 & -\hat{b}_{\bf k} & \ddots & -\hat{b}^\ast_{\bf k} & 0 \\
  \vdots & \quad & 0 & \ddots & \hat{B} & -\hat{b}^\ast_{\bf k} \\
  -\hat{b}^\ast_{\bf k} & \quad & \quad & \cdots & -\hat{b}_{\bf k} & \hat{B}
 \end{array} \right]
 \end{equation}
in the imaginary time variables $n$ and $m$, and each matrix entry is by itself a 
$2\times 2$ matrix:
\begin{equation}
 \hat{b}_{\bf k} = \frac 12 \left( 1-\frac\beta M (\epsilon_{\bf k}+\mu)\right)
 \left( \begin{array}{cc} 1 & {\rm i} \\  -{\rm i} & 1 \end{array} \right)
 \; , \quad \hat{B} = \left(\begin{array}{cc} 1+\frac\beta M \mu & 0\\
  0 & 1-\frac\beta M \mu \end{array}\right) \; .
\end{equation}
The matrix can be diagonalized by using the same unitary transformation
(\ref{FI2_unitary-trafo}), which was applied for the ideal Bose gas. This yields
$$
 (U\hat{A}^{({\bf k})}U^+)_{kn} = 
$$
\begin{equation}
 \delta_{kn} \, \left[ \left( \begin{array}{cc}
 1+\frac\beta M \mu & 0\\ 0 & 1-\frac\beta M \mu \end{array}\right) -
 \left( 1-\frac\beta M (\epsilon_{\bf k}+\mu)\right) \left( \begin{array}{cc} 
 \cos\left(\frac{2\pi}M n\right) & \sin\left(\frac{2\pi}M n\right) \\
 -\sin\left(\frac{2\pi}M n\right) & \cos\left(\frac{2\pi}M n\right) \end{array} \right)
 \right] \; . \label{Bog2_diag-matrix}
\end{equation}
Using the product given in Appendix \ref{App_product}, the determinant of the matrix can be
found as
$$
 \det\, \hat{A}^{({\bf k})} = \left( 1-\frac\beta M\left(\epsilon_{\bf k}+\mu\right)\right)
 \Bigg[ -2+
$$
\begin{equation}
 \left(1 + \frac\beta M \, 
 \sqrt{\epsilon_{\bf k} \left(\epsilon_{\bf k}+ 2\mu\right)} + 
 {\cal O}\left(\frac\beta M\right)^2 \right)^M + \left(1 - \frac\beta M \, 
 \sqrt{\epsilon_{\bf k} \left(\epsilon_{\bf k}+ 2\mu\right)} + 
 {\cal O}\left(\frac\beta M\right)^2 \right)^M \Bigg] \; .
\end{equation}
Thus we obtain the grand canonical partition function of the Bogoliubov Hamiltonian
(after omitting a constant factor):
\begin{equation}
 Z = e^{-A_0} \, \lim_{M\rightarrow\infty} \prod_{{\bf k}\neq 0}
 \left[\det \hat{A}^{({\bf k})}\right]^{-\frac 12} =
 \exp\left(\frac{\beta V\mu^2}{2g}\right) \prod_{{\bf k}\neq 0}
 e^{\frac\beta 2 \left(\epsilon_{\bf k}+\mu\right)} 
 \left[\cosh(\beta E_{\bf k}) -1\right]^{-\frac 12}  \; .
\end{equation}

The distribution function of the particles outside of the condensate is given as
\begin{equation}
 \langle n_{\bf k}\rangle = 
 \langle \delta\phi^\ast_{\bf k}(0) \delta\phi_{\bf k}(0) \rangle =
 \langle \delta\phi'_{\bf k}(0)^2 \rangle + \langle \delta\phi''_{\bf k}(0)^2 \rangle =
 \lim_{M\rightarrow\infty}\frac 12 \left(
 ([\hat{A}^{({\bf k})}]^{-1}_{11})_{nn} + ([\hat{A}^{({\bf k})}]^{-1}_{22})_{nn} \right) \; ,
\end{equation}
with the $11$- and the $22$-component of the matrix with respect to the $2\times 2$ structure.
After inversion of the matrix (\ref{Bog2_diag-matrix}) and the back transformation, the matrix
elements can be found and after performing the limit $M\rightarrow\infty$ we get
\begin{equation}
 \langle n_{\bf k}\rangle = -\frac 12 + \frac{\epsilon_{\bf k}+\mu}{2E_{\bf k}} \,
 \coth\left(\frac\beta 2 E_{\bf k}\right) \; .
 \label{Bog2_nk}
\end{equation}
The quantity
\begin{equation}
 n_{\rm tot}-n_0=\int\langle n_{\bf k}\rangle\, \frac{{\rm d}^3k}{(2\pi)^3}
\end{equation}
is called condensate depletion. Contrary to the ideal Bose gas it is non-zero at zero temperature.

\subsection{Static structure factor}

The static structure factor is given by the fourth-order expectation value
(\ref{FI3_S-without-slice}), which we used for the ideal gas before. 
We replace $\phi_0$ by the order parameter
$\Phi_0$ and for non-zero momenta we replace $\phi_{\bf k}\rightarrow\delta\phi_{\bf k}$.
After splitting the fluctuations into real and imaginary part and applying Wick's theorem
for real variables, we get a similar result as in Eq. (\ref{FI3_S-ideal-final}).
The difference to the ideal Bose gas is, that the anomalous expectation values 
$\langle\phi^\ast_{\bf k}\phi^\ast_{-\bf k}\rangle$ and $\langle\phi_{\bf k}\phi_{-\bf k}\rangle$
also give a contribution here (for simplicity we have dropped the time variable).
The contribution of the anomalous expectation values after splitting it into its
real and imaginary part is
$$
 \left\langle\delta\phi^\ast_{\bf q}\delta\phi^\ast_{-\bf q}\right\rangle +
 \left\langle\delta\phi_{\bf q}\delta\phi_{-\bf q}\right\rangle  = 2 \left(
 \left\langle(\delta\phi'_{\bf k})^2\right\rangle - 
 \left\langle(\delta\phi''_{\bf k})^2\right\rangle \right) \; ,
$$
such that the static structure factor is given as
$$
 S({\bf q}) = 1+2\frac{N_0}{N_{\rm tot}} \langle n_{\bf q} \rangle + \frac{N_0}{N_{\rm tot}}
 \left( \langle\delta\phi^\ast_{\bf q}\delta\phi^\ast_{-\bf q}\rangle +
 \langle\delta\phi_{\bf q}\delta\phi_{-\bf q}\rangle \right) +
 \sum_{{\bf k}\neq\{0,{-\bf q}\}} \langle n_{\bf k} \rangle\langle n_{\bf k+q} \rangle =
$$
\begin{equation}
 1+4\frac{N_0}{N_{\rm tot}} \left\langle(\delta\phi'_{\bf q})^2\right\rangle +
 \sum_{{\bf k}\neq\{0,{-\bf q}\}} \langle n_{\bf q} \rangle\langle n_{\bf k+q} \rangle \; .
 \label{Bog2_S}
\end{equation}
After performing the limit $M\rightarrow\infty$ we find
\begin{equation}
 \left\langle(\delta\phi'_{\bf q})^2\right\rangle = 
 \lim_{M\rightarrow\infty} \frac 12[\hat{A}^{({\bf q})}]^{-1}_{11}=
 -\frac 14  + \frac 14 \frac{\epsilon_{\bf q}}{E_{\bf q}} \,
 \coth\left(\frac\beta 2 E_{\bf q}\right) \; .
\end{equation}
If we neglect the last term in Eq. (\ref{Bog2_S}) which is quadratic in the momentum 
distribution, this expression reduces to 
\begin{equation}
 S({\bf q}) = \frac{\epsilon_{\bf q}}{E_{\bf q}} \, \coth\left(\frac\beta 2 E_{\bf q}\right) \; .
 \label{Bog2_S-final}
\end{equation}
$S({\bf q})=\epsilon_{\bf q}/E_{\bf q}$.

To determine the type of the decay of the density-density correlations for large distances
(i.e. exponentially or algebraically) at zero temperature in $d$ dimensions in the BEC phase,
we Fourier transform the static structure factor for small wave vectors, because they are
relevant for large distances $\bf r$:
\begin{equation}
 D({\bf r}) \sim \int S({\bf q}) e^{{\rm i}\bf q\cdot r} {\rm d}^dq 
 \sim \int \frac{{\bf q}^2}{\sqrt{2(\mu+J){\bf q}^2+{\bf q}^4}} \,
 e^{{\rm i}\bf q\cdot r} {\rm d}^dq
 \sim \int \frac{|{\bf q}|}{\sqrt{2(\mu+J)}} \, e^{{\rm i}\bf q\cdot r} {\rm d}^dq  \; .
 \label{cfm3_D-decay}
\end{equation}
This expression shows an algebraic decay. In $d=1$ the decay is proportional to $1/r^2$
(in agreement with the result (\ref{d12_D-result}) of the one-dimensional system),
in $d=2$ it decays like $1/r^3$, and in $d=3$ like $1/r^4$ (see Appendix \ref{App_corr}).
In the empty phase, all CFs vanish completely at zero temperature.
Thus, the static structure factor is constantly unity.

%++++++++++++++++++++++++++++++++++++++++++++++++++++++++++++++++++++++++++++++++++++++++
%++++++++++++++++++++++++++++++++++++++++++++++++++++++++++++++++++++++++++++++++++++++++
%++++++++++++++++++++++++++++++++++++++++++++++++++++++++++++++++++++++++++++++++++++++++

\section{Strongly interacting bosons in the dense regime\label{Chap_strong}}

\subsection{Paired-fermion model\label{Chap_PF}}

\subsubsection{Bosonic molecules of spin-$1/2$ fermions}

We now introduce a model of hard-core bosons which are constructed by
molecules consisting of pairs of spin-$1/2$ fermions, as an alternative to the
hard-core boson model. In order to distinguish 
it from the latter this model will be referred to as ``paired-fermion model''.

A general model which was introduced to study the dissociation of bosonic molecules into
pairs of fermionic atoms in an optical lattice was proposed in ref. \cite{ziegler3}. 
It is given by the Hamiltonian
\begin{equation}
 \hat{H} - \mu\hat{N}_{\rm tot} = 
 - \frac{\bar{t}}{2d} \sum_{\langle{\bf r,r'}\rangle}\sum_{\sigma=\uparrow,\downarrow}
 \hat{c}^+_{{\bf r},\sigma}\hat{c}_{{\bf r},\sigma} - \frac{J}{2d}\sum_{\langle
 {\bf r,r'}\rangle} \hat{c}^+_{{\bf r}\uparrow} \hat{c}_{{\bf r'}\uparrow}
 \hat{c}^+_{{\bf r}\downarrow} \hat{c}_{{\bf r'}\downarrow}-\mu\sum_{\bf r}
 \sum_{\sigma=\uparrow\downarrow} \hat{c}^+_{{\bf r}\sigma} \hat{c}_{{\bf r}\sigma} \; .
 \label{cfm1_H-mostgeneral}
\end{equation}
The index $\sigma=\uparrow,\downarrow$ denotes the spin. The first term describes tunneling
of individual fermions with rate $\bar t$ and the second term tunneling of local fermion pairs.
Similar Hamiltonians were proposed in a couple of works for homogeneous systems, in order to
study the BEC-BCS crossover \cite{chiofalo1,ohashi1,holland1}. In contrast to the 
lattice-Hamiltonian (\ref{cfm1_H-mostgeneral}) they do not exhibit a Mott insulating phase.

Because the main interest here shall be the model of hard-core bosons, we consider the case
$\bar t=0$ in the following, i.e. we exclude the existence of dissociated fermionic atoms.
Further we will write the index $\sigma=1,2$ as superscript 
instead of the spin indices $\uparrow,\downarrow$.
We write the grand canonical partition function of the system in terms of a fermionic
functional integral of a field of conjugate Grassmann variables as defined in Eq.
(\ref{FI1_Z-fermion-discrete}) with the action
\begin{displaymath}
 A_{\rm ferm}(\bar\psi,\psi) = \sum_{n=1}^M \Bigg\{ \sum_{{\bf r},\sigma}
 \bar\psi^\sigma_{{\bf r},n+1} (\psi^\sigma_{{\bf r},n+1}-\psi^\sigma_{{\bf r},n}) - 
 \frac 12 \,\frac{\beta\mu}M \sum_{{\bf r},\sigma}
 \bar\psi^\sigma_{{\bf r},n+1} \psi^\sigma_{{\bf r},n}
\end{displaymath}
\begin{equation}
 + \frac\beta M \, \sum_{\bf r,r'} \hat{J}_{\bf rr'}
 \bar\psi^1_{{\bf r},n+1} \psi^1_{{\bf r'},n} 
 \bar\psi^2_{{\bf r},n+1} \psi^2_{{\bf r'},n} \Bigg\} \; ,
 \label{cfm1_Sferm}
\end{equation}
with anti-periodic boundary conditions in time.
Here, we have replaced $\mu\rightarrow\mu/2$ due to the fact that the chemical 
potential is associated with the number of {\it paired} fermions (i.e. to the bosonic molecules),
hence the factor $1/2$ in front of the term which contains $\mu$, while in
Eq. (\ref{cfm1_H-mostgeneral}), $\hat{N}_{\rm tot}$ is the particle number operator of 
single fermions.

In the world-line picture, the paired-fermion model
given by $A_{\rm ferm}$ is represented by pairs of fermions with opposite spin $1$ and $2$ whose
world-lines always stay together while they tunnel through the lattice. Tunneling of unpaired
fermions does not exist. The world-lines of two fermions of species 1 and 2 always stick 
together while tunneling through the lattice.

\subsubsection{Hubbard-Stratonovich decoupling}

The idea of a Hubbard-Stratonovich transformation is to decouple a quartic term
of a many-body system by writing it in terms of a Gaussian integral \cite{hubbard1}. The 
original field variables are then only of second order and can be integrated out such that 
the system is represented only by the field variables of the Gaussian integral.

We perform a Hubbard-Stratonovich transformation on the system of paired fermions 
\cite{ziegler3} given by Eq. (\ref{cfm1_Sferm}).
Only the term which describes hopping of fermion pairs is quartic, so we will decouple it.
Contrary to the case of the hard-core boson model, it is not necessary here to decouple 
the entire off-diagonal term, because the term describing the discrete-time derivative and 
the term containing the chemical potential are already of second order.
For the matrix with fermionic boundary conditions we write
\begin{equation}
 \hat{v}^{\rm ferm}_{{\bf rr'};nm} = (\delta_{n,m+1} - \delta_{n1}\delta_{mM})
 \, \frac\beta M \,\hat{J}_{\bf rr'} + s \, \delta_{nm} \; ,
\end{equation}
and insert the identity
\begin{displaymath}
 {\rm const.} \times \exp \left\{ -\frac\beta M\sum_{\bf r,r'} \sum_{n,m=1}^M \hat{J}_{\bf rr'}
 \bar\psi^1_{{\bf r},n+1} \psi^1_{{\bf r'},n} \bar\psi^2_{{\bf r},n+1} \psi^2_{{\bf r'},n} \right\}
\end{displaymath}
\begin{displaymath}
 = \int\exp \Bigg\{ -\frac\beta M \sum_{\bf r,r'} \sum_{n,m} \varphi^\ast_{{\bf r},n} 
   (\hat{v}^{\rm ferm}_{{\bf rr'};nm})^{-1} \varphi_{{\bf r}',m}
 - \frac 1s \sum_{{\bf r},n} \chi^\ast_{{\bf r},n} \chi_{{\bf r},n}
\end{displaymath}
\begin{equation}
 + \sum_{{\bf r},n} \left[ \psi^2_{{\bf r},n}\psi^1_{{\bf r},n} 
 ({\rm i}\varphi^\ast_{{\bf r},n}+\chi^\ast_{{\bf r},n})
 + \bar\psi^1_{{\bf r},n+1}\bar\psi^2_{{\bf r},n+1} 
 ({\rm i}\varphi_{{\bf r},n}+\chi_{{\bf r},n}) \right] \Bigg\} \,
 \prod_ {{\bf r},n} \frac{{\rm d}\varphi^\ast_{{\bf r},n}{\rm d}\varphi_{{\bf r},n}
 {\rm d}\chi^\ast_{{\bf r},n}{\rm d}\chi_{{\bf r},n}}{(2\pi{\rm i})^2} \; .
\end{equation}
The parameter $s$ cares for the convergence of the integral of the complex field
$\varphi$. For $\hat{v}^{\rm ferm}_{{\bf rr'};nm}$ we have the eigenvalues
\begin{equation}
 v^{\rm ferm}_{{\bf k},n} = e^{-{\rm i}\frac{2\pi}M \left(n-\frac 12\right)} \frac\beta M
 \tilde\epsilon_{\bf k} + s \; ,
\end{equation}
therefore one has to choose $s$ large enough such that all eigenvalues are non-negative, but 
besides this condition the choice of $s$ is free. We integrate out the Grassmann field in the
functional integral representation of the partition function, like we did in the previous section:
\begin{equation}
 Z_{\rm ferm} = \int \exp \, 
 [-\tilde{A}_{\rm ferm}(\varphi^\ast,\varphi,\chi^\ast,\chi)]
 \prod_ {{\bf r},n} \frac{{\rm d}\varphi^\ast_{{\bf r},n}{\rm d}\varphi_{{\bf r},n}
 {\rm d}\chi^\ast_{{\bf r},n}{\rm d}\chi_{{\bf r},n}}{(2\pi{\rm i})^2}
 \label{cfm2_Zferm-gen}
\end{equation}
with the action
\begin{equation}
 \tilde{A}_{\rm ferm}(\varphi^\ast,\varphi,\chi^\ast,\chi) =
 \sum_{\bf r,r'} \sum_{n,m} \varphi^\ast_{{\bf r},n} 
  (\hat{v}^{\rm ferm}_{{\bf rr'};nm})^{-1} \varphi_{{\bf r}',m}
 + \frac 1s \sum_{{\bf r},n} \chi^\ast_{{\bf r},n} \chi_{{\bf r},n}
 - \sum_{\bf r} \log \, \det \, \hat{\mathbb{G}}_{\bf r}^{-1} \; ,
 \label{cfm2_Sferm-gen}
\end{equation}
where we have introduced the matrix
\begin{eqnarray}
 \hat{\bf{G}}_{\bf r}^{-1} &=& \delta_{nm} \, \left( \begin{array}{cc}
 {\rm i}\varphi_{{\bf r},n}+\chi_{{\bf r},n} & 1 \\
 1 & -({\rm i}\varphi^\ast_{{\bf r},n}+\chi^\ast_{{\bf r},n}) \end{array}
 \right) \nonumber \\
 && - (\delta_{n,m+1} - \delta_{n1}\delta_{mM})
 \left( \begin{array}{cc} 0 & 1+\frac{\beta\mu}{2M} \\
  1-\frac{\beta\mu}{2M} & 0 \end{array} \right) \; .
 \label{cfm2_matrix-G}
\end{eqnarray}

\subsubsection{Saddle-point expansion}

Under the assumption
\begin{equation}
 \varphi^\ast_{{\bf r},n}\equiv\varphi^\ast_0 \; \quad \varphi_{{\bf r},n}\equiv\varphi_0 \; \quad
 \chi^\ast_{{\bf r},n}\equiv\chi^\ast_0 \; \quad \chi_{{\bf r},n}\equiv\chi_0 \; 
 \label{cfm2_mfs}
\end{equation}
that the mean-field solution is constant in space and time, we can 
Fourier transform the matrix $\hat{\bf G}_{{\bf r},n}^{-1}\equiv\hat{\bf{G}}_n^{-1} $ 
in Eq. (\ref{cfm2_matrix-G}) with respect to the discrete-time index:
\begin{equation}
 \hat{\bf{G}}_n^{-1} = \left( \begin{array}{cc}
 {\rm i}\varphi_0+\chi_0 & 1-e^{-\frac{{\rm i}2\pi}M\left(n-\frac 12\right)}
 \left(1+\frac{\beta\mu}{2M}\right) \\
 1-e^{-\frac{{\rm i}2\pi}M\left(n-\frac 12\right)}\left(1-\frac{\beta\mu}{2M}\right) 
 & -({\rm i}\varphi_0^\ast+\chi_0^\ast)
 \end{array} \right) \; .
\end{equation}
By the use of the identity 
$\sum_{{\bf r'},m}(\hat{v}^{\rm ferm}_{{\bf rr'};nm})^{-1}=(s-\beta J/M)^{-1}$ we have:
\begin{samepage}
\begin{displaymath}
 \frac{\tilde{A}^{\rm ferm}_0}{{\cal N}M} = \frac{\varphi_0^\ast\varphi_0}{
 s+\frac{\beta J}M} + \frac 1s \chi_0^\ast\chi_0
\end{displaymath}
\begin{equation}
 -\frac 1M \sum_{n=1}^M \log\left[ -({\rm i}\varphi_0+\chi_0)({\rm i}\varphi_0^\ast+\chi_0^\ast) -
 1 - e^{-2\frac{{\rm i}2\pi}M\left(n-\frac 12\right)}\left(1-\left(\frac{\beta\mu}{2M}\right)^2
 \right) + 2\, e^{-\frac{{\rm i}2\pi}M\left(n-\frac 12\right)} \right] \; .
\end{equation}
\end{samepage}
From the saddle point conditions
 \begin{equation}
 \frac{\partial\tilde{A}_{\rm ferm}}{\partial\varphi^\ast_{{\bf r},n}}
 = \frac{\partial\tilde{A}_{\rm ferm}}{\partial\varphi_{{\bf r},n}} = 0 \; , \quad
 \frac{\partial\tilde{A}_{\rm ferm}}{\partial\chi^\ast_{{\bf r},n}}
 = \frac{\partial\tilde{A}_{\rm ferm}}{\partial\chi_{{\bf r},n}} = 0 
\end{equation}
we find the mean-field equations
\begin{equation}
 \frac{\chi_0}s = -{\rm i}G \; , \quad \frac{\varphi_0}{s+\frac{\beta J}M} = G \; ,
 \label{cfm2_spe-fermion1}
\end{equation}
where $G$ is calculated in Appendix \ref{App_sumG} and 
the result is
\begin{equation}
 G = \frac{J\varphi_0/s}{\sqrt{\mu^2+\left(\frac{J|\varphi_0|}s\right)^2} } \tanh \left[
 \frac\beta 2 \sqrt{\mu^2+\left(\frac{J|\varphi_0|}s\right)^2} \right] \; .
 \label{cfm2_G-def}
\end{equation}
We find a trivial solution with $\varphi_0=\varphi^\ast_0=\chi_0=\chi^\ast_0=0$ and
a non-trivial solution with broken $U(1)$ symmetry.
For the mean-field action we find (after integrating $G$ with respect to ${\rm i}\varphi_0+\chi_0$):
\begin{equation}
 \tilde{A}^{\rm ferm}_0 = {\cal N} \left[\frac{\beta J}{s^2} |\varphi_0|^2-\frac{\beta\mu} 2-
 \log\, \cosh\left(\frac \beta 2 \sqrt{\mu^2+\left(\frac{J|\varphi_0|}s\right)^2}\right) \right] \; .
 \label{cfm2_Smf}
\end{equation}

The complex fields $\varphi$ and $\chi$ are expected to
fluctuate about the SP solution due to thermal and quantum effects. If we keep our expressions only to the first order of $\tau=\beta/M$, making use of the 
notation $\partial_{\tau}=(\delta_{n,m+1}-\delta_{n,m})/\tau$ and denoting
$\Delta=i\phi+\chi$ and $\bar{\Delta}=i\phi^\ast+\chi^\ast$,
then
\begin{equation}
\hat{\bf{G}}^{-1}=\hat{\bf{G}}^{-1}_{0}+\left(\begin{array}{cc}\delta\Delta
& 0  \\ 0 & -\delta\bar{\Delta} \end{array} \right),
\label{ExpansionGreen}
\end{equation}
where
\[
\hat{\bf{G}}^{-1}_{0}=\left(\begin{array}{cc} \Delta_{0}
&\tau(\partial_{\tau}-\mu)  \\ \tau (\partial_{\tau}+\mu) & -\bar{\Delta}_{0}
\end{array} \right).
\]
Applying the Taylor expansion $\ln(1+x)=x-x^{2}/2+...$ we get
\[
\log \det \hat{\bf{G}}^{-1}=\mbox{tr} \ln
\hat{\bf{G}}^{-1}=\mbox{tr} \ln
\left[\hat{\bf{G}}_{0}^{-1}+\left(\begin{array}{cc} \delta\Delta
& 0  \\ 0 & -\delta\bar{\Delta} \end{array} \right)\right]\approx
\]
\begin{equation}
\approx\mbox{tr} \ln
\hat{\bf{G}}^{-1}_{0}-\frac{1}{2}\mbox{tr}\left[\hat{\bf{G}}_{0}
\left(\begin{array}{cc}
\delta\Delta & 0 \\ 0 & -\delta\bar{\Delta} \end{array}
\right)\right]^{2} .
\label{ExpansionLog}
\end{equation}
Calculating the trace in $p=\{q,\omega\}$ representation we get
\begin{equation}
Z\sim\int D[\delta\varphi]\exp\left[-\delta \tilde{A}^{\rm ferm} \right],
\end{equation}
where  $\delta \tilde{A}^{\rm ferm}$ is given by
\begin{equation}
 \delta \tilde{A}_{\rm ferm} = \sum_{\bf k} \sum_{n,m} 
 \delta\varphi^\ast_{{\bf k},n} \left(\hat{\cal G}_{{\bf k};nm}\right)^{-1}
 \delta\varphi_{{\bf k},m} \; .
\end{equation}
Here, $\hat{\cal G}$ represents the Green's function of quasiparticle fluctuations 
(Appendix \ref{App_pf}).

\subsubsection{Results for the paired-fermion model}

It turns out that even on the mean-field
level, the paired-fermion model shows some interesting physical results. 
The condensate density we get via the definition (\ref{FI1_n0}) and the mean-field approximation
that the CF factorizes for large distances:
\begin{equation}
 n_0 = \lim_{{\bf r-r'}\rightarrow\infty} 
 \left\langle \bar\psi^1_{{\bf r},n+1}\bar\psi^2_{{\bf r},n+1} \psi^2_{{\bf r'},n}\psi^1_{{\bf r'},n} 
 \right\rangle = \langle\bar\psi^1_{{\bf r},n+1}\bar\psi^2_{{\bf r},n+1}\rangle 
 \langle\psi^2_{{\bf r'},n}\psi^1_{{\bf r'},n}\rangle \; .
\end{equation}
Further, the CFs which are of second order in the Grassmann field, are given by the diagonal
elements of the matrix $\hat{\bf{G}}$ whose inverse is given in Eq. (\ref{cfm2_matrix-G}).
These diagonal elements are equal to $G/2$ from Eq. (\ref{cfm2_G-def}):
\begin{equation}
 \langle\bar\psi^1_{{\bf r},n+1}\bar\psi^2_{{\bf r},n+1}\rangle =
 \langle\psi^2_{{\bf r'},n}\psi^1_{{\bf r'},n}\rangle = \frac G2 \; \Longrightarrow \;
 n_0 \equiv \frac{G^2}4 \; .
 \label{cfm3_n0-pf}
\end{equation}
Thus, from the Eqs. (\ref{cfm2_G-def}) and (\ref{cfm3_n0-pf}), together with the $M\rightarrow\infty$
limit of Eq. (\ref{cfm2_spe-fermion1}), one finds a self-consistent equation for the
condensate density:
\begin{equation}
 J = \sqrt{\mu^2+4J^2\, n_0} \, \coth \left[
 \frac\beta 2 \sqrt{\mu^2+4J^2\, n_0} \right] \; .
 \label{cfm3_n0-selfconsistent}
\end{equation}
The total particle density we get from the mean-field action (\ref{cfm2_Smf}) is
\begin{eqnarray}
 n_{\rm tot} &=& -\frac 1{\beta\cal N} \frac{\partial\tilde{A}^{\rm ferm}_0}{\partial\mu} =
 \frac 12 + \frac 12 \, \frac\mu{\sqrt{\mu^2+4J^2\, n_0}} \tanh \left[
 \frac\beta 2 \sqrt{\mu^2+4J^2\, n_0} \right] \nonumber \\ &\quad& \\
 &=& \left\{ \begin{array}{l@{\quad}l}
 \displaystyle \frac 12 \left( 1+\frac \mu J\right) & \mbox{in the condensed phase ($n_0>0$)} \\
 &\quad \\ \displaystyle \frac 12 \left[ 1+ \tanh\left(\frac{\beta\mu}2\right) \right] &
 \mbox{in the non-condensed phase ($n_0=0$).} \end{array} \right.
\end{eqnarray}
It might be interesting to mention that all these mean-field results do not depend on the
parameter $s$ which was introduced in the Hubbard-Stratonovich transformation for the
convergence of the Gaussian integral.
%------------------------------ FIGURE -------------------------------------
\begin{figure}
\centering
\includegraphics{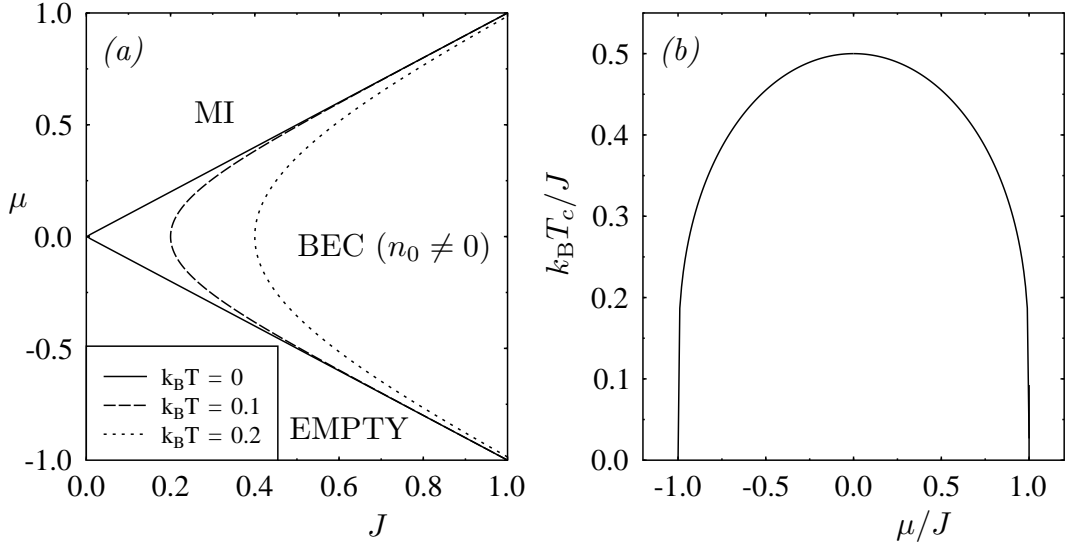}
\caption{{\it (a)} Phase diagram with phase boundaries between the BEC and the non-condensed 
phase for different temperatures. For $k_{\rm B}T\neq 0$ 
there is only one phase boundary between a BEC and a non-condensed phase. The energy unit is 
arbitrary because of a simple scaling behaviour. {\it (b)} Critical temperature of BEC formation.}
\label{Fig_cfm3_cf-phasediagram}
\end{figure}
%---------------------------------------------------------------------------
\begin{figure}
\centering
\includegraphics{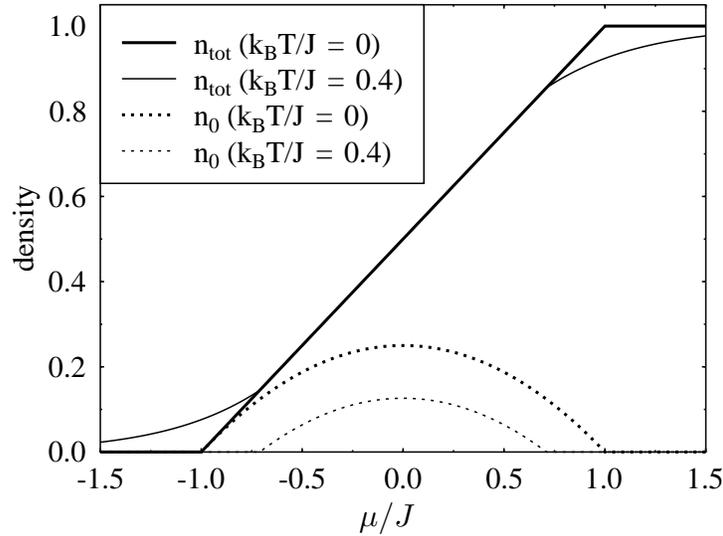}
\caption{Total particle density and condensate density for zero temperature (thick lines, given
by Eqs. (\ref{cfm3_T0n0}) and (\ref{cfm3_T0ntot})) and for
non-zero temperature (thin lines) plotted against chemical potential.}
\label{Fig_cfm3_cf-density}
\end{figure}
%---------------------------------------------------------------------------

The phase boundary between the BEC and the non-condensed phase we get from Eq.
(\ref{cfm3_n0-selfconsistent}). The resulting phase
diagram is depicted in Fig. \ref{Fig_cfm3_cf-phasediagram}. We see in picture {\it (a)}
that for $T>0$ the
phase diagram is separated into two parts, a BEC phase and a non-condensed phase. But at $T=0$
there are three phases: A BEC, an empty phase ($n_{\rm tot}=0$) for $\mu<-J$, and a 
Mott-insulator ($n_{\rm tot}=1$) for $\mu>J$. A density profile of $n_{\rm tot}$ and $n_0$
is plotted in Fig. \ref{Fig_cfm3_cf-density} for different temperatures. At zero temperature 
the sharp transitions between 
the empty phase and the BEC, and the BEC and the MI, can be seen in the plot of the total 
particle density. The zero temperature result is
\begin{equation}
 n_0 = \left\{ \begin{array}{l@{\quad}l}
 \frac 14 \left(1-\frac{\mu^2}{J^2}\right) & \mbox{if } -J<\mu<J \\
 0 & \mbox{else} \end{array} \right. \; ,
 \label{cfm3_T0n0}
\end{equation}
\begin{equation}
 n_{\rm tot} = \left\{ \begin{array}{l@{\quad\mbox{if }}l}
 0 & \mu\le -J \\
 \frac 12 \left(1-\frac\mu J\right) & -J<\mu<J \\
 1 & J\le \mu \end{array} \right. \; .
 \label{cfm3_T0ntot}
\end{equation}
If the temperature increases, the sharp transitions are smeared out.

Calculations for the quasiparticle spectrum by finding the poles of the Green's matrix 
$\hat{\cal G}$ of the Gaussian fluctuations have been made for the zero temperature phase 
diagram \cite{sasha1}. The zero temperature result in the empty phase and in the MI phase is
\begin{equation}
 E_{\bf k} = \epsilon_{\bf k}+|\mu|-J \; ,
 \label{cfm3_E-MI}
\end{equation}
with the gap $\Delta=|\mu|-J$, and in the BEC phase it is
\begin{equation}
 E_{\bf k} = \sqrt{ \epsilon_{\bf k} \left[ J\left(1-\left(\frac\mu J\right)^2\right)+
 \left(\frac\mu J\right)^2 \epsilon_{\bf k} \right] } \; .
\end{equation}
In the dilute regime, i.e. if $\mu=-J+\Delta\mu$, with $\Delta\mu\ll J$, this can be approximated by
\begin{equation}
 E_{\bf k} = \sqrt{\epsilon_{\bf k}(2(\mu+J)+\epsilon_{\bf k})} \; .
\end{equation}
Using the Green's function of quasiparticle fluctuations (see Appendix \ref{App_pf}),
we can calculate the effect of quantum fluctuations on the condensate density:
\begin{equation}
n_{0}=\frac{1}{4}\left(1-\frac{\mu^2}{J^2}\right)+\delta n_0,
\label{no_quantum}
\end{equation}
where the correction  to the mean-field result is
\[
\delta n_{0}=-\frac{(J^2-\mu^2)\mu^2}{J^3}\int \frac{d^d k}{(2\pi)^d} \frac{B_\bold{k}^2
g_\bold{k}}{E_\bold{k}}+\frac{(J^2-\mu^2)}{4 J^3}\int \frac{d^d k}{(2\pi)^d} B_\bold{k}
E_\bold{k}-\frac{(J^2-\mu^2)^2}{4 J^3}\int
\frac{d^d k}{(2\pi)^d}
\frac{B_\bold{k}^2}{E_\bold{k}}+
\]
\begin{equation}
+\frac{3
(J^2-\mu^2)^2\mu^2}{4 J^5}\int \frac{d^d k}{(2\pi)^d} \frac{B_\bold{k}^2 g_\bold{k}}{E_\bold{k}},
\end{equation}
where $B_{\bold{k}}=\dfrac{1}{d}\sum_{j=1}^{d}\cos k_j$, $g_{\bold{k}}=1-B_{\bold{k}}$.
It should be notices that this correction vanishes at the critical point.

It might be interesting to mention that in the zero temperature limit near the phase
transition to the empty phase where $\mu=-J+\Delta\mu$ with $\Delta\mu\ll J$, i.e. in the
dilute regime, it is possible to approximate
\begin{equation}
 n_0 = \frac{\Delta\mu}{2J} + {\cal O}(\Delta\mu^2) = n_{\rm tot} + {\cal O}(\Delta\mu^2) \; .
\end{equation}
This agrees with the
Gross-Pitaevskii result (\ref{Bog2_mu}), if the term of order $\Delta\mu^2$ is neglected,
and the identification $g\equiv 2J$ has been made.

The main correction due to the thermal fluctuations are already included in our mean-field theory,
where the condensed density is given by
\begin{equation}
n_0=\frac{|\varphi_0|^2}{4J^2},
\label{no_thermal}
\end{equation}
and $|\varphi_0|^2$ can be determined from Eqs. (\ref{cfm2_spe-fermion1}-\ref{cfm2_G-def}).

%---------------------------------------------------------------------------
\begin{figure}
\centering
\includegraphics{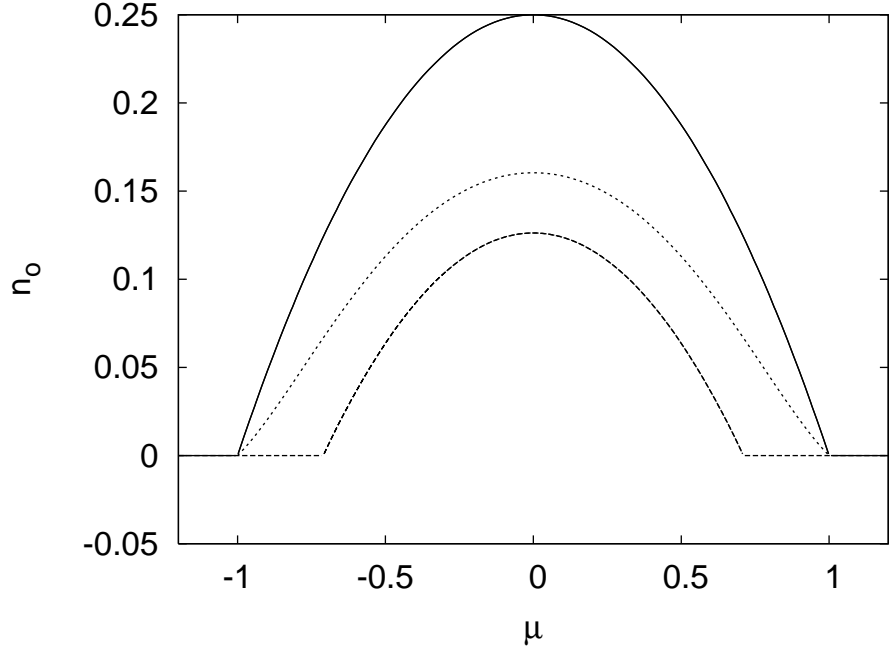}
\caption{Condensate density. The solid, dotted and dashed lines show the mean-field result
at $T=0$, the influence of quantum fluctuations at $T=0$ to the mean-field result and
the mean-field result at $T=0.2$, respectively.}
\label{Fig_fluct}
\end{figure}
%---------------------------------------------------------------------------
The effect of quantum fluctuations and thermal fluctuations is depicted in Fig. \ref{Fig_fluct}.  We see
that both of them lead to a depletion of the
condensate, but the quantum depletion alone does not change the transition points.

The static structure factor for small wave vector $\bold{q}$ and for small temperature $T$ in the BEC
phase reads
\begin{equation}
S(\bold{q})\sim\frac{(J^2-\mu^2)}{J^2n}\frac{Jg_{\bold{q}}}{E_{\bold{q}}}\coth\frac{\beta E_{\bold{q}}}{2},
\end{equation}
where $n$ is a total density of particles.

In the dilute regime, i.e. close to the empty phase, when $n\sim (J+\mu)/J$ and $J-\mu\approx 2J$
we obtain
\begin{equation}
S(\bold{q})\sim \frac{Jg_{\bold{q}}}{E_{\bold{q}}}\coth\frac{\beta E_{\bold{q}}}{2},
\end{equation}
which is in agreement with the well-known result for the weakly interacting Bose gas
(cf. section \ref{Chap_weak}. In the
dense regime, i.e. close to the Mott phase when $n\approx 1$, the static structure factor
vanishes.

In conclusion, we can say that the paired-fermion model has three phases at zero temperature, 
an empty phase, a MI, and a BEC, even on the mean-field level.
However, at non-zero temperatures a new phase emerges from the MI phase and the empty phase,
that is controlled by thermal fluctuations.

%++++++++++++++++++++++++++++++++++++++++++++++++++++++++++++++++++++++++++++++++++++++++
%++++++++++++++++++++++++++++++++++++++++++++++++++++++++++++++++++++++++++++++++++++++++
%++++++++++++++++++++++++++++++++++++++++++++++++++++++++++++++++++++++++++++++++++++++++

\subsection{Slave-boson model\label{Chap_Slave-Bosons}}

\subsubsection{Hamiltonian and functional integral}

In this chapter it shall be shown that a slave-boson approach can be applied to
describe a system of hard-core bosons. The slave-boson representation was originally
developed for fermion systems, e.g. the Hubbard model \cite{kotliar1,woelfle1}. It allows to
account for many aspects of strong correlations even on the mean-field level.
The slave-boson approach to hard-core bosons that will be presented here, has been 
developed in refs. \cite{ziegler0,shukla1,ziegler1,ziegler2}. It is an alternative
to the paired-fermion model which was discussed in the previous chapter.

Again, the starting point is the Hamiltonian (\ref{Int1_H-hardcore}).
We introduce bosonic creation and annihilation operators of empty
($\hat{e}^+_{\bf r}$, $\hat{e}_{\bf r}$) and occupied ($\hat{b}^+_{\bf r}$, $\hat{b}_{\bf r}$) 
sites which act on a fictitious ``vacuum''.
To transfer the Hamiltonian to the extended Fock space, we replace the hard-core boson
operators by
\begin{equation}
 \hat{a}_{\bf r}^+ \rightarrow \hat{b}_{\bf r}^+ \hat{e}_{\bf r} \quad ; \quad 
 \hat{a}_{\bf r} \rightarrow \hat{e}_{\bf r}^+ \hat{b}_{\bf r} \; .
\end{equation}
Then the Hamiltonian (\ref{Int1_H-hardcore}) is replaced by the slave-boson Hamiltonian as
\begin{equation}
 \hat{H}_{\rm hc} \rightarrow \hat{H}_{\rm sb} = -\frac J{2d}
 \sum_{\langle {\bf r,r'}\rangle} \hat{b}_{\bf r}^+ \hat{e}_{\bf r} 
 \hat{e}_{\bf r'}^+ \hat{b}_{\bf r'} + \sum_{\bf r} V_{\bf r} \,
 \hat{b}_{\bf r}^+ \hat{b}_{\bf r} \; .
 \label{sb1_H-sb}
\end{equation}
A hopping process can be understood as a swapping of an occupied site and an empty site.
The occupation number operator of site $\bf r$ is $\hat{b}_{\bf r}^+ \hat{b}_{\bf r}$.
It should be noticed that the external potential acts only on the particles but not on
the empty sites. To assure that a lattice site ${\bf r}$ is either empty or occupied by a boson,
we impose the constraint
\begin{equation}
 \hat{b}^+_{\bf r}\hat{b}_{\bf r} +  \hat{e}^+_{\bf r}\hat{e}_{\bf r} = 1.
\end{equation}

A similar theory for the Bose-Hubbard model has been established in refs. 
\cite{stoof2,lu1}. In this case, an infinite number of operators $(\hat{b}_{\bf r}^\alpha)^+$,
$\hat{b}_{\bf r}^\alpha$ for each occupation number $\alpha$ has to be introduced at each 
lattice site, because multiple occupation is possible. In this respect, the slave-boson approach
for hard-core bosons is much simpler. However, the hard-core boson model describes a projection
of the full Bose-Hubbard model to $n$ and $n+1$ bosons per site, as discussed in the
Introduction \ref{Sec_optical-lattices}.

The grand canonical partition function of the system 
can be expressed as a functional integral with two complex fields
$b_{\bf r}(\tau)$ and $e_{\bf r}(\tau)$.
For the following mean-field calculation, we use the classical approximation here, which only 
takes into account thermal fluctuations but not quantum fluctuations.
This means that for the fields in Matsubara representation
\begin{displaymath}
 b_{\bf r}(\tau) = \frac 1{\sqrt\beta}
 \sum_n b_{{\bf r},n} \, e^{{\rm i}\omega_n\tau} \; ; \quad
 e_{\bf r}(\tau) = \frac 1{\sqrt\beta}
 \sum_n e_{{\bf r},n} \, e^{{\rm i}\omega_n\tau} \; ,
\end{displaymath}
with bosonic Matsubara frequencies $\omega_n$, only the terms
with $\omega_0=0$ are taken into account, if one assumes that
\begin{equation}
 e_{{\bf r},\omega_n} \approx e_{{\bf r},\omega_n} \approx 0 \; , \quad \mbox{if $n\neq 0$.}
\end{equation}
In other words, the time dependence of the fields is neglected.
This is justified if we can assume that
quantum fluctuations (which are neglected in the classical approximation) are small.

The constraint $|b_{\bf r}|^2+|e_{\bf r}|^2=1$ is enforced by a $\delta$-function
in the integration measure:
\begin{equation}
 Z_{\rm sb} = \int {\rm e}^{-A[b,b^\ast,e,e^\ast]} \, {\cal D}[b,b^\ast,e,e^\ast] \; ,
 \label{sb1_Z-eb-full}
\end{equation}
with
\begin{equation}
 {\cal D}[b,b^\ast,e,e^\ast] = \prod_{\bf r}
 \left( |b_{\bf r}|^2 + |e_{\bf r}|^2 -1 \right)
 {\rm d}b_{\bf r} {\rm d}b_{\bf r}^\ast {\rm d}e_{\bf r} {\rm d}e_{\bf r}^\ast
\end{equation}
and the action
\begin{equation}
 A[b,b^\ast,e,e^\ast] = \beta \Bigg\{ -\sum_{\bf r}
 \mu_{\bf r} b^\ast_{\bf r} b_{\bf r}
 -\frac J{2d} \sum_{\langle{\bf r,r'}\rangle} b_{\bf r}^\ast e_{\bf r}
 e_{\bf r'}^\ast b_{\bf r'} \Bigg\} \; .
 \label{sb1_S-continuous-tau}
\end{equation}
Here, we consider a space-dependent chemical potential $\mu_{\bf r}=\mu-V_{\bf r}$.

\subsubsection{Two-fluid theory in classical approximation}

The hopping term of the action is of fourth order in the field variables. Therefore it is 
not possible to perform the integration directly.
However, it is possible to decouple the hopping term by introducing two new fields, a complex
field $\Phi$ and a real field $\varphi$, and perform a Hubbard-Stratonovich transformation.
The fields $b$ and $e$ can be integrated out then, and a mean-field approximation can be
applied to the fields $\Phi$ and $\varphi$ \cite{ziegler2}.

The idea of the Hubbard-Stratonovich decoupling is similar to the one used in the previous
chapter to decouple the fourth order terms of the Grassmann fields.
We insert the identity
\begin{displaymath}
 {\rm const.}\times e^{-A[b,b^\ast,e,e^\ast]} = \int 
 \exp \Bigg\{ -\beta \Bigg[
 \sum_{\bf r,r'} \Phi_{\bf r}^\ast \left[\frac{s-\hat J}{s^2}\right]_{\bf rr'}^{-1}
 \Phi_{\bf r} + s\sum_{\bf r} \varphi_{\bf r}^2
\end{displaymath}
\begin{equation}
 + \sum_{\bf r} (e_{\bf r},b_{\bf r}) \left(
 \begin{array}{cc} 2s\varphi_{\bf r}+s & s\Phi_{\bf r} \\
  s\Phi_{\bf r}^\ast & - \mu_{\bf r} \end{array}\right)\left(
 \begin{array}{c} e_{\bf r}^\ast \\ b_{\bf r}^\ast \end{array}\right) \Bigg] \Bigg\}
 \, {\cal D} [\Phi^\ast,\Phi,\varphi] \; ,
 \label{sb1_decoupling}
\end{equation}
with the integration measure
\begin{equation}
 {\cal D} [\Phi^\ast,\Phi,\varphi] =
 \prod_{\bf r} \frac{{\rm d}\Phi^\ast_{\bf r} {\rm d}\Phi_{\bf r} 
 {\rm d}\varphi_{\bf r}}{(2\pi)^{3/2}} \; .
\end{equation}
Here, $\hat J$ is the hopping matrix (\ref{FI3_hoppingJ}).
The constant factor is of no physical relevance.
Like for the paired-fermion model which was discussed before, the parameter $s$ takes care
of the convergence of the Gaussian integral. It has the unit of an energy and should not be 
too small compared to $J$. Although the exact identity does not depend on $s$, we will see
subsequently that the mean-field equation we will derive, does. This is a difference to
the previously discussed model, where the result which was derived on the mean-field level
and on the level of Gaussian fluctuations, did not depend on the free parameter $s$.

After substituting the identity (\ref{sb1_decoupling}) into the functional integral
(\ref{sb1_S-continuous-tau}), the fields $b$ and $e$
are only of second order and can be integrated out exactly together with the constraint.
This is shown in Appendix \ref{Sec_App_constraint}. The result for the partition function is
\begin{equation}
 Z_{\rm sb} = \int e^{-\tilde{A}(\Phi^\ast,\Phi)} \prod_{\bf r} {\rm d}\Phi_{\bf r} 
 {\rm d}\Phi_{\bf r}^\ast
 \label{sb1_Z-2fluid}
\end{equation}
with the new action
\begin{equation}
 \tilde{A}(\Phi^\ast,\Phi) = \beta \sum_{\bf r,r'} \Phi^\ast_{\bf r} 
 \left[\frac{s-\hat{J}}{s^2}\right]_{\bf rr'}^{-1} \Phi_{\bf r'}
 - \sum_{\bf r} \log \left[Z'_{\bf r} \, e^{\frac{\beta\mu_{\bf r}}4}\right]  \; ,
 \label{sb1_Stilde}
\end{equation}
and the function
\begin{equation}
 Z'_{\bf r} = \int_{-\infty}^\infty {\rm d}\varphi_{\bf r} \,
 \frac{ \sinh\left[\beta\sqrt{\left(\varphi_{\bf r} s+\frac{\mu_{\bf r}}2 \right)^2 
 + s^2|\Phi_{\bf r}|^2}\right]}{ \beta\sqrt{\left(\varphi_{\bf r} s
 + \frac{\mu_{\bf r}}2\right)^2 + s^2|\Phi_{\bf r}|^2}} e^{-\beta s\varphi_{\bf r}^2}\; .
 \label{sb1_Zr}
\end{equation}
Note that the action $\tilde{A}(\Phi^\ast,\Phi)$ does not depend on the real field 
$\varphi$ explicitly, because it appears inside the function $Z'$ only as an
integration variable.

The form (\ref{sb1_Z-2fluid}) of the grand canonical partition function can be 
understood as a two-fluid theory. It is shown in Appendices \ref{Sec_App_SB-n0}
and \ref{Sec_App_SB-ntot}
that the condensate density is related to the field $\Phi$ and is given by the relation
\begin{equation}
 n_0 \approx \frac{s^2}{(s+J)^2} \lim_{{\bf r-r'}\rightarrow\infty}
 \left\langle \Phi_{\bf r}\Phi_{\bf r'}^\ast \right\rangle \; ,
 \label{sb1_n0}
\end{equation}
and that the total particle density at site $\bf r$
is related to the field $\varphi$ by means of the expectation value
\begin{equation}
 n_{\bf r} = \left\langle \varphi_{\bf r} \right\rangle + \frac 12 \; .
 \label{sb1_nr}
\end{equation}

\subsubsection{Mean-field theory}

\begin{figure}
\centering
\includegraphics{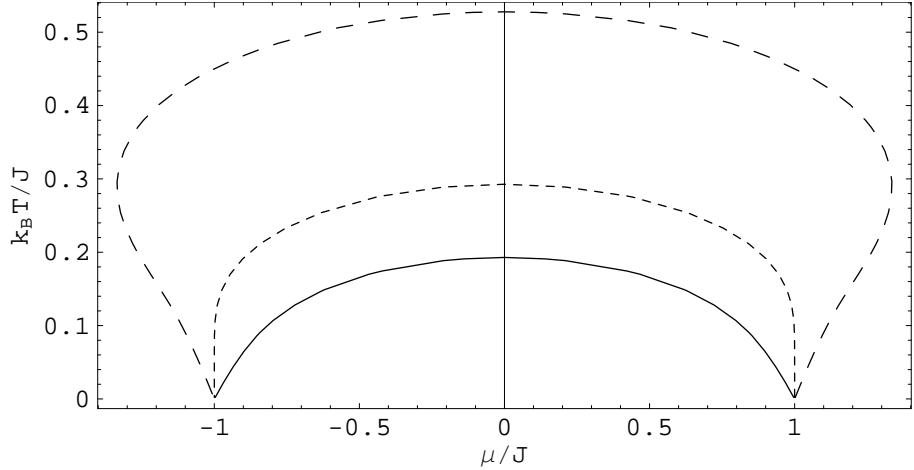}
\caption{Phase boundary between the BEC and the non-condensed phase for $s/J=3$ (long dashes),
$s/J=1$ (short dashes), $s/J=0.2$ (solid line). Compare these graphs with the graph on 
the right hand side
of Fig. \ref{Fig_cfm3_cf-phasediagram}, where the critical temperature of the mean-field result
for the paired-fermion model is plotted.}
\label{Fig_sb1_muT-PD}
\end{figure}

A mean-field solution is found by minimising the action via the variational principle 
$\delta\tilde{A}=0$, which leads to a saddle-point approximation, as
it was done for the paired-fermion model.
Since the field $\varphi$ can be integrated out (e.g. numerically) inside the function
$Z'_{\bf r}$ given in Eq. (\ref{sb1_Zr}), minimization has to be done with respect to the 
complex field $\Phi$ only:
\begin{equation}
 \frac{\partial\tilde{A}}{\partial\Phi_{\bf r}} =
 \frac{\partial\tilde{A}}{\partial\Phi^\ast_{\bf r}} = 0 \; .
\end{equation}
This yields the mean-field equation
\begin{equation}
 \sum_{\bf r'} \left[\frac{s-\hat{J}}{s^2}\right]_{\bf rr'}^{-1} \Phi_{\bf r'}
 - \frac 1\beta \left[ \frac\partial{\partial(|\Phi_{\bf r}|^2)} \log Z'_{\bf r} \right] 
 \Phi_{\bf r} = 0 \; .
 \label{sb1_mean-field-eq}
\end{equation}
In the case of a spatially constant field without external trapping potential, i.e. if
we assume that $\Phi_{\bf r}\equiv\Phi_0$ and $\mu_{\bf r}\equiv\mu$, the mean-field equation is
\begin{equation}
 \frac{s^2}{s+J} - \frac 1\beta \frac\partial{\partial(|\Phi_0|^2)} \log Z' = 0 \; .
 \label{sb1_constant-mean-field}
\end{equation}
If the field $\Phi$ is varying only very slowly between neighbouring lattice sites, 
we can approximate
\begin{equation}
 \sum_{\bf r'} \left[\frac{s-\hat{J}}{s^2}\right]_{\bf rr'}^{-1} \Phi_{\bf r'} \approx
 \frac{s^2}{s+J} \, \Phi_{\bf r} + \frac{s^2}{(s+J)^2} \sum_{\bf r'} 
 \left(J\,\delta_{\bf rr'} + \hat{J}_{\bf rr'}\right) \, \Phi_{\bf r'} \; .
 \label{sb1_hopping-appr}
\end{equation}
In Fig. \ref{Fig_sb1_muT-PD} the phase boundary between the BEC and the non-condensed 
phase is plotted
for different values of $s$. The phase boundary solves Eq. (\ref{sb1_constant-mean-field})
for $\Phi_0=0$, and has been calculated numerically.
%------------------------------ FIGURE -------------------------------------
%---------------------------------------------------------------------------
\begin{figure}
\centering
\includegraphics{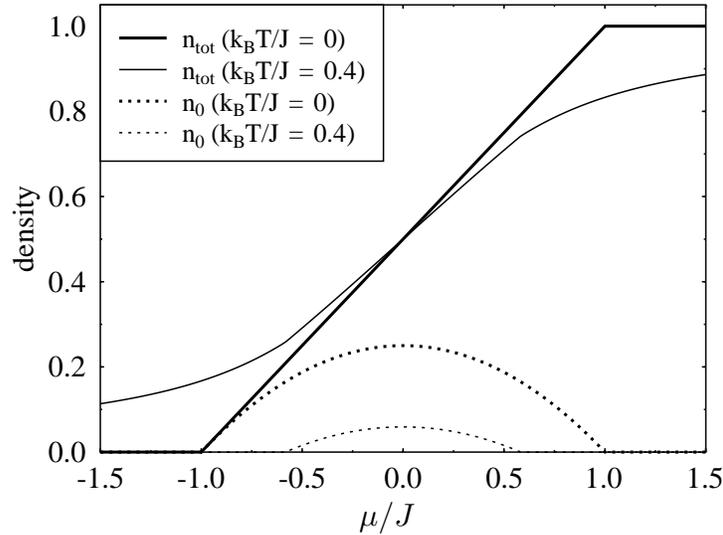}
\caption{Total particle density and condensate density for zero temperature (thick lines) and for
non-zero temperature (thin lines, $s/J=1/5.5$) against chemical potential
\cite{sasha1}. Compare this graph with
the result for the paired-fermion model plotted in Fig. \ref{Fig_cfm3_cf-density}.}
\label{Fig_sb1_density}
\end{figure}
%---------------------------------------------------------------------------

One can see that the BEC phase forms a ``bubble'' in the phase diagram, if $s/J>1$.
This behaviour is unexpected because the BEC phase should become narrower, if
temperature is increased. This means that for too large values of $s/J$ the mean-field theory
seems to be incorrect. However, it turns
out that the absolute minimum of the action with respect to $s$ at constant $J$, $\mu$
and $\beta$ occurs at values of $s/J<1$.

It is possible to find an exact solution for zero temperature, which does {\it not} depend on $s$.
This calculation is shown in Appendix \ref{Sec_App_SB-T0}.
Two phase boundaries are found: A boundary between the BEC and an empty phase with $\mu_c=-J$
and a phase boundary between the BEC and the Mott insulator with $\mu_c=J$.
It is identical to the zero temperature mean-field result in Eqs. (\ref{cfm3_T0n0}) and
(\ref{cfm3_T0ntot}) that was found for the paired-fermion model, and agrees with it qualitatively 
at finite temperatures (see Fig. \ref{Fig_sb1_density}). 
When temperature increases, results strongly depend on $s$.

\subsubsection{Quasiparticle spectrum}

We get the quasiparticle spectrum from the Gaussian fluctuations, the same way as it was done 
for the paired-fermion model. We write
$$
 \Phi_{\bf r} = \Phi_0 + \delta\Phi_{\bf r} \; , \quad
 \Phi^\ast_{\bf r} = \Phi^\ast_0 + \delta\Phi^\ast_{\bf r} \; , \quad
$$
and assume that the fluctuations $\delta\Phi$, $\delta\Phi^\ast$ about the mean-field solution
$\Phi_0$ are small. Substituting this expression into the action (\ref{sb1_Stilde}), and 
expanding it up to second order in the fluctuations, one finds
\begin{equation}
 \tilde{A} = \beta\frac{s^2}{s+J}|\Phi_0|^2 - \log Z'\left(|\Phi_0|^2\right) -
 \frac\beta 2 \sum_{\bf r,r'} \left(\delta\Phi_{\bf r},\delta\Phi^\ast_{\bf r}\right)
 \hat{\cal G}_{\bf rr'}^{-1} \left( \begin{array}{cc} \delta\Phi^\ast_{\bf r'} \\ 
 \delta\Phi_{\bf r'} \end{array} \right) \; ,
\end{equation}
with the matrix
\begin{equation}
 \hat{\cal G}_{\bf rr'}^{-1} = \left( \begin{array}{cc}
 \frac{J\,\delta_{\bf rr'}+\hat{J}_{\bf rr'}}{s+J} + 
 (\tilde{a}_2+|\Phi_0|^2\tilde{a}_4)\delta_{\bf rr'}
 & (\Phi_0^\ast)^2 \, \tilde{a}_4\delta_{\bf rr'} \\
 \Phi_0^2 \, \tilde{a}_4\delta_{\bf rr'} &  \frac{J\,\delta_{\bf rr'}+\hat{J}_{\bf rr'}}{s+J} +
 (\tilde{a}_2+|\Phi_0|^2\tilde{a}_4)\delta_{\bf rr'}
 \end{array} \right) \; .
\end{equation}
Here, we have introduced the abbreviations
\begin{eqnarray}
 \tilde{a}_2 &:=& -\left. \frac 1\beta \frac\partial{\partial(|\Phi|^2)} \log Z'
 \right|_{\Phi=\Phi_0} + \frac{s^2}{s+J} \; , \label{sb1_a2} \\
 \tilde{a}_4 &:=& -\left. \frac 1\beta \frac{\partial^2}{\partial(|\Phi|^2)^2} 
 \log Z' \right|_{\Phi=\Phi_0} \; , \label{sb1_a4}
\end{eqnarray}
and used the approximation in Eq. (\ref{sb1_hopping-appr}).
The matrix $\hat{\cal G}$ has no time-structure because of the classical approximation. To find
the Green's function of quasiparticles, we {\it artificially} introduce the imaginary time
by writing
\begin{equation}
 \hat{\cal G}_{\bf rr'}^{-1} = \left( \begin{array}{cc}
 \frac{J\,\delta_{\bf rr'}+\hat{J}_{\bf rr'}+\hbar\frac\partial{\partial\tau}}{s+J} + 
 \tilde{a}_2 + |\Phi_0|^2\tilde{a}_4 \delta_{\bf rr'} &
 (\Phi_0^\ast)^2 \, \tilde{a}_4\delta_{\bf rr'} \\
 \Phi_0^2 \, \tilde{a}_4\delta_{\bf rr'} &  
 \frac{J\,\delta_{\bf rr'}+\hat{J}_{\bf rr'}- \hbar\frac\partial{\partial\tau}}{s+J} + 
 \tilde{a}_2 + |\Phi_0|^2\tilde{a}_4 \delta_{\bf rr'} \end{array} \right) \; ,
\end{equation}
in analogy with the Bogoliubov theory. After a Fourier transformation it leads 
to the Green's function
\begin{equation}
 \hat{\cal G}^{-1}({\bf k},\omega_n) = \frac{s^2}{(s+J)^2} \left( \begin{array}{cc}
 \epsilon_{\bf k}+\frac{(s+J)^2}{s^2}\tilde{a_2} & {\rm i}\hbar\omega_n \\
 {\rm i}\hbar\omega_n & \epsilon_{\bf k} + \frac{(s+J)^2}{s^2} \left( \tilde{a}_2+
 2\tilde{a}_4|\Phi_0|^2 \right) \end{array} \right) \; ,
\end{equation}
which is equivalent to the matrix (\ref{Bog2_qp-Greens}), and $\epsilon_{\bf k}$
is the lattice dispersion (\ref{FI2_dispersion-latt}).
The quasiparticle spectrum is given by the poles of $\hat{\cal G}$, and can be found
by performing the analytic continuation ${\rm i}\hbar\omega_n\rightarrow E_{\bf k}$ and
solving the equation $\det \hat{\cal G}^{-1}=0$. We find solutions for both the 
BEC phase and the non-condensed phase:

In the BEC phase, where $|\Phi_0|^2>0$, the coefficient $\tilde{a}_2$ vanishes,
because $\Phi_0$ solves the mean-field equation (\ref{sb1_constant-mean-field}), which is
equivalent to $\tilde{a}_2=0$. The solution is
\begin{equation}
 E_{\bf k} = \sqrt{ \epsilon_{\bf k} \left( 2\, \frac{(s+J)^2}{s^2}\, \tilde{a}_4\, 
 |\Phi_0|^2 +\epsilon_{\bf k} \right) } \; .
\end{equation}
It is gapless and agrees with the Bogoliubov spectrum (\ref{Bog1_bogol-spectrum}), 
when we identify the condensate density with
$n_0=s^2|\Phi_0|^2/(s+J)^2$, and the interaction constant with $g=(s+J)^4\tilde{a}_4/s^4$. 
The coefficient $\tilde{a}_4$ depends
on both temperature and chemical potential. Its zero-temperature result is given in
Eq. (\ref{App3_a4}) of Appendix \ref{Sec_App_SB-T0}. In the dilute gas (i.e. near the phase 
transition to the empty phase) where $n_0\ll 1$, we find at
zero temperature for the interaction constant the result $g\approx 2J$.

In the non-condensed phase, where $|\Phi_0|^2=0$ and $\tilde{a}_2\neq 0$, the
quasiparticle spectrum is gapped, in agreement with the findings of the 
paired-fermion model:
\begin{equation}
 E_{\bf k} = \epsilon_{\bf k} + \Delta \; ,
\end{equation}
with the gap $\Delta=(s+J)^2\tilde{a}_2/s^2$.
At zero temperature and near the phase transitions, we find the result 
$\Delta=|\mu-\mu_c|+{\cal O}((\mu-\mu_c)^2)$
which is identical to the zero-temperature result (\ref{cfm3_E-MI})
for the paired-fermion model.

\subsubsection{Renormalized Gross-Pitaevskii equation}

In this section we will derive a mean-field equation which is appropriate to describe
the BEC as well as the Mott insulator in a strongly interacting Bose gas, and which is similar
to the stationary Gross-Pitaevskii equation. 
The mean-field equation for a hard-core Bose gas in an optical lattice within the 
slave-boson approach is given by
\begin{equation}
 \frac{s^2}{(s+J)^2} \sum_{\bf r'} 
 \left(J\,\delta_{\bf rr'} + \hat{J}_{\bf rr'}\right) \, \Phi_{\bf r'} +
 \frac{s^2}{s+J} \, \Phi_{\bf r}
 - \frac 1\beta \left[ \frac\partial{\partial(|\Phi_{\bf r}|^2)} \log Z'_{\bf r} \right] 
 \Phi_{\bf r} = 0 \; .
\end{equation}
This we get by applying the approximation (\ref{sb1_hopping-appr}) in Eq. 
(\ref{sb1_mean-field-eq}).
However, it also possible to describe a system of strongly interacting bosons without lattice
potential within this approximation. Therefore we perform a continuum approximation of the 
hopping term: If the lattice constant $a$ is so small that the order parameter $\Phi_{\bf r}$
varies only slowly over neighbouring lattice sites, we can treat the $3$-dimensional lattice
approximately as a continuum:
\begin{equation}
 \sum_{\bf r'} \left(J\,\delta_{\bf rr'} + \hat{J}_{\bf rr'}\right) \, \Phi_{\bf r'}
 = -\frac{Ja^2}{6} \sum_{j=1}^3 \frac{ \Phi_{{\bf r}+a{\bf e}_j} - 2\Phi_{\bf r} +
 \Phi_{{\bf r}-a{\bf e}_j} }{a^2} \approx
 - \frac{Ja^2}{6} \nabla^2 \Phi_{\bf r} \; .
\end{equation}
When working on the continuum, we rescale the order parameter by
\begin{equation}
 \Phi({\bf r}):=a^{-3/2}\Phi_{\bf r} \; , 
 \label{sb2_continuum-renorm}
\end{equation}
such that the action (\ref{sb1_Stilde}) can be written as
\begin{eqnarray}
 \tilde{A}(\Phi^\ast,\Phi) &=& \frac{\beta s^2}{(s+J)^2} \int 
 \bigg\{ - \frac{Ja^2}{6}
 \Phi^\ast({\bf r}) \nabla^2 \Phi({\bf r}) + (s+J) |\Phi({\bf r})|^2 \nonumber \\
 && - \frac{(s+J)^2}{\beta s^2}
 \log \left[Z'({\bf r}) \, e^{\frac{\beta\mu({\bf r})}4}\right] \bigg\} {\rm d}^3r \; .
\end{eqnarray}
The order parameter is normalied to the number of condensed particles by
\begin{equation}
 N_0 = \frac{s^2}{(s+J)^2} \int |\Phi({\bf r})|^2 {\rm d}^3r \; .
 \label{sb2_N0}
\end{equation}
The replacement (\ref{sb2_continuum-renorm}) has also to be made inside the function $Z'$, 
of course. The corresponding mean-field equation for the continuum is
\begin{equation}
 \left[ - \frac{Ja^2}{6} \nabla^2 + (s+J) - \frac{(s+J)^2}{\beta s^2} \,
 \frac\partial{\partial(a^3|\Phi({\bf r})|^2)} \log Z'({\bf r}) \right] \Phi({\bf r}) = 0 \; .
 \label{sb2_full-meanfield}
\end{equation}
The parameters can be identified with those of the
conventional GP equation: The mass $m$ of the particles is given by the hopping constant 
$J$ and the original lattice constant $a$ via
\begin{equation}
 \frac{\hbar^2}{2m}\equiv\frac{Ja^2}6 \; .
 \label{sb2_mass-id}
\end{equation}
In the continuum $a$ looses its identity as lattice constant, but describes a
characteristic length scale that can be interpreted as
the spacial extension of a boson. Thus, it should be of the same order of magnitude as
the $s$-wave scattering length $a_s$.

If the order parameter $\Phi$ is small, we can expand the potential part of
the action up to fourth order:
\begin{displaymath}
 (s+J) a^3|\Phi({\bf r})|^2 - \frac{(s+J)^2}{\beta s^2}
 \log \left[Z'({\bf r}) \, e^{\frac{\beta\mu({\bf r})}4}\right]
\end{displaymath}
\begin{equation}
 = a_0 - \mu_{\rm R} |\Phi({\bf r})|^2 + 
 \frac{g_{\rm R}}2 \, \frac{s^2}{(s+J)^2} |\Phi({\bf r})|^4 + {\cal O}(|\Phi|^6) \; ,
\end{equation}
where we have introduced the coefficients
\begin{eqnarray}
 a_0 &=& - \frac{(s+J)^2}{\beta s^2} \left. \log Z'({\bf r}) \right|_{\Phi=0} \\
 \mu_{\rm R} &=& - (s+J) +\frac{(s+J)^2}{\beta s^2} 
 \left. \frac\partial{\partial(a^3|\Phi({\bf r})|^2)} \log Z'({\bf r}) \right|_{\Phi=0} 
 \label{sb2_muR} \\
 g_{\rm R} &=& - \frac{a^3(s+J)^4}{\beta s^4} \left. 
 \frac{\partial^2}{\partial(a^3|\Phi({\bf r})|^2)^2} \log Z'({\bf r}) \right|_{\Phi=0} \; .
 \label{sb2_gR}
\end{eqnarray}
They depend on $\mu$, $J$, $\beta$, and $|\Phi({\bf r})|^2$. 
Further, we introduce the rescaled order parameter
\begin{equation}
 \Phi_{\rm R}({\bf r}) = \frac{s}{s+J} \, \Phi({\bf r}) \; .
\end{equation}
With these coefficients, the full mean-field equation (\ref{sb2_full-meanfield}) can be
approximated by the equation
\begin{equation}
 \left[ - \frac{Ja^2}{6} \nabla^2 - \mu_{\rm R} + g_{\rm R} |\Phi_{\rm R}({\bf r})|^2 \right]
 \Phi_{\rm R}({\bf r}) = 0 \; ,
\end{equation}
%------------------------------ FIGURE -------------------------------------
\begin{figure}
\includegraphics{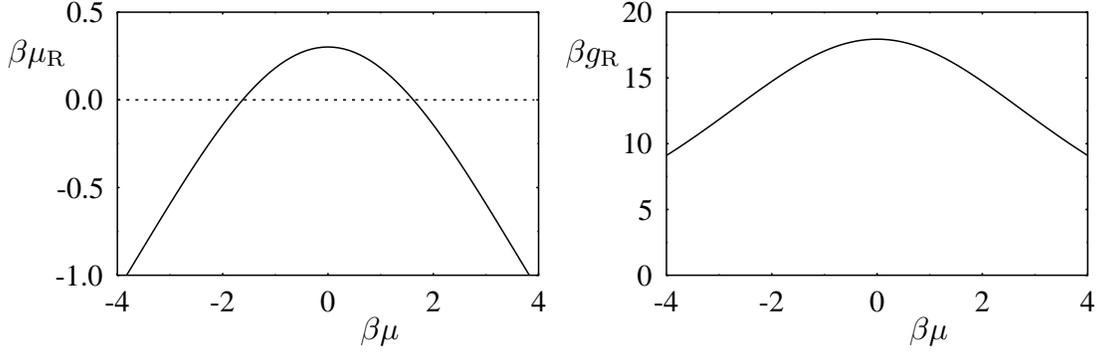}
\caption{Coefficients $\mu_{\rm R}$ and $g_{\rm R}$ of the renormalized GP theory
plotted against the chemical potential $\mu$. All parameters are normalised by the inverse
temperature $\beta$. The tunneling rate was chosen to be $\beta J=5.5$ and the free parameter
was chosen as $s=k_{\rm B}T$.}
\label{Fig_sb2_muR-gR}
\end{figure}
%---------------------------------------------------------------------------
This equation has the same form as the conventional stationary GP equation, where $\mu_{\rm R}$
and $g_{\rm R}$ play the role of a renormalised chemical potential and a renormalised interaction
constant, respectively. Their dependence on $\mu$ is shown in Fig. \ref{Fig_sb2_muR-gR}.
Therefore we refer to this equation as a 
``renormalised GP equation'' \cite{SBmanuscript}.
The zero temperature limits of the coefficients are calculated in Appendix \ref{Sec_App_SB-T0}, see Eq. (\ref{App3_coefficients}). Near the phase transition to the empty phase, i.e. in the 
dilute regime, where $\mu=-J+\Delta\mu$, $\Delta\mu\ll J$, we find 
$\mu_{\rm R}=\Delta\mu+{\cal O}(\Delta\mu^2)$. Thus, in the limiting case of a dilute BEC
and zero temperature, the renormalised GP equation goes over to the conventional GP equation
with the interaction parameter $g=g_{\rm R}=2a^3J$.
While $g_{\rm R}$ is always positive, $\mu_{\rm R}$ can change sign. A BEC exists if
$\mu_{\rm R}>0$, otherwise the order parameter vanishes. The phase transition between 
the BEC and the non-condensate phase 
is given by the relation $\mu_{\rm R}=0$, which is equivalent to Eq. 
(\ref{sb1_constant-mean-field}) in a translational-invariant system.
Inside the BEC phase, $\mu_{\rm R}$ increases linearly with increasing $\mu$, reaches a maximum
and decreases again until the condensate is destroyed totally due to strong interaction effects.

%++++++++++++++++++++++++++++++++++++++++++++++++++++++++++++++++++++++++++++++++++++++++
%++++++++++++++++++++++++++++++++++++++++++++++++++++++++++++++++++++++++++++++++++++++++
%++++++++++++++++++++++++++++++++++++++++++++++++++++++++++++++++++++++++++++++++++++++++

\section{Discussion}

\subsection{Comparison of the results}

The main results that we found for the one-dimensional model, 
the paired-fermion model, and the slave-boson model, will be summarized and discussed
in this section. All three models give more or less the same physics at zero temperature, with 
an empty phase, a phase with a particle number per lattice site between $0$ and $1$, and
a Mott insulator. Their common features and differences shall be pointed out in detail.

\subsubsection{Phase diagram, total density and condensate density}

At zero temperature, the exact solution of the one-dimensional model exhibits three phases
in the translational invariant case, as shown in Fig. \ref{Fig_d11_phase-diagram1} in the 
$J$-$\mu$ plane: An empty phase which contains no particles in equilibrium (physically speaking,
it costs energy to put a particle into the system), an incommensurate
phase with a particle number per lattice site $n_{\rm tot}$ between $0$ and $1$, and a 
Mott-insulator with $n_{\rm tot}=1$. The same zero-temperature phase diagram has been found 
for the paired-fermion model (see picture {\it (a)} in Fig. \ref{Fig_cfm3_cf-phasediagram}) 
and the slave-boson model on the mean-field level. The only difference is that for the
three-dimensional models, the incommensurate phase is a BEC, whereas in the case of the 
one-dimensional model there is no BEC but only a long range correlated phase.
This is a consequence of the Mermin-Wagner theorem \cite{merminwagner,hohenberg}.
At non-zero temperatures, the empty phase and the MI are affected by thermal fluctuations, 
and they have no clear phase boundary any more. However, the
three-dimensional systems still have a single phase boundary between a BEC with a non-zero
order parameter, and a non-condensed phase where the order parameter vanishes. The shape
of this phase boundary depends on temperature (see picture on the right hand side of Fig.
\ref{Fig_cfm3_cf-phasediagram} for the paired-fermion model).

For the one dimensional model, the total particle density at $T=0$ and $T>0$ 
is shown in Fig. \ref{Fig_d11_ntot1}. At $T=0$, the derivative 
$\partial n_{\rm tot}/\partial\mu$ diverges at the phase transitions
between the BEC and the empty phase and the BEC and the MI phase. The sharp transitions are 
``washed out'' at finite temperatures.

The zero temperature mean-field results for the total particle density and the condensate
density of the paired-fermion model and the slave-boson model agree with each other
and are given in the Eqs. (\ref{cfm3_T0n0}) and (\ref{cfm3_T0ntot}).
We find a total particle density which increases linearly with $\mu$.
In the dilute regime the condensate density is given by $n_0=n_{\rm tot}-{\cal O}(n_{\rm tot}^2)$.
If we neglect the terms of order $n_{\rm tot}^2$, this is in agreement with Gross-Pitaevskii 
theory which assumes that all particles are condensed in this regime. 
In the absence of a trapping potential, a solution 
of the stationary GP equation is given by
\begin{equation}
 n_0 = \frac\mu g \; .
\end{equation}
This describes a linearly increasing condensate density $n_0$ 
with respect to the chemical potential. Although it takes the repulsion into account
by a factor $1/g$ which is decreasing  with increasing interaction constant $g$, the
saturation of $n_0$
cannot be seen in this solution. From the physical point of view, in a realistic
description for large densities, the particle density must saturate because there is
a finite scattering volume around each particle. Furthermore, for increasing particle
density, the condensate density should reach a maximum and for even larger densities,
decrease again until its total destruction, because of the increasing interparticle
interaction. This is the behaviour that we found for the slave-boson and the paired-fermion model in 
mean-field approximation.
A similar behaviour has also  been found by variational perturbation theory 
\cite{kleinert2}, and diffusion Monte Carlo calculations \cite{dubois1}.
In order to describe condensates at higher densities, the second order term in the
low-density expansion of the energy density has been taken into account which leads to a
modified GP theory \cite{Bpitaevskii,dubois1,fabrocini1,nilsen1}.

At non-zero temperatures the phase boundaries of the empty phase and the MI are not well
defined any more,
like in the one-dimensional case. The region of BEC shrinks and the condensate density decreases.
Non-zero temperature results of the paired-fermion model and the slave-boson model are 
very similar but not identical (compare the figs. \ref{Fig_cfm3_cf-density} 
and \ref{Fig_sb1_density}). This is a consequence of the different mean-field approaches.
The effect of quantum fluctuations on the zero-temperature result has been studied
for the paired-fermion model. A condensate depletion was found, but the critical points were
not affected (see Fig. \ref{Fig_fluct}).

\subsubsection{Excitation spectrum\label{Sec_qp}}

The spectrum of quasiparticle excitations is found on the level of Gaussian fluctuations.
For the paired-fermion model, and the slave-boson model, 
the expressions for the quasiparticle
spectra $E_{\bf k}$ are summarised in the subsequent table:
\begin{center}
\begin{tabular}{|l|l|l|}
 \hline
 && \\[-1.5ex]
 $E_{\bf k}$ & in the BEC phase & in the non-condensed phases\\ \hline\hline
%  && \\[-1.5ex]
%  hard-core Bose & $\sqrt{\epsilon_{\bf k}\left(2(\mu+J)+\epsilon_{\bf k}\right)}$ & 
%  $\epsilon_{\bf k} + |\mu+J|$ \\ 
 && \\[-1.5ex] \hline
 && \\[-1.5ex]
 paired-fermion model & $\sqrt{\epsilon_{\bf k} \left[ J\left(1-\left(\frac\mu J\right)^2\right)+
 \left(\frac\mu J\right)^2 \epsilon_{\bf k} \right]}$ & $\epsilon_{\bf k}+|\mu|-J$ \\ \hline
 && \\[-1.5ex]
 slave-boson model & $\sqrt{ \epsilon_{\bf k} \left( 2\, \frac{(s+J)^2}{s^2}\, \tilde{a}_4\, 
 |\Phi_0|^2 + \epsilon_{\bf k}\right) }$ & $\epsilon_{\bf k}+(s+J)^2\tilde{a}_2/s^2$ \\ \hline
\end{tabular} 
\end{center}
Here, $\epsilon_{\bf k}$ is the free-particle dispersion relation in the optical lattice,
given by Eq. (\ref{FI2_dispersion-latt}).
We find a spectrum which is linear for small wave vectors $\bf k$ in the BEC phase, whereas the
spectrum has a gap in the non-condensed phases. The gapless spectrum in the BEC phase is caused
by a Goldstone mode due to a broken global $U(1)$ symmetry \cite{goldstone}. 
The result given for the
paired-fermion model is only valid at zero temperature. The gapped spectrum is found both in the
empty phase and in the MI phase.
The result for the slave-boson model depends implicitly on temperature via the coefficients
$\tilde{a}_2$ and $\tilde{a}_4$ given in Eqs. (\ref{sb1_a2}) and (\ref{sb1_a4}), and it 
also depends on the non-physical parameter $s$.

We have shown that the zero-temperature results of all three models 
inside the BEC phase and near the phase boundary to the empty phase ($\mu+J\ll J$),
agree with the Bogoliubov result
\begin{displaymath}
 E_{\bf k} = \sqrt{\epsilon_{\bf k} \left( 2\mu+\epsilon_{\bf k} \right)} \; .
\end{displaymath} 
The only difference is that the 
chemical potential is shifted ($\mu\rightarrow\mu+J$), because the phase transition in
Bogoliubov theory is given by $\mu=0$ instead of $\mu=-J$ for the two
three-dimensional models.
The region near the phase transition to the empty phase is the weakly interacting regime,
therefore Bogoliubov theory is applicable there. The interaction constant was identified
as $g\equiv 2a^3J$ (where the lattice constant $a$ was set to $1$ in the lattice models). 

The gapped spectrum in the MI that was found in the paired-fermion and slave-boson models 
is of the form 
\begin{equation}
 E_{\bf k}=\epsilon_{\bf k}+\Delta \; .
 \label{con1_gapped-spectrum}
\end{equation}
We have shown that in the MI phase, near the phase transition to the BEC phase, 
the gap is given by $\Delta=\mu-J$. 

For the one-dimensional system, the excitation spectrum in the incommensurate phase can be
found indirectly by means of the Feynman relation and is given in Eq. (\ref{d12_Ek}).
It is linear for small wave-vectors $\bf k$, like in the BEC phase of the three-dimensional 
systems discussed above.

\subsubsection{Static structure factor}

The static structure factor is defined as the Fourier transform of the equal-time
density-density CF, as it is defined in Eq. (\ref{FI1_dd-FI}). At zero temperature it is 
related to the quasiparticle excitation spectrum via the Feynman relation
$$
 S({\bf q}) = \frac{Ja^2{\bf q}^2}{2d\,E_{\bf q}} \; ,
$$
where the identification $\hbar^2/2m\equiv Ja^2/2d$ can be considered for a lattice system
(in this case $m=m^\ast$ is the band mass as defined in Eq. (\ref{FI2_band-mass})).
For the weakly interacting Bogoliubov gas the density-density CF was calculated explicitly 
on the level of a Gaussian approximation. It shows an algebraic decay with $1/r^{d+1}$, 
where $d$ is the dimension. The result for the static structure factor agrees with the 
Feynman relation. For the one-dimensional system the density-density CF, and therefore the 
static structure factor, were calculated exactly in the
incommensurate phase, and agree with results from the literature. In the 
MI phase it vanishes.

\subsection{Comparison with results from the Bose-Hubbard model}

In previous calculations, performed on the Bose-Hubbard model, each phase requires its
own specific mean-field approach \cite{stoof1,liu} or a single one close to the phase boundary
\cite{huber1}. Within a Bogoliubov approximation to the Bose-Hubbard model
the quasiparticle spectrum in the BEC phase was found as \cite{stoof1,liu}
\begin{displaymath}
 \epsilon_{q}=\sqrt{J^2 g_{q}^2+2U n_{0} J g_{q}},
\end{displaymath}
where $U$ is the interaction parameter and $n_{0}$ is the condensate density. 
In contrast to this expression,
we found for the spectrum the expressions in the table in section \ref{Sec_qp}. These expressions 
do not agree in the limit $U\to\infty$. Thus our hard-core Bose gas cannot be described within 
the Bogoliubov approximation to
the Bose-Hubbard model by simply sending $U$ to infinity. On the other hand, our results are in good
agreement with a variational Schwinger-boson mean-field approach to the Bose-Hubbard model, which
describe the phases near the phase transition, by sending $U$ to infinity \cite{huber1}.
In the large-$U$ limit of the Bose-Hubbard model, multiple occupation of lattice
sites is prohibited because it cost a large amount of energy. Therefore one can
assume that in this case, the bosons behave like hard-core bosons. 

The results for the excitation spectrum in the Mott-insulating phase from the 
paired-fermion model and the slave-boson model are consistent with the spectrum that
was found for the Bose-Hubbard model in the large-$U$ limit. Inside the first Mott lobe, 
which is the equivalent to
the MI with filling $n_{\rm tot}=1$ for hard-core bosons, the latter is given by the expression
\cite{stoof1,stoof2,huber1}
\begin{equation}
 E^{\rm qp/qh}_{\bf k} = \pm\left(-\mu+\frac U2 - \frac{J-\epsilon_{\bf k}}2\right)
 + \frac 12 \sqrt{(J-\epsilon_{\bf k})^2-6U(J-\epsilon_{\bf k})+U^2} \; ,
\end{equation}
which describes two branches: One (``$+$'' sign) is assigned to quasiparticles and one
(``$-$'' sign) to quasiholes. It depends on the interaction parameter $U$.
For our hard-core bosons, only the quasihole branch can exist,
because the hard-core condition prohibits multiple occupation of lattice sites, in contrary
to the Bose-Hubbard model, where multiple occupation is possible and allows the creation
of particle-hole pairs.
For large values of $U$ the square root term can be written as
$$
 \frac 12 \sqrt{(J-\epsilon_{\bf k})^2-6U(J-\epsilon_{\bf k})+U^2} =
 \frac U2 - \frac 32\left(J-\epsilon_{\bf k}\right) + {\cal O}\left(U^{-1}\right) \; ,
$$
such that we find for the two branches the large-$U$ results
\begin{eqnarray}
 E^{\rm qp}_{\bf k} &=& \epsilon_{\bf k}+U-\left(\mu+2J\right)+{\cal O}\left(U^{-1}\right)  \; , \\
 E^{\rm qh}_{\bf k} &=& \epsilon_{\bf k}+\left(\mu-J\right)+{\cal O}\left(U^{-1}\right) \; .
\end{eqnarray}
The gap of the quasiparticle branch is of the order of $U$, and in the $U\rightarrow\infty$ 
limit it goes to infinity, because the energy to occupy a site with two particles is
infinitely large. On the other hand, the terms which are proportional to $U$ cancel for the
quasihole branch, and its $U\rightarrow\infty$ limit is identical to the result given
in Eq. (\ref{con1_gapped-spectrum}).
Particle-hole excitations cannot be created for hard-core bosons, so the creation of an 
elementary excitation is associated to removing a particle out of the Mott-insulator.
This is possible
in the grand-canonical ensemble, where only the average number is fixed but the 
number of particles fluctuates.
Inside the empty phase, the same quasiparticle spectrum was found as for the Mott-insulator,
due to the particle-hole symmetry.
Here, the creation of an excitation is interpreted by putting an additional particle
into the system.

%++++++++++++++++++++++++++++++++++++++++++++++++++++++++++++++++++++++++++++++++++++++++
%++++++++++++++++++++++++++++++++++++++++++++++++++++++++++++++++++++++++++++++++++++++++
%++++++++++++++++++++++++++++++++++++++++++++++++++++++++++++++++++++++++++++++++++++++++

\section{Conclusion}

In this review, the many-particle problem of strongly interaction bosons in a
lattice potential was investigated. This is motivated by recent experiments on
Bose-Einstein condensates in optical lattices which showed the phase transition from a BEC to
a Mott-insulator. Three different models are discussed,
which allow the calculation of the phase diagram, and experimentally observable
physical quantities like the total density, the condensate density, the quasiparticle
spectrum, and the static structure factor.
All these models have in common that they simulate a strong repulsive interaction
by imposing a hard-core condition on the bosons, which prohibits a multiple occupation 
of lattice sites. They are defined by means of the functional integral method.

The first model is a special construction which describes
non-interacting impenetrable fermions in a one-dimensional lattice. 
We exploited the well-known fact that such a fermionic system is equivalent to impenetrable
bosons in one dimension, and that the static structure factors of the fermionic and the 
bosonic system are identical.
As the fermions are non-interacting, the model can be integrated out exactly.
We calculated the local particle density, the
density-density correlation function and the static structure factor in a translational 
invariant system as well as in a system with a harmonic trap potential.
In the translational invariant case, the static structure factor, which is experimentally 
accessible in Bragg scattering experiments, increases
linearly for small wave vectors, until it reaches unity and remains constant. The
density-density correlation function shows characteristic oscillations and decays 
like $1/r^2$.

The other two models were applied on a Bose gas in a three dimensional lattice.
They were treated in mean-field theory. 
The first one, which was called the paired-fermion model, 
was constructed by a field of pairs of Grassmann variables.
It can be seen as an interacting fermionic model. The second one was based on a slave-boson 
approach. A Hubbard-Stratonovich 
transformation allows to integrate out the original fields in both models.
This transformation leads to new fields, which are connected to the condensate order parameter.
A saddle-point approximation provides both a mean-field solution and Gaussian fluctuations.
The latter contain the information about quasiparticle excitations.
For a three-dimensional lattice, the total particle density and the condensate density can be
calculated in mean-field theory, and the quasiparticle spectrum and the static structure
factor was calculated on the level of Gaussian fluctuations.
The saddle point approximations of the two models lead to qualitatively the same results.

Our results for the one-dimensional model, the paired-fermion model, and the slave-boson 
model, show a particle hole symmetry. At zero temperature, 
they have a common phase diagram,
with one phase boundary between the empty phase and the incommensurate phase, and one
between the incommensurate phase and the Mott-insulating phase. If the temperature is non-zero,
there is no clear phase transition between the
empty phase and the Mott-insulator due to thermal fluctuations.
While there is no
Bose-Einstein condensation in the one-dimensional system, the incommensurate
phase is a BEC in the paired-fermion and slave-boson model in three dimensions.
For the latter two models, the mean-field results for the total density and the
condensate density agree exactly at zero temperature, at higher temperature they agree
qualitatively. It was shown that they lead to the Gross-Pitaevskii result in the
limit of low temperature, if the density is small compared to the lattice
constant. At higher temperatures, we have shown that the slave-boson model leads to
a renormalised Gross-Pitaevskii equation with temperature dependent coefficients.
A similar theory could in principle be derived on the mean-field level from the 
paired-fermion model as well. It could be compared to the renormalised 
Gross-Pitaevskii theory which was derived from the slave-boson model.

The quasiparticle spectra which were found for both three-dimensional models, are gapless
(Goldstone mode) in the BEC phase. In the dilute regime,
they agree with the well-known Bogoliubov result.
In the empty phase and the Mott-insulator, the quasiparticle spectrum is gapped.
Our results agree with results which were derived for the Bose-Hubbard model, if
the on-site interaction constant $U$ is very large.
The Goldstone mode in the BEC phase of the paired-fermion model was found as the quasiparticle 
pole of only one eigenvalue of the $4\times 4$ quasiparticle Green's function.
Additional massive modes may be found from the remaining eigenvalues.

At zero temperature, the elementary excitations are connected to the static structure factor
via the Feynman relation.
In the empty phase and the Mott-insulator, the static structure factor vanishes because of
the absence of density fluctuations.

%++++++++++++++++++++++++++++++++++++++++++++++++++++++++++++++++++++++++++++++++++++++++
%++++++++++++++++++++++++++++++++++++++++++++++++++++++++++++++++++++++++++++++++++++++++
%++++++++++++++++++++++++++++++++++++++++++++++++++++++++++++++++++++++++++++++++++++++++

%++++++++++++++++++++++++++++++++++++++++++++++++++++++++++++++++++++++++++++++++++++++++
%++++++++++++++++++++++++++++++++++++++++++++++++++++++++++++++++++++++++++++++++++++++++
%++++++++++++++++++++++++++++++++++++++++++++++++++++++++++++++++++++++++++++++++++++++++

\appendix

\section{Finite sums and products\label{Sec_App1-sums}}
\subsection{Bosonic sum}
For bosonic systems, which have a periodic structure in the imaginary time variable, 
we have to perform sums of the type
$$
 \sum_{n=1}^M \frac 1M \frac{e^{-\frac{2\pi\rm i}M nm}}{1-a\, e^{\frac{2\pi\rm i}M n}} \; .
$$
This sum is performed by finding the common denominator, which is
given by $1-a^M$. The numerator then is
$$
 \mbox{numerator}= \sum_{n=1}^M e^{-\frac{2\pi\rm i}M nm}
 \prod_{k\ne n} \left( 1-a\, e^{+\frac{2\pi\rm i}M k} \right)
$$
where
$$
 \prod_{k\ne n} \left( 1-a\, e^{\frac{2\pi i}M k} \right) =
 \frac{1-a^M}{1-a\, e^{\frac{2\pi\rm i}M n}} =
 1 + a\, e^{\frac{2\pi\rm i}M n} + a^2 e^{\frac{2\pi\rm i}M 2n} + \ldots +
 a^{M-1} e^{\frac{2\pi\rm i}M (M-1)n} \; .
$$
Therefore we find
$$
 \mbox{numerator}= \sum_{n=1}^M e^{-\frac{2\pi\rm i}M nm}
 \sum_{l=1}^M a^{l-1} e^{\frac{2\pi\rm i}M (l-1)n} =
 \sum_{n,l=1}^M a^{l-1} e^{-\frac{2\pi\rm i}M n(m-l+1)} =
$$
$$
 M \sum_{l=1}^M a^{l-1} \delta'_{l,m+1}, \quad \mbox{where }
 \delta'_{l,k} := \sum_{j=-\infty}^\infty \delta_{l,k+jM} \; .
$$
With the restriction $m=-(M-1),\ldots,M-1$ the ``enhanced'' Kronecker symbol $\delta'$
contributes for the two cases
$$
 \begin{array}{l@{\quad \mbox{if} \quad}l}
 l=m+1 & m\ge 0 \\
 l=M+m+1 & m<0 \; .
 \end{array}
$$
Finally, this leads to the components of the inverse matrix:
\begin{equation}
 \sum_{n=1}^M \frac 1M \frac{e^{-\frac{2\pi\rm i}M nm}}{1-a\, e^{\frac{2\pi\rm i}M n}} =
 \frac 1{1-a^M} \times \left\{ \begin{array}{l@{\quad \mbox{if} \quad}l}
 a^m & m\ge 0 \\ a^{M+m} & m<0 \end{array} \right. \; . \vspace{0.6cm}
 \label{App1_sum-bosonic}
\end{equation}

\subsection{Fermionic sum}
For fermionic systems, which have an anti-periodic structure in the imaginary time variable, 
we have to perform sums of the type
\begin{equation}
 \sum_{n=1}^M \frac 1M \frac{e^{-\frac{2\pi\rm i}M \left(n-\frac 12\right)m}}{1-a\, 
 e^{\frac{2\pi\rm i}M \left(n-\frac 12\right)}} = \frac 1{1+a^M} \times
 \left\{ \begin{array}{l@{\quad \mbox{if} \quad}l}
 a^m & m\ge 0 \\ -a^{M+m} & m<0 \end{array} \right. \; .
 \label{App1_sum-fermionic}
\end{equation}
This sum differs from the sum given in Eq. (\ref{App1_sum-bosonic}) only by the
substitution $a\rightarrow a\, e^{-\pi{\rm i}m/M}$ and a multiplication by the factor
$e^{\pi{\rm i}m/M}$, so the result can be verified easily.

\subsection{Sums with cosines}
The following two sums require the condition $|b|>1$:
\begin{equation}
 \sum_{n=1}^M \frac 1M \frac 1{\cos\left(\frac{2\pi}M n\right) -b} =
 \frac 1{\sqrt{b^2-1}} \, \frac{\left(b-\sqrt{b^2-1}\right)^M+
 \left(b+\sqrt{b^2-1}\right)^M+2}{\left(b-\sqrt{b^2-1}\right)^M-\left(b+\sqrt{b^2-1}\right)^M}
 \label{App1_sum-cos1}
\end{equation}
\begin{equation}
 \sum_{n=1}^M \frac 1M 
 \frac{\cos{\left(\frac{2\pi}M n\right)}}{\cos\left(\frac{2\pi}M n\right) -b} =
 \frac 1{\sqrt{b^2-1}} \, \frac{\left(b-\sqrt{b^2-1}\right)^{M-1}+
 \left(b+\sqrt{b^2-1}\right)^{M-1}+2b}{\left(b-\sqrt{b^2-1}\right)^M-\left(b+\sqrt{b^2-1}\right)^M}
 \label{App1_sum-cos2}
\end{equation}
To perform these two sums the following identities were used:
$$
 \frac 1{\cos(x)-\frac{a^2+1}{2a}} = \frac{2a^2}{a^2-1} \left[
 \frac 1{e^{{\rm i}x}-a} - \frac 1a \, \frac 1{a\, e^{{\rm i}x}-1} \right]
$$
$$
 \frac{\cos(x)}{\cos(x)-\frac{a^2+1}{2a}} = \frac{a^2}{a^2-1} \left[
 \frac 1{a\, e^{{\rm i}x}-1} - \frac 1a \, \frac 1{e^{{\rm i}x}-a} -
 \frac 1a \, \frac 1{e^{-{\rm i}x}-a} + \frac 1{a\, e^{-{\rm i}x}-1} \right]
$$
All separate terms can be traced back to the sum given in Eq. (\ref{App1_sum-bosonic}).

\subsection{Sum for $C(k)$ in Eq. (\ref{d11_Ck})\label{App_sumCk}}
We perform the sum
\begin{equation}
 C(k) = \lim_{M\rightarrow\infty}
 \sum_{l=1}^M \frac 1M \frac{\left[ -e^{\frac{2\pi\rm i}M l}+e^{\frac{\pi\rm i}M}\left(
 1-\frac\beta M \mu\right)\right] e^{\frac{\pi i}M}}{
 \left( e^{\frac{2\pi\rm i}M l}-e^{\frac{\pi\rm i}M}\left(1-\frac\beta M \mu\right)\right)^2 -
 e^{\frac{2\pi\rm i}M l} e^{\frac{\pi\rm i}M}\left(\frac\beta M J\right)^2 \cos^2 \frac k2} \; .
\end{equation}
Make the following substitutions:
$$
 a:=-\left(1-\frac\beta M \mu\right)e^{\frac{\pi i}M} \quad ;
 \quad b=\frac\beta M J e^{\frac{\pi i}{2M}} \cos\frac k2 \; ,
$$
$$ 
 f(z):=\frac{z+a}{(z+a)^2-b^2 z} \; .
$$
With these definitions, the sum is given as
$$
 C(k) = - \lim_{M\rightarrow\infty}
 \sum_{l=1}^M \frac 1M \, e^{\frac{\pi\rm i}M} \, f\left(e^{\frac{2\pi\rm i}M l}\right) \; .
$$
The roots of the denominator of $f(z)$ are
$$
 z^\pm = \frac{b^2}2-a\pm \frac b2 \sqrt{b^2-4a} \; .
$$
We perform an expansion into partial fraction and find
$$
 f(z) = \frac A{z-z^+} + \frac B{z-z^-} = \frac{(A+B)z-(Az^- + Bz^+)}{(z-z^+)(z-z^-)}
$$
with
$$
 A=\frac 12 + \frac b{2\sqrt{b^2-4a}} \quad ; \quad B=\frac 12 - \frac b{2\sqrt{b^2-4a}} \; .
$$
To perform the sum, we use the following identity which can be traced back to Eq.
(\ref{App1_sum-bosonic}):
$$
 \sum_{l=1}^M \frac 1M \, \frac 1{e^{\frac{2\pi i}M l}-z^\pm} =
 -\frac 1{z^\pm} \, \frac 1{1-\left(\frac 1{z^\pm}\right)^M}
$$
$$
 \Longrightarrow 
 -\sum_{l=1}^M \frac 1M \, e^{\frac{\pi\rm i}M} \, f\left(e^{\frac{2\pi\rm i}M l}\right)
 = \left[\frac A{z^+} \, \frac 1{1-\left(\frac 1{z^+}\right)^M}
 +\frac B{z^-} \, \frac 1{1-\left(\frac 1{z^-}\right)^M}\right]e^{\frac{\pi i}M} \; .
$$
The limit $M\rightarrow\infty$ can now be performed, by the help of the identities
$$
 \lim_{M\rightarrow\infty}(z^\pm)^M = e^{\pi i} \,
 \lim_{M\rightarrow\infty} \left(1+\left(\pm J\cos\frac k2 -\mu\right)\frac\beta M+
 {\cal O}\left(\frac 1{M^2}\right)\right)^M =
 - e^{\beta\left(\pm J\cos\frac k2 -\mu\right)}
$$
$$
 \lim_{M\rightarrow\infty} z^\pm = 1 \quad ; \quad
 \lim_{M\rightarrow\infty} A,B = \frac 12 \; .
$$
The result is given in Eq. (\ref{d11_sum-result}).

\subsection{Sum for $G$ in Eq. (\ref{cfm2_G-def})\label{App_sumG}}
We perform the sum
\begin{equation}
 G = \frac 1M \sum_{n=1}^M \frac{{\rm i}({\rm i}\varphi_0+\chi_0)}{
 1+({\rm i}\varphi_0+\chi_0)({\rm i}\varphi_0^\ast+\chi_0^\ast)-
 2\, e^{-\frac{{\rm i}2\pi}M\left(n-\frac 12\right)} + 
 \left(1-\left(\frac{\beta\mu}{2M}\right)^2\right)
 \, e^{-2\frac{{\rm i}2\pi}M\left(n-\frac 12\right)} } \; .
 \label{cfm2_G-sum}
\end{equation} 
We define
$$
 a := 1+({\rm i}\varphi+\chi)({\rm i}\varphi^\ast+\chi^\ast) \; , \quad
 b := 1-\left(\frac{\beta\mu}{2M}\right)^2 \; ,
$$
$$
 f(z) = \frac 1{a-2z+bz^2} \; .
$$
The roots of the denominator of $f(z)$ are
$$
 z^\pm = \frac 1b \left(1\pm\sqrt{1-ab}\right) \; .
$$
An expansion into partial fraction leads to
$$
 f(z) = A\left( \frac 1{z-z^+} - \frac 1{z-z^-} \right) \; , \quad \mbox{where} \quad
 A = \frac 1{2\sqrt{1-ab}} \; .
$$
To perform the sum, we use the following identity which can be traced back to Eq.
(\ref{App1_sum-fermionic}):
$$
 \sum_{l=1}^M \frac 1M \, \frac 1{e^{\frac{2\pi i}M\left(l+\frac 12\right)}-z^\pm} =
 -\frac 1{z^\pm} \, \frac 1{1+\left(\frac 1{z^\pm}\right)^M}
$$
$$
 \Longrightarrow 
 -\sum_{l=1}^M \frac 1M f\left(e^{\frac{2\pi\rm i}M\left(l+\frac 12\right)}\right)
 = A \left[\frac 1{z^+} \, \frac 1{1+\left(\frac 1{z^+}\right)^M}
 -\frac 1{z^-} \, \frac 1{1+\left(\frac 1{z^-}\right)^M}\right] \; .
$$

\subsection{Product to calculate the determinant of Eq. (\ref{Bog2_diag-matrix})\label{App_product}}
We want to perform a product of the type
$$
 \prod_{n=1}^M \left(b-\cos\left(\frac{2\pi}M n\right)\right)\; , \quad |b|>1 \; .
$$
This can be verified to be equal to
$$
 \prod_{n=1}^M \left[ \frac 12\left(b+\sqrt{b^2-1}\right)\left(
 1-\left(b-\sqrt{b^2-1}\right)e^{{\rm i}\frac{2\pi}M n}\right)
 \left(1-\left(b-\sqrt{b^2-1}\right)e^{-{\rm i}\frac{2\pi}M n}\right) 
 \right] \; ,
$$
such that the identity 
\begin{equation}
 \prod_{n=1}^M \left(1-a\, e^{\frac{2\pi\rm i}M n}\right) = 1-a^M\; ,
\end{equation}
can be applied. As a result we find
\begin{equation}
  \prod_{n=1}^M \left(b-\cos\left(\frac{2\pi}M n\right)\right) =
  2^{-M} \left( \left(b+\sqrt{b^2-1}\right)^M + \left(b-\sqrt{b^2-1}\right)^M - 2\right) \; .
  \label{App1_prod2}
\end{equation}

\section{Coherent states for bosons and fermions\label{App_coherentstates}}

The functional integral representation for bosonic and fermionic systems is constructed 
of coherent states \cite{Bnegele}. We denote bosonic 
operators by $\hat{a}^+_\alpha$, $\hat{a}_\alpha$, and the
fermionic operators by $\hat{c}^+_\alpha$, $\hat{c}_\alpha$. The commutation relations are
\begin{eqnarray}
 \left[ \hat{a}_\alpha,\hat{a}^+_{\alpha'} \right]_- &=& \delta_{\alpha\alpha'} \; , \\
 \left[ \hat{c}_\alpha,\hat{c}^+_{\alpha'} \right]_+ &=& \delta_{\alpha\alpha'} \; . 
\end{eqnarray}
The vacuum state, i.e. the state containing no
particle, we call $|0\rangle$. We define coherent states for
\begin{itemize}
\item bosons by means of complex field variables $\phi^\ast_{\alpha}$, $\phi_{\alpha}$:
\begin{equation}
 \left| \phi \right\rangle = e^{\sum_{\alpha} \phi_{\alpha} \hat{a}^+_{\alpha}} 
 |0\rangle \; , \quad
 \left\langle \phi \right| = \langle 0| \, 
 e^{\sum_{\alpha} \phi^\ast_{\alpha} \hat{a}_{\alpha}} \; .
\end{equation}
\item fermions by means of conjugate Grassmann variables $\bar\psi_{\alpha}$, $\psi_{\alpha}$,
where we require, that the Grassmann variables anticommute with the fermionic operators:
\begin{equation*}
 \left| \psi \right\rangle = e^{-\sum_\alpha \psi_{\alpha} \hat{c}^+_{\alpha}} |0\rangle =
 \prod_{\alpha} \left( 1-\psi_{\alpha} \hat{c}^+_{\alpha} \right) |0\rangle \; ,
\end{equation*}
\begin{equation}
 \left\langle \psi \right| = \langle 0| \, 
 e^{\sum_{\alpha} \bar\psi_{\alpha} \hat{c}_\alpha} =
 \langle 0| \, \prod_{\alpha} \left( 1+\bar\psi_{\alpha} \hat{c}_\alpha \right) \; .
\end{equation}
\end{itemize}
For the construction of the coherent state functional integral, the following properties
are relevant. They can be checked by using the previous definitions and the integration
properties of complex, Grassmannian and nilpotent variables:
\begin{itemize}
\item Coherent states are eigenvalues of annihilation operators:
\begin{equation}
 \hat{x}_\alpha | \xi \rangle = \xi_\alpha | \xi \rangle \; , \quad
 \langle \xi | \hat{x}^+_\alpha = \langle \xi | \bar\xi_\alpha \; ,
\end{equation}
where $\hat{x}=\hat{a}$, $\xi=\phi$, $\bar\xi=\phi^\ast$ for bosons, and
$\hat{x}=\hat{c}$, $\xi=\psi$, $\bar\xi=\bar\phi$ for fermions.
\item  Scalar product, where the operator $\hat X$ is built of bosonic, fermionic, or hard-core
operators, respectively:
\begin{equation}
 \langle\xi|\hat{X}(\hat{x}^+,\hat{x})|\xi'\rangle = 
 e^{\sum_\alpha \bar\xi_\alpha \xi'_\alpha} X(\bar\xi_\alpha,\xi'_\alpha) \; ,
\end{equation}
where $\hat{x}$, $\xi$, $\bar\xi$ have to be chosen as mentioned above.
\item Closure relation (the unity operator is denoted by $\bf 1$):
\begin{eqnarray}
 {\bf 1} &=& \int e^{-\sum_\alpha \phi^\ast_\alpha \phi_\alpha}
 \left| \phi \right\rangle \left\langle \phi \right| \prod_\alpha 
 \frac{{\rm d}\phi^\ast_\alpha{\rm d}\phi_\alpha}{2\pi\rm i} \\
 {\bf 1} &=& \int e^{-\sum_\alpha \bar\psi_\alpha \psi_\alpha}
 \left| \psi \right\rangle \left\langle \psi \right| \prod_\alpha 
 {\rm d}\bar\psi_\alpha{\rm d}\psi_\alpha \; .
\end{eqnarray}
\item Trace of an operator $\hat X$:
\begin{eqnarray}
 {\rm Tr}\; \hat{X}(\hat{a}^+_\alpha,\hat{a}_\alpha) &=& 
 \int e^{-\sum_\alpha \phi^\ast_\alpha \phi_\alpha}
 \langle\phi|\hat{X}|\phi\rangle \prod_\alpha 
 \frac{{\rm d}\phi^\ast_\alpha{\rm d}\phi_\alpha}{2\pi\rm i} \\
 {\rm Tr}\, \hat{X}(\hat{c}^+_\alpha,\hat{c}_\alpha) &=& 
 \int e^{-\sum_\alpha \bar\psi_\alpha \psi_\alpha}
 \langle-\psi|\hat{X}|\psi\rangle \prod_\alpha 
 {\rm d}\bar\psi_\alpha{\rm d}\psi_\alpha \; .
\end{eqnarray}
\end{itemize}
Using these identities, the functional integral of the grand canonical partition function
$$
 Z = {\rm Tr}\; e^{-\beta(\hat{H}(\hat{x}^+_\alpha,\hat{x}_\alpha)-
 \mu\hat{N}(\hat{x}^+_\alpha,\hat{x}_\alpha))}
$$
with the Hamiltonian $\hat H$ is constructed in the following manner: 
We apply the relation for the trace and insert the closure relation $M-1$ times. Introducing the 
discrete-imaginary-time index $n=1,\ldots,M$ we have
\begin{equation}
 Z = \int e^{ \sum_{\alpha,n} \bar\xi_{\alpha,n} \xi_{\alpha,n}}
 \langle\sigma\bar\xi_{1}|e^{-\frac\beta M(\hat{H}-\mu\hat{N})}
 |\xi_{M}\rangle 
 \prod_{n=2}^{M} \langle\bar\xi_{n}|e^{-\frac\beta M(\hat{H}-\mu\hat{N})}
 |\xi_{n-1}\rangle
 \prod_{\alpha,n} \frac{{\rm d}\bar\xi_\alpha{\rm d}\xi_\alpha}{\cal N} \; ,
\end{equation}
where $\sigma=+1$ for bosons and $-1$ for fermions, and ${\cal N}=2\pi\rm i$ for bosons and
$1$ for fermions.
The minus sign inside the scalar product in the fermionic trace gives rise to the anti-periodicity
of the fermionic field variables. The different sign in the exponent of the hard-core bosonic
trace is the reason that the diagonal term in the action for hard-core bosons is different from
bosonic and fermionic actions.

The operator in the exponent 
$\hat{H}(\hat{x}^+_\alpha,\hat{x}_\alpha)-\mu\hat{N}(\hat{x}^+_\alpha,\hat{x}_\alpha)$
can be replaced by its normal ordered from by making an error of the order $(\beta/M)^2$
which vanishes for $M\rightarrow\infty$. Applying the eigenvalue property and the
product property yields
\begin{equation}
 Z = \lim_{M\rightarrow\infty} \int e^{-A(\bar\xi,\xi)}
 \prod_{n=1}^M \prod_\alpha \frac{{\rm d}\bar\xi_{\alpha,n}{\rm d}\xi_{\alpha,n}}{\cal N}
\end{equation}
with the action
\begin{equation}
 A(\bar\xi,\xi) = \frac\beta M \sum_{n=1}^{M} \left\{ \sum_\alpha
 \sigma_1 \bar\xi_{\alpha,n+1} \left[ \frac M\beta \left(\xi_{\alpha,n+1}-\xi_{\alpha,n}\right) -
 \mu \xi_{\alpha,n} \right] + H(\xi^\ast_{\alpha,n+1},\xi_{\alpha,n}) \right\}
\end{equation}
and the boundary condition $\xi_{\alpha,1}=\sigma_2\xi_{\alpha,M+1}$, $\bar\xi_{\alpha,1}=\sigma_2\bar\xi_{\alpha,M+1}$.

\section{Expectation values and Wick's theorem\label{App_Wick}}

An expectation value of an expression in terms of real/complex/Grassmann variables is
defined by means of Eq. (\ref{FI1_expectation-FI}).
A {\it second order expectation value} provides the matrix element of the (inverse) 
Green's matrix $\hat{\cal G}$: \\
% \begin{center}
\begin{equation}
 \begin{array}{ll}
 \mbox{Real variables:} & \langle\phi_j\phi_k\rangle = \frac 12 \hat{\cal G}^{-1}_{jk}\\
 \mbox{Complex conjugate variables:} & \langle\phi^\ast_j\phi_k\rangle = \hat{\cal G}^{-1}_{jk}\\
 \mbox{Conjugate Grassmann variables:} & \langle\bar\psi_j\psi_k\rangle = \hat{\cal G}_{jk}
 \end{array}
\end{equation}
% \end{center}
{\it Forth order expectation values} can be calculated
via the application of Wick's theorem \cite{Bpopov,Bnegele}.
It can be split into products of second-order expectation values
and a sum has to be performed over all possible pairings (including a sign for Grassmann 
variables): \\
\begin{equation}
 \begin{array}{ll}
 \mbox{Real var.:} & \langle\phi_j\phi_k\phi_l\phi_m\rangle =
 \langle\phi_j\phi_k\rangle \langle\phi_l\phi_m\rangle +
 \langle\phi_j\phi_l\rangle \langle\phi_k\phi_m\rangle +
 \langle\phi_j\phi_m\rangle \langle\phi_k\phi_l\rangle \\
 \mbox{C. conj. var.:} & \langle\phi^\ast_j\phi^\ast_k\phi_l\phi_m\rangle =
 \langle\phi^\ast_j\phi_m\rangle \langle\phi^\ast_k\phi_l\rangle +
 \langle\phi^\ast_j\phi_l\rangle \langle\phi^\ast_k\phi_m\rangle \\
 \mbox{Conj. Gr. var.:} & \langle\bar\psi_j\bar\psi_k\psi_l\psi_m\rangle =
 \langle\bar\psi_j\psi_m\rangle \langle\bar\psi_k\psi_l\rangle -
 \langle\bar\psi_j\psi_l\rangle \langle\bar\psi_k\psi_m\rangle \\
 \end{array} \label{App1_wick}
\end{equation}

\section{Correlations\label{App_corr}}

The decay of the density-density CF given in Eq. (\ref{cfm3_D-decay}) is investigated in
$d=1,2,3$ dimensions. For convenience we write $c:=\sqrt{2(\mu+J)}$. We use a cut-off
at $|{\bf q}|=Q$ for the integrals.
\begin{itemize}
\item {\bf One dimension:}
$$
 D(r) = \int_{-Q}^{Q} \frac{|q|}{c} \, e^{{\rm i}qr} \, {\rm d}q = 
 \frac 2{cr^2} \int_{0}^{Qr} q' \cos(q') {\rm d}q' \sim \frac 1{r^2}
$$
The anti-symmetrical part which is $\sim \sin(q')$ does not contribute.
\item {\bf Two dimensions} with polar coordinates $(q,\phi)$:
$$
 D(r) = \int_0^Q {\rm d}q \, q\int_0^{2\pi}{\rm d}\phi \, \frac qc \, e^{{\rm i}qr\cos\phi} =
 \frac 1{cr^3} \int_0^{2\pi}{\rm d}\phi \, \frac 1{\cos^{3}\phi} \int_0^{rQ} q'^2
 \cos(q') {\rm d}q' \sim \frac 1{r^3}
$$
\item {\bf Three dimensions} with spherical coordinates $(q,\theta,\phi)$:
$$
 D(r) = \int_0^Q {\rm d}q \, q^2 \int_0^{2\pi}{\rm d}\phi
 \int_1^{-1} {\rm d}(\cos\theta) \, \frac qc \, e^{{\rm i}qr\cos\theta} 
$$
$$
 = \frac{2\pi}{cr^3} \int_1^{-1} {\rm d}(\cos\theta) \frac 1{\cos^{4}\phi} \int_0^{rQ} q'^3
 \cos(q') {\rm d}q' \sim \frac 1{r^4}
$$
\end{itemize}

\section{Calculations to the paired-fermion model\label{App_pf}}

In this Appendix we write out the expression for the Green's function
in both cases $|\phi|=0$ and $|\phi|\ne 0$.\\ % We denote ${\bf p}=\{{\bf k},\omega\}$.
Case: $|\phi|=0$\\
Deviation of the effective  action due to fluctuations is
\begin{equation}
\delta A_{\rm eff}=\sum_{{\bf k},\omega}
(\begin{array}{cc} \delta \phi_{{\bf k},\omega} \ \ \delta\chi_{{\bf k},\omega}  
\end{array})\overbrace{\left(\begin{array}{cc} v^{-1}_{\bf k}-D(\omega) & {\rm i} D(\omega) \\  
{\rm i} D(\omega) & \frac{1}{2J}+D(\omega) 
\end{array}\right)}^{\displaystyle \hat{\cal G}^{-1}}\left(\begin{array}{c}
\delta\phi^\ast_{{\bf k},\omega} \\ 
\delta\chi^\ast_{{\bf k},\omega}   
\end{array}\right),
\end{equation}
where
\[
D(\omega)= \frac{1}{|\mu|-{\rm i}\omega}, \ \ v^{-1}_{\bf k}=\frac{1}{J(3-\epsilon_{\bf k})}.
\]
The determinant of the Green's function reads
\begin{equation}
\det
\hat{\cal G}^{-1}=\frac{v^{-1}_{\bf k}}{2J}-D(\omega)\left(\frac{1}{2J}-v^{-1}_{\bf k}
\right).
\end{equation}
Case: $|\phi|\ne 0$\\
Deviation of the effective action due to fluctuations is
\begin{equation}
\delta A_{\rm eff}
=\sum_{{\bf k},\omega}(\begin{array}{cccc} \delta \phi_{{\bf k},\omega}, \delta\chi_{{\bf k},\omega}, 
\delta\phi^\ast_{-{\bf k},-\omega}, \delta\chi^\ast_{-{\bf k},-\omega}  
\end{array})\hat{\cal G}^{-1}\left(\begin{array}{c}\delta\phi^\ast_{{\bf k},\omega} \\ 
\delta\chi^\ast_{{\bf k},\omega}\\ \delta\phi_{-{\bf k},-\omega}\\ \delta\chi_{-{\bf k},-\omega}  
\end{array}\right)
\end{equation}
with the Green's function
\begin{equation}
\hat{\cal G}^{-1}=\left(\begin{array}{cccc} 
v^{-1}_{\bf k}-D(\omega) & {\rm i} D(\omega) & -a & {\rm i} a \\ {\rm i} D(\omega) & 
\frac{1}{2J}+D(\omega) & {\rm i} a & a \\
-a & {\rm i} a &  v^{-1}_{\bf k}-D(\omega) & {\rm i} D(\omega)  \\ {\rm i} a & a &  
{\rm i} D(\omega) & \frac{1}{2J}+D(\omega)  
\end{array}\right),
\end{equation}
where
\[
D(\omega)=\frac{1}{2}\cdot\frac{\mu^{2}+J^{2}+2{\rm i}\mu\omega}{J(J^{2}+\omega^{2})},
\]
\[
D(-\omega)=\frac{1}{2}\cdot\frac{\mu^{2}+J^{2}-2{\rm i}\mu\omega}{J(J^{2}+\omega^{2})},
\]
\[
a=-\frac{1}{2}\cdot\frac{|\Phi|^{2}/9}{J(J^{2}+\omega^{2})}.
\]
The determinant of the Green's function is
\begin{equation}
\det \hat{\cal G}^{-1}=\frac{1}{[2J^{2}(3-\epsilon_{\bf k})]^{2}(J^{2}+\omega^{2})}
\cdot [\omega^{2}+(J^{2}-\mu^{2})\epsilon_{\bf k}+\mu^{2}\epsilon_{\bf k}^{2}].
\end{equation}

\section{Calculations to the slave-boson model}

\subsection{Integration of the constraint\label{Sec_App_constraint}}

We perform the integration of the complex fields $b$ and $e$.
The integral factorises such that it can be performed for each lattice site $\bf r$
independently. Therefore we will drop the index $\bf r$ here temporarily and evaluate 
the expression
\begin{equation}
 \int \exp \left\{ - \beta s \varphi^2 - \beta (e,b) \left(
 \begin{array}{cc} 2s\varphi+s & s\Phi \\ s\Phi^\ast & -\mu
 \end{array}\right)\left( \begin{array}{c}
 e^\ast \\ b^\ast \end{array}\right) \right\}
 \delta(|b|^2 + |e|^2 -1) {\rm d}e^\ast {\rm d}e \, {\rm d}b^\ast {\rm d}b \; .
 \label{App3_constraint}
\end{equation}
The eigenvalues of the $2\times 2$ matrix are
$$
 \lambda_\pm = \beta s\left(\varphi+\frac 12\right)-\beta\frac\mu 2 \pm \beta \sqrt{
 \left[\left(\varphi+\frac 12\right)s+\frac\mu 2\right]^2 + s^2 |\Phi|^2} \; .
$$
A unitary transformation can be applied to the vector $(e,b)$ such that the matrix has
diagonal form. This does not affect the constraint, because the expression $|b|^2+|e|^2=1$
remains unchanged after a unitary transformation. Therefore the integral is equal to
$$
 \int {\rm d}e^\ast {\rm d}e \, {\rm d}b^\ast {\rm d}b \exp
 \left[ -\beta s\varphi^2 - \lambda_1 |e|^2 - \lambda_2 |b|^2 \right]
 \delta(|b|^2 + |e|^2 -1)
$$
$$
 = (2\pi)^2 \frac 12 \int_0^1 {\rm d}\rho \, \rho \exp
 \left[ -\beta s\varphi^2 -\lambda_1 \rho^2 - \lambda_2\left(1-\rho^2\right)\right]
$$
$$
 = 2 \pi^2 e^{-\beta s\varphi^2} \, 
 \frac{{\rm e}^{-\lambda_1}-{\rm e}^{-\lambda_2}}{\lambda_1-\lambda_2}
$$
$$
 = 4 \pi^2 \exp\left[-\beta s\varphi^2-\beta s\left(\varphi+\frac 12\right)+\beta\frac\mu 2\right]
 \frac{\sinh\left[\beta\sqrt{\left[\left(\varphi+\frac 12\right)s+\frac\mu 2\right]^2
 +s^2|\Phi|^2}\right]}{
 \beta\sqrt{\left[\left(\varphi+\frac 12\right)s+\frac\mu 2\right]^2+
 s^2|\Phi|^2}} \; .
$$
After performing the shift $\varphi+1/2\rightarrow\varphi$ and using the index $\bf r$ again,
the integral (\ref{App3_constraint}) gives the result
\begin{equation}
 \int_{-\infty}^\infty {\rm d}\varphi_{\bf r} \,
 \frac{ \sinh\left[\beta\sqrt{\left(\varphi_{\bf r} s+\frac{\mu_{\bf r}}2 \right)^2 
 + s^2|\Phi_{\bf r}|^2}\right]}{ \beta\sqrt{\left(\varphi_{\bf r} s
 + \frac{\mu_{\bf r}}2\right)^2 + s^2|\Phi_{\bf r}|^2}} 
 e^{-\beta s\varphi_{\bf r}^2+\frac{\beta\mu_{\bf r}}4}\; .
\end{equation}

\subsection{Condensate density\label{Sec_App_SB-n0}}

In a Bose system in an optical lattice, which is described by a complex field 
$\phi_{\bf r}(\tau)$, the condensate density is defined by the expression (\ref{FI1_n0})
via the concept of off-diagonal long range order.
In classical approximation, the field does not depend on imaginary time $\tau$, and in the
slave-boson approach, we replace
$$
 \phi^\ast_{\bf r} \rightarrow b^\ast_{\bf r} e_{\bf r} \; ; \quad
 \phi_{\bf r} \rightarrow e^\ast_{\bf r} b_{\bf r} \; ,
$$
thus we use the definition
\begin{equation}
 n_0 = \lim_{{\bf x-x'}\rightarrow\infty}
 \left\langle b^\ast_{\bf x} e_{\bf x} e_{\bf x'}^\ast b_{\bf x'} \right\rangle \; .
 \label{App3_exp-value}
\end{equation}
for the condensate density.
Here, the expectation value is given by
\begin{equation}
 \langle\cdots\rangle = \frac 1{Z_{\rm sb}} \int\cdots\exp[\ldots]\
 {\cal D} [\Phi^\ast,\Phi,\varphi] \, {\cal D}[b,b^\ast,e,e^\ast] \; .
\end{equation}
We are interested in the connection between the correlation function
$\left\langle\Phi_{\bf x} \Phi_{\bf x'}^\ast\right\rangle$ and the condensate density.
For this purpose we integrate out the field $\Phi$ to transform the correlation
function of the field $\Phi$ back to a correlation function of the fields $b$ and $e$.
Therefore, we write
$$
 \hat{v}_{\bf rr'} := \frac{s\delta_{\bf rr'}-\hat{J}_{\bf rr'}}{s^2}
$$
for simplicity and perform the integration
$$
 \beta^2 s^2 \int \Phi_{\bf x }\Phi_{\bf x'}^\ast \exp\left[ 
 \beta \sum_{\bf r,r'} \Phi^\ast_{\bf r} \hat{v}^{-1}_{\bf rr'} \Phi_{\bf r'}
 + \beta s \sum_{\bf r} \Phi_{\bf r} b_{\bf r}^\ast e_{\bf r} 
 + \beta s \sum_{\bf r} \Phi_{\bf r}^\ast e_{\bf r}^\ast b_{\bf r} \right]
 \prod_{\bf r} {\rm d}\Phi_{\bf r} {\rm d}\Phi_{\bf r}^\ast =
$$
$$
 \frac\partial{\partial(b_{\bf x}^\ast e_{\bf x})} \, 
 \frac\partial{\partial(b_{\bf x'} e_{\bf x'}^\ast)}
 \int \exp\left[ \beta \sum_{\bf r,r'} \Phi^\ast_{\bf r} \hat{v}^{-1}_{\bf rr'} \Phi_{\bf r'}
 + \beta s\sum_{\bf r} \Phi_{\bf r} b^\ast_{\bf r} e_{\bf r} 
 + \beta s\sum_{\bf r} \Phi^\ast_{\bf r} e^\ast_{\bf r} b_{\bf r} \right]
 \prod_{\bf r} {\rm d}\Phi_{\bf r} {\rm d}\Phi_{\bf r}^\ast =
$$
$$
 \frac\partial{\partial(b_{\bf x}^\ast e_{\bf x})} \, 
 \frac\partial{\partial(b_{\bf x'} e_{\bf x'}^\ast)}
 \det \left(\frac{\hat{v}}\beta\right) \, \exp\left[ \beta s^2
 \sum_{\bf r,r'} b^\ast_{\bf r} e_{\bf r} \hat{v}_{\bf rr'} e^\ast_{\bf r'} b_{\bf r'} \right] =
$$
$$
 \beta s^2 \det \left(\frac{\hat{v}}\beta\right) \left[\hat{v}_{\bf xx'} + \beta s^2
 \sum_{\bf r,r'} b_{\bf r}^\ast e_{\bf r} e_{\bf r'}^\ast b_{\bf r'}
 \hat{v}_{\bf rx} \hat{v}_{\bf x'r'} \right]
 \exp\left[ \beta s^2\sum_{\bf r,r'} b_{\bf r}^\ast e_{\bf r} 
 \hat{v}_{\bf rr'} e_{\bf r'}^\ast b_{\bf r'} \right] \; .
$$
Since we are interested in the limit ${\bf x-x'}\rightarrow\infty$, 
and the matrix $\hat{J}_{\bf xx'}$
includes nearest-neighbour hopping only, the term $\hat{v}_{\bf xx'}$ vanishes.
This yields for far distant lattice sites $\bf x,x'$ the expression
$$
 \left\langle\Phi^\ast_{\bf x} \Phi_{\bf x'}\right\rangle = s^2 \sum_{\bf r,r'} 
 \left\langle b_{\bf r}^\ast e_{\bf r}
 e_{\bf r'}^\ast b_{\bf r'} \right\rangle\hat{v}_{\bf rx} \hat{v}_{\bf x'r'} \; .
$$
Further we can assume that
$\langle b_{\bf r}^\ast e_{\bf r} e_{\bf r'}^\ast b_{\bf r'} \rangle
= \langle b_{\bf x}^\ast e_{\bf x} e_{\bf x'}^\ast b_{\bf x'} \rangle$ for
$\bf r,x$ and $\bf r',x'$ nearest neighbours. Using
$$
 \sum_{\bf r} \hat{v}_{\bf rx} = \sum_{\bf r'} \hat{v}_{\bf x'r'} = \frac{s+J}{s^2} \; ,
$$
we get
$$
 \lim_{{\bf x-x'}\rightarrow\infty} \langle \Phi^\ast_{\bf x}\Phi_{\bf x'} \rangle =
 \frac{(s+J)^2}{s^2} \lim_{{\bf x-x'}\rightarrow\infty}
 \left\langle b_{\bf x}^\ast e_{\bf x} e_{\bf x'}^\ast b_{\bf x'} \right\rangle
$$
and therefore
$$
 n_0 = \frac{s^2}{(s+J)^2} \lim_{{\bf x-x'}\rightarrow\infty}
 \langle \Phi^\ast_{\bf x} \Phi_{\bf x'} \rangle \; .\vspace{0.7cm}
$$

\subsection{Total particle density\label{Sec_App_SB-ntot}}

The total particle density at site $\bf r$ is given as
\begin{equation}
 n_{\bf r} = 1 - \left\langle |e_{\bf r}|^2 \right\rangle \; ,
 \label{App3_nr}
\end{equation}
where $e$ is the field associated to empty sites. 
It is possible to express the expectation value of the complex field $e$ in terms of an
expectation value of the real field $\varphi$.
To achieve that, let us regard the integration over the fields $b$, $e$, and $\varphi$.
After performing the substitution $\varphi+1/2\rightarrow\varphi$ and dropping the index $\bf r$,
we have
$$
 \int {\rm d}\varphi \, {\rm e}^{-\beta s\left(\varphi-\frac 12\right)^2}
 \int {\cal D}[b,b^\ast,e,e^\ast] \, |e|^2 \exp \left\{ - \beta (e,b) \left(
 \begin{array}{cc} 2s\varphi & s\Phi \\ s\Phi^\ast & -\mu
 \end{array}\right)\left( \begin{array}{c}
 e^\ast \\ b^\ast \end{array}\right) \right\}
$$
$$
 =- \frac 1{2s\beta} \int {\rm d}\varphi \, {\rm e}^{-\beta s\left(\varphi-\frac 12\right)^2}
 \frac\partial{\partial\varphi} \int {\cal D}[b,b^\ast,e,e^\ast] \, 
 \exp \bigg\{ \ldots \bigg\} \; .
$$
Partial integration leads to
$$
 \frac 1{2s\beta} \int {\rm d}\varphi \, \left[-2 \beta s\left(\varphi-\frac 12\right)\right] \,
 {\rm e}^{-\beta s\left(\varphi-\frac 12\right)^2}
 \int {\cal D}[b,b^\ast,e,e^\ast] \, \exp \bigg\{ \ldots \bigg\} \; .
$$
Therefore we find
$$
 \left\langle |e|^2 \right\rangle = \left\langle -\left(\varphi-\frac 12\right) \right\rangle \; .
$$
Together with Eq. (\ref{App3_nr}) we find for the local total particle density the expression
\begin{equation}
 n_{\bf r} = \left\langle \varphi_{\bf r} \right\rangle + \frac 12 \; .
\end{equation}

\subsection{Zero temperature limit\label{Sec_App_SB-T0}}

We want to integrate out the function $Z'$ (we drop the index $\bf r$) given in Eq. 
(\ref{sb1_Zr}) for zero temperature, i.e. in the limit $\beta\rightarrow\infty$.
For simplicity we write $\tilde\beta:=\beta s$ and perform the limit
$\beta\rightarrow\infty$ instead. Further we write $a:=\mu/2s$, and $x:=|\Phi|^2$. 
The function $Z'$ we write as
$$
 Z'=\frac 1{2\tilde\beta} (Z_- - Z_+) \; ,
$$
where
$$
 Z_\pm = \int_{-\infty}^\infty \frac{
 e^{-\tilde\beta f_\pm(\varphi,x)} }{ \sqrt{(\varphi+a)^2+x} } \, {\rm d}\varphi
$$
and
$$
 f_\pm(\varphi,x) = \varphi^2 \pm \sqrt{(\varphi+a)^2+x} \; .
$$
In the limit $\tilde\beta\rightarrow\infty$ we can calculate the $\varphi$-integral $Z_\pm$ 
exactly by means of a saddle-point integration. This is done by expanding the functions
$f_\pm$ in second order about their minimum with respect to $\varphi$. We need partial derivatives
\begin{eqnarray*}
 \frac{\partial f_\pm(\varphi,x)}{\partial\varphi} &=& 2\varphi \pm \frac{\varphi+a}{
   \sqrt{(\varphi+a)^2+x} } \\
 \frac{\partial^2 f_\pm(\varphi,x)}{\partial\varphi^2} &=& 2 \pm
   \frac x{\left[ (\varphi+a)^2+x \right]^{\frac 32} } \; .
\end{eqnarray*}
We determine the extrema of $f_\pm$:
\begin{equation}
 \frac{\partial f_\pm(\varphi_0,x)}{\partial\varphi}=0 
 \quad \Rightarrow \quad \sqrt{(\varphi_0+a)^2+x} = \mp \frac{\varphi_0+a}{2\varphi_0} \; ,
 \label{App3_varphiSP}
\end{equation}
which is equivalent to
\begin{equation}
 x = (\varphi_0+a)^2 \left( \frac 1{4\varphi_0^2} -1 \right) \; .
 \label{App3_varphi0}
\end{equation}
Thus the saddle point approximation for large values of $\tilde\beta$ is
$$
 Z_\pm \approx \int_{-\infty}^\infty \frac{e^{-\tilde\beta \left[ f_\pm(\varphi_0,x) +
 \frac 12 \frac{\partial^2 f_\pm}{\partial\varphi^2}(\varphi_0,x)
 (\varphi-\varphi_0)^2 \right]}}{\sqrt{(\varphi_0+a)^2+x}} \, {\rm d}\varphi 
$$
$$
 = \sqrt{\frac\pi{(\varphi_0+a)^2+x}} \, \frac{e^{-\tilde\beta f_\pm(\varphi_0,x)}}{
 \sqrt{ \frac{\tilde \beta}2 \frac{\partial^2 f_\pm(\varphi_0,x)}{\partial\varphi^2} }} \; .
$$
From Eq. (\ref{App3_varphiSP}) we get
$$
 f_\pm(\varphi_0) = \varphi_0^2 - \frac 12 - \frac a{2\varphi_0} \; ; \;
 \frac{\partial^2 f_\pm(\varphi_0)}{\partial\varphi^2} = 2 - 
 \frac{8x(\varphi_0) \varphi_0^3}{(\varphi_0+a)^3} \; ,
$$
where $x$ itself depends on $\varphi_0$ independently via Eq. (\ref{App3_varphi0}).
For given $x$ there are two solutions for $\varphi_0$, but only the one which is the 
absolute minimum contributes to $Z'$ for large values of $\tilde\beta$. Therefore:
$$
 \log Z' = \log(\varphi_0) - \log(\varphi_0+a) -
 \frac 12 \log\left(\frac{\partial^2 f_\pm(\varphi_0)}{\partial\varphi^2}\right) -
 \tilde\beta f_\pm(\varphi_0) + \mbox{const} \; .
$$
The term that is proportional to $\tilde\beta$ dominates all the others, and in the limit
$\tilde\beta\rightarrow\infty$ we find the {\it exact} result
$$
 \lim_{\tilde\beta\rightarrow\infty} \frac 1{\tilde\beta} \log Z' = - f_\pm(\varphi_0)
$$
$$
 \Rightarrow \quad \lim_{\tilde\beta\rightarrow\infty} \frac 1{\tilde\beta} \, 
 \frac\partial{\partial x} \log Z' =
 - \left[ \frac{{\rm d}f_\pm(\varphi_0)}{{\rm d}\varphi_0} \right] \,
 \frac{{\rm d}{\varphi_0}}{{\rm d} x} \; .
$$
The derivative of $\varphi_0$ with respect to $x$ we get from Eq. (\ref{App3_varphi0}) by
means of the implicit function theorem:
$$
 \frac{{\rm d}\varphi_0}{{\rm d}x} = 
 \frac{-2\varphi_0^3}{(\varphi_0+a)(4\varphi_0^3+a)} \, .
$$
Therefore:
$$
 \lim_{\tilde\beta\rightarrow\infty} \frac 1{\tilde\beta} \, \frac\partial{\partial x} \log Z' =
 \frac{\varphi_0}{\varphi_0+a} \, .
$$
Together with the mean-field equation (\ref{sb1_constant-mean-field}), we find the
zero temperature result in the condensed phase (i.e. where $x>0$):
$$
 \frac{s}{s+J} - \frac{\varphi_0}{\varphi_0+a} = 0 \quad \Rightarrow \quad
 \varphi_0 = \frac\mu{2J} \, .
$$
For the order parameter we find from Eq. (\ref{App3_varphi0}) in the condensed phase:
$$
 |\Phi|^2 = x = \frac 14 \left( \frac{s+J}{Js} \right)^2
 \left( J^2 - \mu^2 \right) \; .
$$
Thus the condensate density by the definition in Eq. (\ref{sb1_n0}) is:
\begin{equation}
 n_0 = \frac{s^2}{(s+J)^2} |\Phi|^2 = \left\{ \begin{array}{l@{\quad}l}
 \frac 14 \left(1-\frac{\mu^2}{J^2}\right) & \mbox{if } -J<\mu<J \\
 0 & \mbox{else} \; , \end{array} \right.
\end{equation}
and because of $\langle\varphi\rangle=\varphi_0$ the total particle density by the definition
(\ref{sb1_nr}) is:
\begin{equation}
 n_{\rm tot} = \varphi_0+\frac 12 = \left\{ \begin{array}{l@{\quad\mbox{if }}l}
 0 & \mu\le -J \\
 \frac 12 \left(1-\frac\mu J\right) & -J<\mu<J \\
 1 & J\le \mu \; . \end{array} \right.
\end{equation}

To determine the coefficient $\tilde{a}_4$ in Eq. (\ref{sb1_a4}), we need the
second derivative of $\log Z$ with respect to $x$:
\begin{eqnarray*}
 \lim_{\tilde\beta\rightarrow\infty} \frac 1{\tilde\beta} \, 
 \frac{\partial^2}{\partial x^2} \log Z' &=&
 \left[ \frac{\rm d}{{\rm d}\varphi} \,
 \lim_{\tilde\beta\rightarrow\infty} \frac 1{\tilde\beta} \, 
 \frac\partial{\partial x} \log Z' \right] \frac{{\rm d} \varphi_0}{{\rm d} x} \; , \\
 &=& \frac 1s \frac{-\mu \varphi_0^3}{(\varphi_0+a)^3 (4\varphi_0^3+a)} \; .
\end{eqnarray*}
With the above solution this yields
\begin{equation}
 \lim_{\beta\rightarrow\infty} \frac 1{\beta} \, 
 \frac{\partial^2}{\partial x^2} \log Z' = 2J \frac{s^4}{(s+J)^4} \left[
 1- 4\, \frac{s}{s+J} \, n_0 \right] \; . \label{App3_a4}
\end{equation}
With these results we also find the zero temperature expressions for the renormalised
coefficients (\ref{sb2_muR}) and (\ref{sb2_gR}):
\begin{equation}
 \mu_{\rm R} = -(s+J)+\frac{(s+J)^2}{s+|\mu|} \; ; \quad
 g_{\rm R} = 2a^3J \; .
 \label{App3_coefficients}
\end{equation}

\bibliographystyle{prsty}
% \bibliographystyle{prsty}
% \nocite{*}
\bibliography{literature}

\end{document}